\documentclass[prx, twocolumn, superscriptaddress, longbibliography]{revtex4-1}
\usepackage{bm, amsmath, amsfonts, amssymb, braket}
\usepackage{multirow}
\usepackage[dvipdfmx]{graphicx}
\usepackage{float}
\usepackage{color}
\usepackage{comment}

\begin{document}

\newcommand{\ii}{\text{i}}
\newcommand{\U}{U}
\newcommand{\V}{V}
\newcommand{\BZ}{\left[ 0, 2\pi \right]}
\newcommand{\CZ}{\left[ 0, 1 \right]}

\title{Higher-order non-Hermitian skin effect}

\author{Kohei Kawabata}
	\email{kawabata@cat.phys.s.u-tokyo.ac.jp}
	\affiliation{Department of Physics, University of Tokyo, 7-3-1 Hongo, Bunkyo-ku, Tokyo 113-0033, Japan}

\author{Masatoshi Sato}
	\email{msato@yukawa.kyoto-u.ac.jp}
	\affiliation{Yukawa Institute for Theoretical Physics, Kyoto University, Kyoto 606-8502, Japan}

\author{Ken Shiozaki}
	\email{ken.shiozaki@yukawa.kyoto-u.ac.jp}
	\affiliation{Yukawa Institute for Theoretical Physics, Kyoto University, Kyoto 606-8502, Japan}

\date{\today}

\begin{abstract}
The non-Hermitian skin effect is a unique feature of non-Hermitian systems, in which an extensive number of boundary modes appear under the open boundary conditions. Here, we discover higher-order counterparts of the non-Hermitian skin effect that exhibit new boundary physics. In two-dimensional systems with the system size $L \times L$, while the conventional (first-order) skin effect accompanies $\mathcal{O}\,( L^{2} )$ skin modes, the second-order skin effect accompanies $\mathcal{O}\,( L )$ corner skin modes. This also contrasts with Hermitian second-order topological insulators, in which only $\mathcal{O}\,( 1 )$ corner zero modes appear. Moreover, for the third-order skin effect in three dimensions, $\mathcal{O}\,( L )$ corner skin modes appear from all $\mathcal{O}\,( L^{3} )$ modes. We demonstrate that the higher-order skin effect originates from intrinsic non-Hermitian topology protected by spatial symmetry. We also show that it accompanies the modification of the non-Bloch band theory in higher dimensions.
\end{abstract}

\maketitle
%%%%% Introduction %%%%%
\section{Introduction}
	\label{sec: introduction}

Topology plays an important role in characterization of phases of matter~\cite{Kane-review, Zhang-review, Schnyder-Ryu-review}. The central principle of topological phases is the bulk-boundary correspondence: boundaries host anomalous gapless modes arising from bulk topology under the open boundary conditions. The number of these boundary modes is $\mathcal{O}\,( L^{d-1} )$ in $d$-dimensional systems with the system size $L^d$. For example, $\mathcal{O}\,( 1 )$ zero-energy modes appear at the two ends of a chiral-symmetric chain such as the Su-Schrieffer-Heeger model~\cite{SSH-79}, and $\mathcal{O}\,( L )$ chiral (helical) modes appear at the edges of the quantum Hall (quantum spin Hall) insulator~\cite{Haldane-88, Kane-Mele-05-QSH, *Kane-Mele-05-Z2}.

Recently, higher-order counterparts of topological phases were revealed and investigated extensively~\cite{BBH-17, *BBH-17B, Langbehn-17, Song-17, Fang-Fu-19, Kunst-18, Schindler-18S, Khalaf-18B, Khalaf-18X, Yan-18, Matsugatani-18, Trifunovic-19, Liu-19, Benalcazar-19}. Higher-order topological phases are protected by spatial symmetry such as inversion, mirror, and rotation symmetry. Importantly, the nature of the bulk-boundary correspondence is changed. In two dimensions, second-order topology leads to $\mathcal{O}\,( 1 )$ zero modes localized at the corners, which is sharply contrasted with $\mathcal{O}\, ( L )$ chiral or helical edge modes accompanied by first-order topology. Similarly, in three dimensions, third-order topology leads to $\mathcal{O}\, ( 1 )$ zero modes localized at the corners, instead of $\mathcal{O}\, ( L^2 )$ surface modes in first-order topological insulators. Higher-order topology was observed in various experiments~\cite{Serra-Garcia-18, Imhof-18, Peterson-18, Schindler-18N, Xue-19, Ni-19, Mittal-19, Hassan-19, Peterson-20}, and may lead to unique phenomena and functionalities due to its boundary physics.

Topological phases and their boundary physics are enriched also by non-Hermiticity~\cite{Ota-review, Bergholtz-review}. In general, non-Hermiticity arises from nonconservation of energy or particles and ubiquitously appears, for example, in nonequilibrium open systems~\cite{Konotop-review, Christodoulides-review}. The interplay of topology and non-Hermiticity gives rise to new physics both in theory~\cite{Rudner-09, Sato-11, *Esaki-11, Hu-11, Schomerus-13, Malzard-15, Lee-16, Leykam-17, Xu-17, Xiong-18, Shen-18, *Kozii-17, Takata-18, MartinezAlvarez-18, Gong-18, *Kawabata-19, YW-18-SSH, *YSW-18-Chern, Kunst-18NH, KSU-18, McDonald-18, Lee-19, Jin-19, Budich-19, Okugawa-19, Liu-19NHSOTI, Yoshida-19, *Kimura-19, Zhou-19, Lee-Li-Gong-19, Ezawa-19, Kunst-19, Edvardsson-19, KSUS-19, ZL-19, Herviou-19, Zhang-19NHSOTI, Zirnstein-19, Borgnia-19, KBS-19, YM-19, Luo-19, McClarty-19, Okuma-19, Song-19-Lindblad, Song-19-real, Bergholtz-19, Rui-19, Schomerus-20, Imura-19, Herviou-19-ES, Chang-20, Zhang-19, OKSS-20, Longhi-20, Li-19, Wojcik-20, Wang-20, Yoshida-20, Scheibner-20, Yokomizo-20, Yi-Yang-20, KOS-20, Terrier-20, Budich-20, McDonald-20, Yu-20, Denner-20} and experiments~\cite{Poli-15, Zeuner-15, Zhen-15, Weimann-17, Xiao-17, St-Jean-17, Parto-17, Bahari-17, Zhao-18, Zhou-18, Harari-18, *Bandres-18, Cerjan-19, Zhao-19, Brandenbourger-19-skin-exp, *Ghatak-19-skin-exp, Helbig-19-skin-exp, *Hofmann-19-skin-exp, Xiao-19-skin-exp, Weidemann-20-skin-exp}. One of the unique features of non-Hermitian systems is the non-Hermitian skin effect. This is the extreme sensitivity of non-Hermitian systems to boundary conditions, and an extensive number of boundary modes appear under the open boundary conditions. In particular, an extensive number of [i.e., $\mathcal{O}\, ( L )$] skin modes appear in one dimension, which is impossible in Hermitian systems. Although the skin effect invalidates the conventional Bloch band theory, researchers formulated a non-Bloch band theory that works even under arbitrary boundary conditions~\cite{YW-18-SSH, YM-19}. Moreover, the skin effect was found to originate from intrinsic non-Hermitian topology~\cite{Zhang-19, OKSS-20}. Symmetry further enriches the skin effect and gives rise to new types of the skin effect originating from symmetry-protected non-Hermitian topology.

Despite the rich physics of non-Hermitian topological systems, little research has hitherto addressed non-Hermitian topological phenomena in higher dimensions. In particular, the non-Hermitian skin effect has been investigated mainly in one dimension. Few exceptions include the skin effect in reciprocal non-Hermitian systems in two dimensions~\cite{OKSS-20}; there, only $\mathcal{O} \left( L \right)$ skin modes appear at edges although the total number of the modes is $\mathcal{O}\,( L^{2} )$. In three dimensions, surface skin modes can have a single exceptional point~\cite{Terrier-20, Denner-20}, which is forbidden in the bulk. Still, the skin effect has remained largely unknown in higher dimensions. Similarly, the non-Bloch band theory in Refs.~\cite{YW-18-SSH, YM-19} is applicable only to one dimension, and its validity in higher dimensions has been unclear.

\begin{figure}[t]
\centering
\includegraphics[width=86mm]{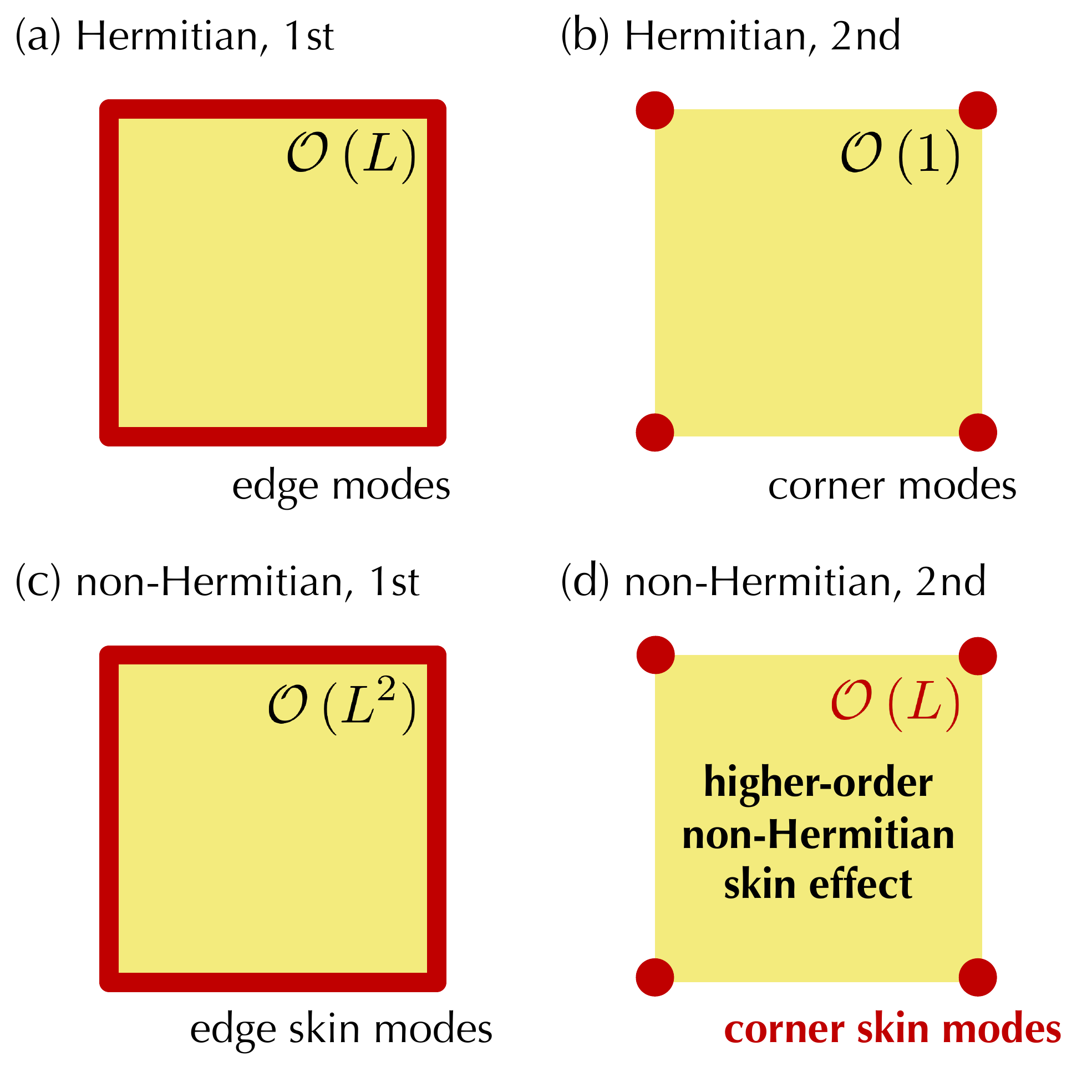} 
\caption{Higher-order non-Hermitian skin effect. Boundary modes (red lines or dots) are shown in a two-dimensional system with the system size $L \times L$. (a)~Hermitian first-order topological insulator. At the edges, $\mathcal{O}\,(L)$ chiral or helical modes appear. (b)~Hermitian second-order topological insulator. At the corners, $\mathcal{O}\,(1)$ zero modes appear. (c)~First-order non-Hermitian skin effect. At arbitrary boundaries, $\mathcal{O}\,(L^2)$ skin modes appear because of intrinsic non-Hermitian topology. (d)~Second-order non-Hermitian skin effect. At the corners, $\mathcal{O}\,(L)$ skin modes appear because of intrinsic non-Hermitian topology.}
	\label{fig: HOSE}
\end{figure}

In this work, we discover higher-order counterparts of the non-Hermitian skin effect. They give rise to new types of boundary modes as a result of higher-order non-Hermitian topology (Fig.~\ref{fig: HOSE}). In two-dimensional systems with the system size $L \times L$ and open boundaries along both directions, the conventional skin effect accompanies $\mathcal{O}\,( L^{2} )$ skin modes at arbitrary boundaries [Fig.~\ref{fig: HOSE}\,(c)]. For the second-order skin effect, by contrast, $\mathcal{O}\,( L )$ skin modes appear at the corners [Fig.~\ref{fig: HOSE}\,(d)]. This is also distinct from Hermitian second-order topological insulators, in which only $\mathcal{O}\,( 1 )$ corner modes appear as a result of Hermitian topology [Fig.~\ref{fig: HOSE}\,(b)]. We demonstrate that the higher-order skin effect cannot be described by the conventional non-Bloch band theory, which implies its inevitable modification in higher dimensions.

Notably, the higher-order non-Hermitian skin effect is distinct from non-Hermitian extensions of higher-order topological insulators~\cite{Liu-19NHSOTI, Lee-Li-Gong-19, Edvardsson-19, Zhang-19NHSOTI, Luo-19, Yu-20}. There, even in the presence of non-Hermiticity, the corner modes have the same topological nature as the Hermitian counterparts. Consequently, the number of these corner modes is $\mathcal{O}\,( 1 )$. Moreover, the other modes typically exhibit the first-order skin effect and are also localized at boundaries.  For the second-order skin effect, by contrast, the corner skin modes originate from intrinsic non-Hermitian topology that has no counterparts in Hermitian systems. Almost all the $\mathcal{O}\,( L^2 )$ modes are delocalized through the bulk, and only $\mathcal{O}\,( L )$ skin modes appear at the corners.

This work is organized as follows. In Sec.~\ref{sec: FOSE}, we review the conventional (first-order) skin effect. In Sec.~\ref{sec: SOSE}, we consider the second-order skin effect. In Sec.~\ref{sec: SOSE-model}, we introduce a model exhibiting the second-order skin effect on the basis of a Hermitian second-order topological insulator. The spectrum and the eigenstates are investigated in Sec.~\ref{sec: SOSE-spectrum}. Then, in Sec.~\ref{sec: SOSE-WZ}, we identify the topological invariant for the second-order skin effect as the Wess-Zumino term protected by spatial symmetry. We discuss the implications for the non-Bloch band theory in Sec.~\ref{sec: SOSE-band}. Furthermore, in Sec.~\ref{sec: TOSE}, we investigate the third-order skin effect. We conclude this work in Sec.~\ref{sec: conclusion}.

%%%%%%%%%%
\section{First-order non-Hermitian skin effect}
	\label{sec: FOSE}

We begin with reviewing the conventional non-Hermitian skin effect that has the first-order nature. It accompanies the emergence of an extensive number of skin modes localized at arbitrary boundaries; $\mathcal{O}\,( L^{d} )$ skin modes appear in $d$ dimensions. Such anomalous boundary modes are unique to non-Hermitian systems and originate from intrinsic non-Hermitian topology. This sharply contrasts with Hermitian systems, in which the bulk is insensitive to boundary conditions, and there appear $\mathcal{O}\,( L^{d-1} )$ boundary modes under the open boundary conditions.

%%%%%%%%%%
\subsection{Hatano-Nelson model}
	\label{sec: Hatano-Nelson}

A prototypical model that exhibits the first-order skin effect is the Hatano-Nelson model~\cite{Hatano-Nelson-96, *Hatano-Nelson-97}:
\begin{equation}
\hat{H}_{\rm HN} = \sum_{n} \left[ \left( t-g \right) \hat{c}_{n+1}^{\dag} \hat{c}_{n} + \left( t+g \right) \hat{c}_{n}^{\dag} \hat{c}_{n+1} \right],
\end{equation}
where $t, g \in \mathbb{R}$ are the hopping amplitudes, and $\hat{c}_{n}$ ($\hat{c}_{n}^{\dag}$) annihilates (creates) a particle on site $n$. We assume $t \geq g \geq 0$ for simplicity. The corresponding Bloch Hamiltonian reads
\begin{eqnarray}
H_{\rm HN} \left( k \right) 
&=& \left( t-g \right) e^{-\ii k} + \left( t+g \right) e^{\ii k} \nonumber \\
&=& 2t \cos k + 2\ii g \sin k.
	\label{eq: Hatano-Nelson-Bloch}
\end{eqnarray}
Under the periodic boundary conditions, the system is described by $H_{\rm HN} \left( k \right)$ with real wavenumbers $k \in \BZ$. The spectrum forms a loop in the complex-energy plane, and the eigenstates are delocalized through the bulk. 

Under the open boundary conditions, by contrast, the system is no longer described by $H_{\rm HN} \left( k \right)$. To understand this, let us consider the following similarity transformation (imaginary gauge transformation~\cite{Hatano-Nelson-96, *Hatano-Nelson-97}):
\begin{equation}
\hat{V}_{r}^{-1} \hat{c}_{i} \hat{V}_{r} = r^{-i} \hat{c}_{i},\quad
\hat{V}_{r}^{-1} \hat{c}_{i}^{\dag} \hat{V}_{r} = r^{i} \hat{c}_{i}^{\dag}
\end{equation}
for $r \in \left( 0, \infty \right)$. The Hamiltonian $\hat{H}_{\rm HN}$ transforms into
\begin{equation}
\hat{V}_{r}^{-1} \hat{H}_{\rm HN} \hat{V}_{r} = \sum_{n} \left[ r \left( t-g \right) \hat{c}_{n+1}^{\dag} \hat{c}_{n} + \frac{t+g}{r} \hat{c}_{n}^{\dag} \hat{c}_{n+1} \right].
\end{equation}
In particular, using $r_{\times} := \sqrt{\left( t+g\right)/\left( t-g\right)}$, we have
\begin{equation}
\hat{V}_{r_{\times}}^{-1} \hat{H}_{\rm HN} \hat{V}_{r_{\times}} = \sqrt{t^{2}-g^{2}}\sum_{n} \left( \hat{c}_{n+1}^{\dag} \hat{c}_{n} + \hat{c}_{n}^{\dag} \hat{c}_{n+1} \right),
\end{equation}
which is Hermitian. Importantly, this transformation does not change the spectrum since it is a similarity transformation under the open boundary conditions. Hence, the non-Hermitian Hamiltonian $\hat{H}_{\rm HN}$ with open boundaries has the same spectrum as the Hermitian Hamiltonian $\hat{V}_{r_{\times}}^{-1} \hat{H}_{\rm HN} \hat{V}_{r_{\times}}$, which is given as
\begin{equation}
E \left( k \right) = 2 \sqrt{t^{2}-g^{2}} \cos k,\quad
k  \in \BZ.
\end{equation}
The spectrum lies on the real axis in the complex-energy plane. Since $\hat{V}_{r_{\times}}^{-1} \hat{H}_{\rm HN} \hat{V}_{r_{\times}}$ has delocalized eigenstates, all the eigenstates of $\hat{H}_{\rm HN}$ are localized at the left edge as $\sim e^{-n/\xi}$ with the localization length $\xi = \left( \log r_{\times} \right)^{-1}$. Clearly, the spectrum and the eigenstates of the bulk are dramatically sensitive to the boundary conditions, which is impossible in Hermitian systems. This is the non-Hermitian skin effect in the Hatano-Nelson model. 

In a similar manner, the skin effect generally occurs in non-Hermitian systems. In $d$ dimensions, the skin modes can appear at arbitrary boundaries including edges and corners. Still, $\mathcal{O}\,( L^{d} )$ skin modes usually accompany the first-order skin effect. However, different types of skin effects can occur in the presence of symmetry or in higher dimensions. In particular, $\mathcal{O}\,( L^{d-1} )$ [$\mathcal{O}\,( L^{d-2} )$] skin modes accompany the second-order (third-order) skin effect, which we focus on in this work.

%%%%%%%%%%
\subsection{Non-Hermitian topology}
	\label{sec: FOSE-topology}

The skin effect originates from intrinsic non-Hermitian topology~\cite{Zhang-19, OKSS-20}. In one dimension, the topological invariant is given as a winding number $W \left( E \right) \in \mathbb{Z}$ defined for complex energy $E \in \mathbb{C}$ and the Bloch Hamiltonian $H \left( k \right)$~\cite{Gong-18, KSUS-19}:
\begin{equation}
W \left( E \right) := \oint_{0}^{2\pi} \frac{dk}{2\pi\ii} \frac{d}{dk} \log \det \left[ H \left( k \right) - E \right].
	\label{eq: winding}
\end{equation}
This topological invariant is well defined as long as the spectrum of $H \left( k \right)$ does not cross given $E$ [i.e., $H \left( k \right)$ is point-gapped in terms of a reference point $E$~\cite{Gong-18, KSUS-19}]. If $W \left( E\right)$ is nonzero, the skin effect occurs; otherwise, no skin effect occurs.

The non-Hermitian topology of $H \left( k \right)$ can also be understood on the basis of the extended Hermitian Hamiltonian
\begin{equation}
\tilde{H} \left( k, E \right) := \begin{pmatrix}
0 & H \left( k \right) - E \\
H^{\dag} \left( k \right) - E^{*} & 0
\end{pmatrix}.
	\label{eq: extended-Hermitian}
\end{equation}
By construction, $\tilde{H} \left( k, E \right)$ respects chiral symmetry
\begin{equation}
\sigma_{z} \tilde{H} \left( k, E \right) \sigma_{z}^{-1} 
= - \tilde{H} \left( k, E \right)
\end{equation}
with a Pauli matrix $\sigma_{z}$. If the non-Hermitian Hamiltonian $H \left( k \right)$ is topologically nontrivial for $E$ and the skin effect occurs, the extended Hermitian Hamiltonian $\tilde{H} \left( k, E \right)$ is also topologically nontrivial and has zero-energy edge modes under the open boundary conditions.

For the Hatano-Nelson model, we have $W \left( E \right) = \mathrm{sgn} \left( g \right)$ as long as $E$ is inside the loop described by Eq.~(\ref{eq: Hatano-Nelson-Bloch}). The extended Hermitian Hamiltonian in Eq.~(\ref{eq: extended-Hermitian}) is similar to the Su-Schrieffer-Heeger model~\cite{SSH-79}. The skin modes in the Hatano-Nelson model correspond to a pair of zero modes in the Su-Schrieffer-Heeger model. 

Importantly, the topological invariant $W \left( E \right)$ is intrinsic to non-Hermitian systems~\cite{Gong-18, KSUS-19}. In fact, without symmetry protection, no topological invariant is well defined in Hermitian systems in one dimension~\cite{Kane-review, Zhang-review, Schnyder-Ryu-review}. Such intrinsic non-Hermitian topology is the origin of the skin effect, which is also an intrinsic non-Hermitian topological phenomenon. This sharply contrasts with topologically-protected boundary modes in Hermitian systems.

In the presence of symmetry, different types of topological invariants can be defined, and consequently, different types of skin effects can occur. In one-dimensional systems with symplectic reciprocity, for example, a $\mathbb{Z}_{2}$ topological invariant is well defined, although the $\mathbb{Z}$ invariant in Eq.~(\ref{eq: winding}) vanishes~\cite{KSUS-19}. In contrast to the conventional skin effect, the nontrivial $\mathbb{Z}_{2}$ topological invariant leads to the reciprocal skin effect~\cite{OKSS-20}. There, some skin modes are localized at one end and other skin modes are localized at the other end, both of which form Kramers pairs.

In higher dimensions, $W \left( E \right)$ can still be well defined as a weak topological invariant. In two dimensions, for example, $W \left( E \right)$ can be obtained by $H \left( k_{x}, k_{y} \right)$ for each $k_{x}$ or $k_{y}$ [see Eq.~(\ref{eq:2d_1dWNs}) for details]. In contrast to the weak topological invariants, strong topological invariants in higher dimensions can result in different types of skin effects that are unique to higher-dimensional systems. For example, in reciprocal non-Hermitian systems in two dimensions, a $\mathbb{Z}_{2}$ topological invariant is well defined in terms of both $k_{x}$ and $k_{y}$~\cite{KSUS-19}. If this invariant is nontrivial, $\mathcal{O}\,(L)$ skin modes appear at the edges~\cite{OKSS-20}. The three-dimensional winding number also results in a new skin effect in three dimensions~\cite{Terrier-20, Denner-20}. Similarly, the higher-order skin effect is a new type of skin effects. In contrast to the skin effects in Refs.~\cite{OKSS-20, Terrier-20, Denner-20}, the higher-order skin effect is characterized by spatial-symmetry-protected higher-order topology, as discussed in Sec.~\ref{sec: SOSE-WZ}.

%%%%%%%%%%
\subsection{Non-Bloch band theory}

Because of the non-Hermitian skin effect, the conventional Bloch band theory is not generally applicable in non-Hermitian systems. In fact, the Bloch band theory works only under the periodic boundary conditions, and the skin effect invalidates it under the open boundary conditions. To overcome this difficulty, recent works have developed a non-Bloch band theory that works even under the open boundary conditions~\cite{YW-18-SSH, YM-19}.

The non-Bloch band theory is formulated as follows. Let $\beta_{i}$'s ($i = 1, 2, \cdots, 2M$; $\left| \beta_{1} \right| \leq \left| \beta_{2} \right| \leq \cdots \leq \left| \beta_{2M} \right|$) be the solutions to the characteristic equation $\det \left[ H \left( \beta \right) - E \right] = 0$ for a given eigenenergy $E \in \mathbb{C}$. Here, the bulk Hamiltonian $H \left( \beta \right)$ is obtained by replacing $k$ with $\beta := e^{\ii k}$ for the Bloch Hamiltonian $H \left( k \right)$. Then, the bulk bands are formed by $H \left( \beta \right)$ with the trajectory of $\beta_{M}$ and $\beta_{M+1}$ satisfying
\begin{equation}
\left| \beta_{M} \right| = \left| \beta_{M+1} \right|.
	\label{eq: Yao-Wang-Yokomizo-Murakami}
\end{equation}

For example, in the Hatano-Nelson model, the bulk Hamiltonian reads
\begin{equation}
H \left( \beta \right) = \left( t-g \right) \beta^{-1} + \left( t+g \right) \beta.
\end{equation}
The characteristic equation $\det \left[ H \left( \beta \right) - E \right] = 0$ forms the quadratic equation
\begin{equation}
\left( t+g \right) \beta^{2} - E\beta + t-g = 0.
\end{equation}
Since $\beta_{1}$ and $\beta_{2}$ are the two solutions to this quadratic equation, we have
\begin{equation}
\beta_{1} + \beta_{2} = \frac{E}{t+g},\quad
\beta_{1} \beta_{2} = \frac{t-g}{t+g}.
\end{equation}
Then, the condition~(\ref{eq: Yao-Wang-Yokomizo-Murakami}) leads to
\begin{equation}
\left| \beta_{1} \right| = \left| \beta_{2} \right| = \sqrt{\frac{t-g}{t+g}} = r_{\times}^{-1},
\end{equation}
which reproduces the skin modes in Sec.~\ref{sec: Hatano-Nelson}.

Notably, the above non-Bloch band theory can break down in the presence of symmetry. For example, it is modified in the symplectic class~\cite{Yi-Yang-20, KOS-20}, which accounts for the $\mathbb{Z}_{2}$ reciprocal skin effect~\cite{OKSS-20}. Furthermore, the non-Bloch band theory is not directly applicable if the open boundary conditions are imposed more than one direction. Thus, the non-Bloch band theory can be modified in higher dimensions. In Sec.~\ref{sec: SOSE-band}, we demonstrate that such modification in higher dimensions indeed arises and underlies the higher-order skin effect.

%%%%%%%%%%
\section{Second-order non-Hermitian skin effect}
	\label{sec: SOSE}

For the conventional non-Hermitian skin effect discussed in the preceding section, an extensive number of eigenstates are localized at boundaries. More precisely, $\mathcal{O}\,( L^{d} )$ skin modes appear in $d$-dimensional systems with the system size $L^{d}$. For the higher-order non-Hermitian skin effect, by contrast, most of the eigenstates remain delocalized and form bulk bands. Still, a part of the eigenstates exhibit the skin effect. For the second-order skin effect in two dimensions, which we focus on in the present section, $\mathcal{O}\,( L^{2} )$ bulk modes and $\mathcal{O} \left( L \right)$ corner skin modes simultaneously appear in a two-dimensional system with the system size $L \times L$. This also contrasts with Hermitian second-order topological insulators, in which $\mathcal{O} \left( 1 \right)$ corner modes appear.

In Sec.~\ref{sec: SOSE-model}, we introduce a non-Hermitian model in two dimensions that exhibits the second-order skin effect [Eq.~(\ref{eq: NH-BBH-2D})]. This model is systematically constructed on the basis of a Hermitian second-order topological insulator~\cite{BBH-17}. The spectra and the wavefunctions of this system are investigated in Sec.~\ref{sec: SOSE-spectrum}. Then, in Sec.~\ref{sec: SOSE-WZ}, we identify the topological origin of the second-order non-Hermitian skin effect as the Wess-Zumino term~\cite{WZ-71}. This topological invariant is protected by four-fold-rotation-type symmetry in Eqs.~(\ref{eq: NH-BBH-C4-rotation}) and (\ref{eq:C4Sym}). Remarkably, the second-order skin effect requires modification of the non-Bloch band theory, as demonstrated in Sec.~\ref{sec: SOSE-band}.

%%%%%%%%%%
\subsection{Model and symmetry}
	\label{sec: SOSE-model}
	
\begin{figure*}[t]
\centering
\includegraphics[width=172mm]{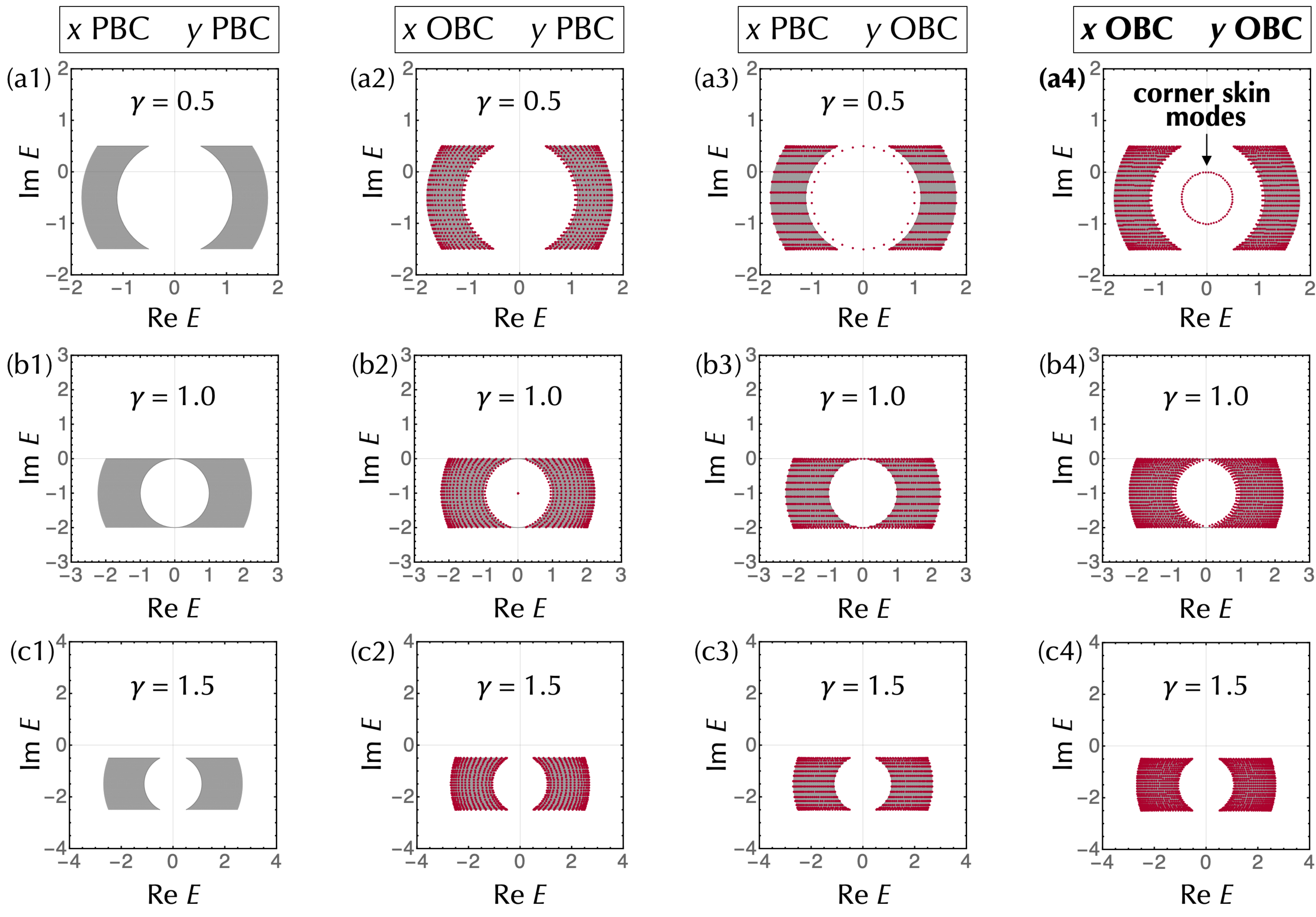} 
\caption{Second-order non-Hermitian skin effect. The complex spectra of the non-Hermitian model in two dimensions [Eq.~(\ref{eq: NH-BBH-2D})] are shown for $30 \times 30$ sites. The parameters are given as $\lambda = 1.0$, as well as (a1, a2, a3, a4)~$\gamma = 0.5$, (b1, b2, b3, b4) $\gamma = 1.0$, or (c1, c2, c3, c4)~$\gamma = 1.5$. The open boundary conditions are imposed along none of the directions for (a1, b1, c1), only along the $x$ direction for (a2, b2, c2), only along the $y$ direction for (a3, b3, c3), and both of the directions for (a4, b4, c4). The spectra for the periodic boundary conditions are shown as the grey regions, while the spectra for the open boundary conditions are shown as the red dots. For $\left| \gamma/\lambda \right| < 1$, the corner skin modes appear under the open boundary conditions along all the directions, as shown in (a4). The spectrum of these corner skin modes is given as $E = -\ii \gamma \left( 1 + e^{\ii \theta} \right)$ with $\theta \in \BZ$.}
	\label{fig: spectrum-SOSE}
\end{figure*}

We provide a model that exhibits the second-order non-Hermitian skin effect. The Bloch Hamiltonian reads	
\begin{eqnarray}
&&H \left( \bm{k} \right) = -\ii \left( \gamma + \lambda \cos k_{x} \right) + \lambda \left( \sin k_{x} \right) \sigma_{z} \nonumber \\
&&\qquad\qquad\quad + \left( \gamma + \lambda \cos k_{y} \right) \sigma_{y} + \lambda \left( \sin k_{y} \right) \sigma_{x},
	\label{eq: NH-BBH-2D}
\end{eqnarray}
where $\gamma$ and $\lambda$ are real parameters, and $\sigma_{i}$'s ($i = x, y, z$) are Pauli matrices. As discussed in Sec.~\ref{sec: FOSE-topology}, the Hatano-Nelson model is closely related to the Su-Schrieffer-Heeger model. Similarly, this model is constructed on the basis of a Hermitian second-order topological insulator. In fact, the extended Hermitian Hamiltonian is given as
\begin{eqnarray}
\tilde{H}_{\rm BBH} \left( \bm{k} \right)
&=& \begin{pmatrix}
0 & H \left( \bm{k} \right) \\
H^{\dag} \left( \bm{k} \right) & 0 
\end{pmatrix} \nonumber \\
&=& \left( \gamma + \lambda \cos k_{x} \right) \tau_{y} + \lambda \left( \sin k_{x} \right) \sigma_{z} \tau_{x} \nonumber \\
&&~+ \left( \gamma + \lambda \cos k_{y} \right) \sigma_{y} \tau_{x}  + \lambda \left( \sin k_{y} \right) \sigma_{x} \tau_{x},\qquad
\end{eqnarray}
where $\tau_{i}$'s ($i = x, y, z$) are Pauli matrices that describe the additional degrees of freedom. This Hermitian Hamiltonian is a prototypical model of a second-order topological insulator that was first introduced by Benalcazar, Bernevig, and Hughes~\cite{BBH-17}. There, no edge modes appear under the open boundary conditions solely along one direction. Nevertheless, under the open boundary conditions along both directions, zero-energy modes appear at the corners for $\left| \gamma/\lambda \right| < 1$.

Spatial symmetry plays a crucial role in the second-order topological phase of $\tilde{H}_{\rm BBH} \left( \bm{k} \right)$ and the second-order non-Hermitian skin effect of $H \left( \bm{k} \right)$. First, both $\tilde{H}_{\rm BBH} \left( \bm{k} \right)$ and $H \left( \bm{k} \right)$ respect spatial-inversion (parity) symmetry:
\begin{eqnarray}
\sigma_{y} \tilde{H}_{\rm BBH} \left( \bm{k} \right) \sigma_{y}^{-1} &=& \tilde{H}_{\rm BBH} \left( - \bm{k} \right),\\
\sigma_{y} H \left( \bm{k} \right) \sigma_{y}^{-1} &=& H \left( - \bm{k} \right).
	\label{eq: NH-BBH-inversion}
\end{eqnarray}
In addition, $\tilde{H}_{\rm BBH} \left( \bm{k} \right)$ respects mirror symmetry:
\begin{eqnarray}
\left( \sigma_{z} \tau_{y} \right) \tilde{H}_{\rm BBH} \left( k_{x}, k_{y} \right) \left( \sigma_{z} \tau_{y} \right)^{-1} 
&=& \tilde{H}_{\rm BBH} \left( -k_{x}, k_{y} \right),\quad\\
\left( \sigma_{x} \tau_{y} \right) \tilde{H}_{\rm BBH} \left( k_{x}, k_{y} \right) \left( \sigma_{x} \tau_{y} \right)^{-1} 
&=& \tilde{H}_{\rm BBH} \left( k_{x}, -k_{y} \right).\quad
\end{eqnarray}
Correspondingly, $H \left( \bm{k} \right)$ respects
\begin{eqnarray}
\sigma_{z} H^{\dag} \left( k_{x}, k_{y} \right) \sigma_{z}^{-1} 
&=& - H \left( -k_{x}, k_{y} \right),
    \label{eq: NH-BBH-mirror-x} \\
\sigma_{x} H^{\dag} \left( k_{x}, k_{y} \right) \sigma_{x}^{-1} 
&=& -H \left( k_{x}, -k_{y} \right).
    \label{eq: NH-BBH-mirror-y}
\end{eqnarray}
They also respect the following transposition-associated mirror symmetry
\begin{eqnarray}
\sigma_{x} \tilde{H}_{\rm BBH}^{T} \left( k_{x}, k_{y} \right) \sigma_{x}^{-1} &=& \tilde{H}_{\rm BBH} \left( - k_{x}, k_{y} \right), \\
\sigma_{z} \tilde{H}_{\rm BBH}^{T} \left( k_{x}, k_{y} \right) \sigma_{z}^{-1} &=& \tilde{H}_{\rm BBH} \left( k_{x}, -k_{y} \right),
\end{eqnarray}
and
\begin{eqnarray}
\sigma_{x} H^{T} \left( k_{x}, k_{y} \right) \sigma_{x}^{-1} &=& H \left( - k_{x}, k_{y} \right), 
    \label{eq: NH-BBH-transposition-x} \\
\sigma_{z} H^{T} \left( k_{x}, k_{y} \right) \sigma_{z}^{-1} &=& H \left( k_{x}, -k_{y} \right).
	\label{eq: NH-BBH-transposition-y}
\end{eqnarray}
The combination of Eqs.~(\ref{eq: NH-BBH-mirror-x}) and (\ref{eq: NH-BBH-mirror-y}), or the combination of Eqs.~(\ref{eq: NH-BBH-transposition-x}) and (\ref{eq: NH-BBH-transposition-y}) reduces to Eq.~(\ref{eq: NH-BBH-inversion}). As shown in Sec.~\ref{sec: SOSE-band}, the symmetry in Eqs.~(\ref{eq: NH-BBH-transposition-x}) and (\ref{eq: NH-BBH-transposition-y}) vanishes the first-order skin effect in $H \left( \bm{k} \right)$ along the $x$ and $y$ directions, respectively. Furthermore, $\tilde{H}_{\rm BBH} \left( \bm{k} \right)$ respects four-fold-rotation symmetry:
\begin{equation}
\mathcal{R}_{4} \tilde{H}_{\rm BBH} \left( k_{x}, k_{y} \right) \mathcal{R}^{-1}_{4}
= \tilde{H}_{\rm BBH} \left( - k_{y}, k_{x} \right),
\end{equation}
where $\mathcal{R}_{4}$ is a unitary matrix given as
\begin{equation}
\mathcal{R}_{4} = \begin{pmatrix}
0 & -\ii \sigma_{y} \\
1 & 0
\end{pmatrix}.
\end{equation}
Correspondingly, $H \left( \bm{k} \right)$ respects
\begin{equation}
-\ii \sigma_{y} H^{\dag} \left( k_{x}, k_{y} \right) = H \left( - k_{y}, k_{x} \right)
	\label{eq: NH-BBH-C4-rotation}
\end{equation}
This rotation-type symmetry protects the second-order skin effect, as shown in Sec.~\ref{sec: SOSE-WZ}.

%%%%%%%%%%
\subsection{Corner skin effect}
	\label{sec: SOSE-spectrum}

\begin{figure}[t]
\centering
\includegraphics[width=86mm]{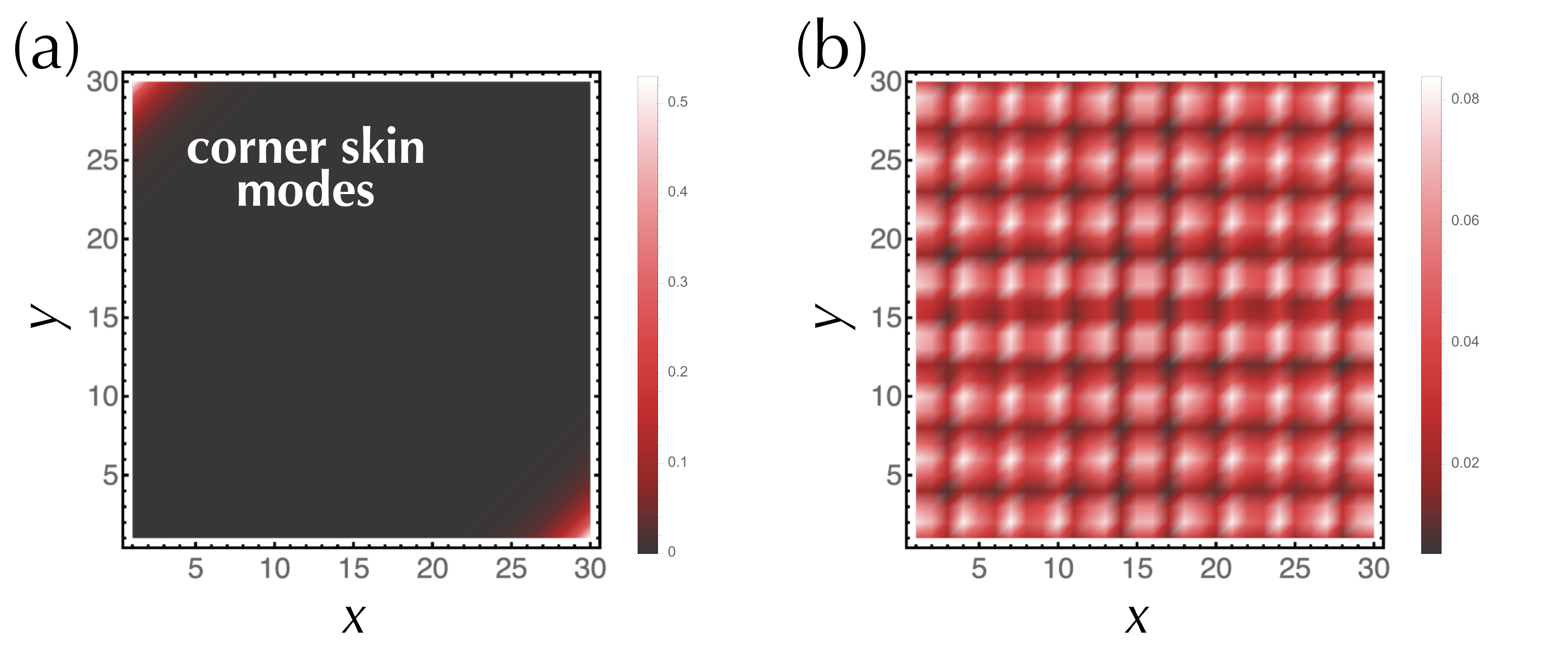} 
\caption{Wavefunctions for the second-order non-Hermitian skin effect. Under the open boundary conditions along both $x$ and $y$ directions, eigenstates of the non-Hermitian model in two dimensions [Eq.~(\ref{eq: NH-BBH-2D})] are shown for $L = 30$, $\gamma = 0.5$, and $\lambda = 1.0$. (a)~Corner skin modes ($E=-0.027-0.0008\ii$). (b)~Delocalized bulk modes ($E = -1.64 - 0.94\ii$).}
	\label{fig: wavefunction-SOSE}
\end{figure}

We numerically obtain the complex spectrum of the non-Hermitian model under various boundary conditions, as shown in Fig.~\ref{fig: spectrum-SOSE}. Under the periodic boundary conditions, eigenstates are delocalized through the bulk and form two bands [Fig.~\ref{fig: spectrum-SOSE}\,(a1, b1, c1)]; the bulk spectrum is given as
\begin{eqnarray}
&&E \left( \bm{k} \right) = \pm \sqrt{\lambda^2 \sin^2 k_{x} + \left( \gamma + \lambda \cos k_{y} \right)^2 + \lambda^2 \sin^2 k_{y}} \nonumber \\
&&\qquad\qquad\qquad\qquad\qquad\qquad\quad-\ii \left( \gamma + \lambda \cos k_{x} \right).
\end{eqnarray}
The complex-energy gap between the two bands is closed at $\left| \gamma \right| = \left| \lambda \right|$. Similarly, under the open boundary conditions solely along the $x$ direction [Fig.~\ref{fig: spectrum-SOSE}\,(a2, b2, c2)] or solely along the $y$ direction [Fig.~\ref{fig: spectrum-SOSE}\,(a3, b3, c3)], no skin effect occurs, in general. This corresponds to the absence of zero modes in $\tilde{H}_{\rm BBH}$ under these boundary conditions.

Under the open boundary conditions in both directions, by contrast, skin modes appear for $\left| \gamma \right| < \left| \lambda \right|$ [Fig.~\ref{fig: spectrum-SOSE}\,(a4)]. These skin modes are not included in the bulk spectrum and localized at boundaries. In particular, the skin modes are localized at the corners, while the other bulk modes are delocalized (Fig.~\ref{fig: wavefunction-SOSE}). From all the $2L^2$ eigenstates, the number of the corner skin modes is $2L$, while the number of the delocalized bulk modes is $2L \left( L-1 \right)$. Notably, the skin spectrum forms a loop in the complex-energy plane even under the open boundary conditions, which is forbidden for the conventional skin effect~\cite{Zhang-19, OKSS-20}.

This model can be solved also in an analytical manner (see Appendix~\ref{appendix: SOSE} for details). In particular, for sufficiently large $L$, the spectrum of the corner skin modes is given as
\begin{equation}
E = -\ii \gamma \left( 1 + e^{\ii \theta} \right),\quad
\theta \in \BZ,
\end{equation}
and their localization lengths $\xi_{x}$ and $\xi_{y}$ along the $x$ and $y$ directions are given as
\begin{equation}
\xi_{x} = \xi_{y} = \left( \log \left| \frac{\lambda}{\gamma} \right| \right)^{-1}.
    \label{eq: localization-length}
\end{equation}
These analytical results are consistent with the numerical results.

The corner skin modes are a new type of boundary modes unique to non-Hermitian systems in higher dimensions. They are distinct from both $\mathcal{O}\,( L^2 )$ skin modes for the conventional skin effect and $\mathcal{O}\,( 1 )$ corner modes in Hermitian second-order topological insulators. We call this new type of the skin effect the second-order skin effect. It originates from second-order non-Hermitian topology protected by spatial symmetry, as shown in Sec.~\ref{sec: SOSE-WZ}.

Notably, Ref.~\cite{Lee-Li-Gong-19} provided another non-Hermitian model in two dimensions that exhibits corner skin modes. Similarly to our model, the $\mathcal{O}\,( L )$ skin modes are localized at the corners, while the other $\mathcal{O}\,( L^{2} )$ modes are delocalized through the bulk. However, this model is characterized by the bulk Chern number, and under the open boundary conditions, there appear chiral edge modes closing the line gap. Non-Hermiticity further pushes these chiral edge modes to the corners, resulting in the corner skin modes. Thus, the corner skin modes in Ref.~\cite{Lee-Li-Gong-19} arise from the combination of Hermitian topology (i.e., Chern number) and non-Hermiticity. On the other hand, the Chern number vanishes for our model in Eq.~(\ref{eq: NH-BBH-2D}). Instead, the corner skin modes in our model are characterized solely by intrinsic non-Hermitian topology in terms of a point gap, as discussed in Sec.~\ref{sec: SOSE-WZ}.

We also note in passing that different types of boundary modes appear for other boundary conditions. Even under the periodic boundary conditions along the $x$ direction, there appear $\mathcal{O} \left( L \right)$ modes away from the bulk bands as long as the open boundary conditions are imposed along the $y$ direction, as shown in Fig.~\ref{fig: spectrum-SOSE}\,(a3). These modes are localized at both edges with the localization length $\xi_{y}$ in Eq.~(\ref{eq: localization-length}) (see Appendix~\ref{appendix: edge-OBCy} for details), and their spectrum is given as
\begin{equation}
    E = - \ii \gamma - \ii \lambda e^{-\ii k_{x}}.
\end{equation}
These edge modes do not touch the bulk bands, which contrasts with chiral edge states in Chern insulators. On the other hand, even if the open boundary conditions are imposed along the $x$ direction, no edge modes generally appear under the periodic boundary conditions along the $y$ direction. Even under these boundary conditions, boundary modes appear for $\left| \gamma/\lambda  \right| = 1$ [Fig.~\ref{fig: spectrum-SOSE}\,(b2)]. Anomalously, they belong to the same wavenumber $k_{y} = \pi$ and the same eigenenergy $E = -\ii \gamma$, and form exceptional points; $2L$ eigenstates coalesce into only $2$ eigenstates, one of which is localized at the left edge and the other of which is localized at the right edge (see Appendix~\ref{appendix: edge-OBCx} for details). The relationship between these boundary modes and the corner skin modes may merit further investigation.

%%%%%%%%%%
\subsection{Wess-Zumino term}
	\label{sec: SOSE-WZ}
	
The second-order non-Hermitian skin effect originates from the $\mathbb{Z}_2$-quantized Wess-Zumino (WZ) term introduced shortly. As discussed in Sec.~\ref{sec: FOSE-topology}, in two dimensions, we can define the one-dimensional winding numbers 
\begin{align}
W_{i} := \oint_0^{2\pi} \frac{dk_{i}}{2\pi\ii} \frac{\partial}{\partial k_{i}} \log \det \left[ H \left( k_x, k_y \right) \right],\quad i =x,y, 
	\label{eq:2d_1dWNs}
\end{align}
along the $x$ and $y$ directions, respectively. As shown below, given a non-Hermitian Hamiltonian $H \left( k_x, k_y \right)$ which is invertible and has no one-dimensional winding numbers $W_x=W_y=0$, we can define a geometric quantity ${\rm WZ} \left[ H \right]$ called the WZ term which takes a value in the circle $\CZ$. The absence of the one-dimensional winding numbers ensures the existence of a smooth path of invertible Hamiltonians $H \left( k_x, k_y, t \right)$ from the original one
\begin{equation}
H \left( k_x, k_y, t=0 \right) := H \left( k_x, k_y \right) 
\end{equation}
to another constant one 
\begin{equation}
H \left( k_x, k_y, t=1 \right) := H_{\rm const} 
\end{equation}
at the end. The WZ term is defined by~\cite{WZ-71}
\begin{align}
{\rm WZ} \left[ H \right] := \frac{1}{24\pi^2} \oint_{\BZ^{2} \times \CZ} \mathrm{tr} \left[ H^{-1} dH \right]^3.
	\label{eq:WZterm}
\end{align}
While the WZ term is a real number, it is not quantized in the absence of symmetry. 

Although the extension $H \left( k_x, k_y \right) \to H \left( k_x, k_y, t \right)$ is not unique, the difference ${\rm WZ} \left[ H \right] - {\rm WZ} \left[ H' \right]$ between the two extensions $H \left( k_x, k_y, t \right)$ and $H' \left( k_x, k_y, t \right)$ is nothing but the integer-valued three-dimensional winding number of the third homotopy class $\pi_3 \left( \mathrm{GL}_N \left( \mathbb{C} \right) \right) = \mathbb{Z}$, where $N\geq 2$ is the dimension of the matrix $H \left( k_x, k_y \right)$. Thus, the WZ term in Eq.~\eqref{eq:WZterm} does not depend on extensions of $H \left( k_x, k_y \right)$ as a quantity in the circle $\CZ$. It is a two-dimensional analog of the Berry-phase formula of the electric polarization~\cite{Vanderbilt-textbook}.

Spatial symmetry can quantize the WZ term, similarly to the quantization of the electric polarization due to spatial-inversion symmetry. Here, we focus on the following four-fold-rotation-type symmetry:
\begin{align}
\U H^{\dag} \left( k_x, k_y \right) \V^{-1} = H \left( -k_y, k_x \right),\quad \left( \U\V \right)^2 = 1, 
	\label{eq:C4Sym}
\end{align}
where $\U$ and $\V$ are unitary matrices that are, in general, independent of each other. The two-dimensional model in Eq.~(\ref{eq: NH-BBH-2D}) respects this symmetry with $\U = -\ii e^{\ii\pi/4} \sigma_{y}$ and $\V = e^{\ii\pi/4}$ [i.e., Eq.~(\ref{eq: NH-BBH-C4-rotation})]. We show that this  rotation-type symmetry indeed quantizes the WZ term to the $\mathbb{Z}_2$ values 
\begin{equation}
{\rm WZ} \left[ H \right] \in \left\{0, \frac{1}{2} \right\}. 
\end{equation}
Given an extension $H \left( k_x, k_y \right) \to H \left( k_x, k_y, t \right)$ for $t \in [0,1]$, we introduce a different extension by 
\begin{align}
H' \left( k_x, k_y, t \right) := \U H^\dag \left( k_y, -k_x, t \right) \V^{-1},\quad t \in [0,1].
\end{align}
Thanks to rotation-type symmetry in Eq.~\eqref{eq:C4Sym}, $H' \left( k_x, k_y, t \right)$ at $t = 0$ coincides with the original Hamiltonian:
\begin{align}
H' \left( k_x, k_y, t=0 \right) = H \left( k_x, k_y \right).
\end{align}
In a straightforward manner, we can also show
\begin{equation}
{\rm WZ} \left[ H' \right] = - {\rm WZ} \left[ H \right],
\end{equation}
and
\begin{equation}
2 {\rm WZ} \left[ H \right] = {\rm WZ} \left[ H \right] - {\rm WZ} \left[ H' \right].
\end{equation}
The right-hand side of this equation gives the integer-valued three-dimensional winding number, which proves that the WZ term ${\rm WZ} \left[ H \right]$ is quantized to the $\mathbb{Z}_2$ value. For our model in Eq.~\eqref{eq: NH-BBH-2D}, the WZ term takes the nontrivial value ${\rm WZ} \left[ H \right] = 1/2$ for $\left|\gamma/\lambda \right| < 1$. Thus, the $\mathbb{Z}_2$-quantized WZ term is a meaningful topological invariant of two-dimensional non-Hermitian systems, as long as four-fold-rotation-type symmetry in Eq.~\eqref{eq:C4Sym} is respected.

In general, the WZ term is quantized to the $\mathbb{Z}_2$ value when either rotation-type symmetry 
\begin{align}
    U H^{\dag} \left( \bm{k} \right) V^{-1} = H \left( c_n\bm{k} \right)
        \label{eq:NH_rot_sym}
\end{align}
or reflection symmetry 
\begin{align}
    U H \left( \bm{k} \right) V^{-1} = H \left( m\bm{k} \right)
\end{align}
is respected, where $\bm{k} \mapsto c_n\bm{k}$ is an $n$-fold rotation and $\bm{k} \mapsto m\bm{k}$ is a reflection on an axis. 
It can be proven in the same way as four-fold-rotation-type symmetry in Eq.~\eqref{eq:C4Sym}.

It should also be noted that four-fold-rotation-type symmetry in Eq.~\eqref{eq:C4Sym} vanishes the one-dimensional winding numbers in Eq.~\eqref{eq:2d_1dWNs}. In fact, we have
\begin{eqnarray}
W_{x}
&=& \oint_0^{2\pi} \frac{dk_x}{2\pi\ii} \frac{\partial}{\partial k_x} \log \det \left[ H^{\dag} \left( -k_y, k_x \right) \right]
= - W_{y},\qquad
\end{eqnarray}
and on the other hand, we have
\begin{eqnarray}
W_{y} &=& \oint_0^{2\pi} \frac{dk_y}{2\pi\ii} \frac{\partial}{\partial k_y} \log \det \left[ H^{\dag} \left( -k_y, k_x \right) \right]
= W_{x}.\quad
\end{eqnarray}
These equations result in
\begin{equation}
W_{x} = W_{y} = 0.
\end{equation}

The quantization of the WZ term is closely related to the corner skin effect. This can be understood in view of the topological invariant in momentum space and the adiabatic parameter by Teo and Kane~\cite{Teo-Kane-10}. Let us consider a point defect and a circle $S^1$ that encloses this point defect. We consider a non-Hermitian Hamiltonian $H \left( k_{x}, k_{y}, s \right)$ and the corresponding extended Hermitian Hamiltonian $\tilde H \left( k_x, k_y, s \right)$ defined with the adiabatic parameter $s \in S^1$ that characterizes the spatial modulation of the Hamiltonians far from the point defect. The zero modes of $\tilde H \left( k_x, k_y, s \right)$ at the point defect are detected by the three-dimensional winding number $W_3$, which is in turn given as the winding of the WZ term 
\begin{align}
W_3 = \oint_{0}^{1} ds \frac{d}{ds} {\rm WZ} \left[ H \left( s \right) \right]. 
	\label{eq:W3}
\end{align}
In $\tilde H \left( k_x, k_y, s \right)$, there appear $W_3$ zero modes localized at the point defect. In a similar manner to the Hatano-Nelson model, these zero modes accompany the skin modes at the same defect in the original non-Hermitian Hamiltonian $H \left( k_{x}, k_{y}, s \right)$.

In the following, we show that the nonzero WZ term $\mathrm{WZ} \left[ H \right]$ leads to the presence of the corner zero modes in the extended Hermitian Hamiltonian, and consequently, the presence of the corner skin modes in the original non-Hermitian Hamiltonian. Let us impose the open boundary conditions along both $x$ and $y$ directions. Near the edges, no zero modes appear because of the vanishing one-dimensional winding numbers in Eq.~\eqref{eq:2d_1dWNs}, allowing us to consider adiabatic changes of the microscopic Hamiltonian near the edges into a slowly-varying Hamiltonian while keeping the topological phase. In doing so, we can define a family of Hamiltonians $\tilde H \left( k_x, k_y, s \right)$ for each edge such that $\tilde H \left( k_x, k_y, s = 0 \right)$ is the Hamiltonian deep inside the bulk and that $\tilde H \left( k_x, k_y, s = 1 \right)$ is outside the finite system. For example, $\tilde H \left( k_x, k_y, s = 1 \right)$ can be chosen as the vacuum Hamiltonian $\tilde H_{\rm vac}$. Let the families of the edge Hamiltonians be $\tilde{H}_{\rm l} \left( k_x, k_y, s \right)$, $\tilde{H}_{\rm r} \left( k_x, k_y, s \right)$, $\tilde{H}_{\rm u} \left( k_x, k_y, s \right)$, and $\tilde{H}_{\rm d} \left( k_x, k_y, s \right)$ for the left, right, up, and down edges, respectively. We assume that the edge Hamiltonians, as well as the bulk Hamiltonian, enjoy four-fold-rotation-type symmetry, meaning that they are related to each other in the four-fold-symmetric way. For example, the up-edge Hamiltonian is related to the right-edge one by 
\begin{equation}
H_{\rm u} \left( k_x, k_y, s \right) = \U H_{\rm r}^\dag \left( k_x, k_y, s \right) \V^{-1}
\end{equation} 
for the off-diagonal parts. 

\begin{figure}[t]
\centering
\includegraphics[width=86mm]{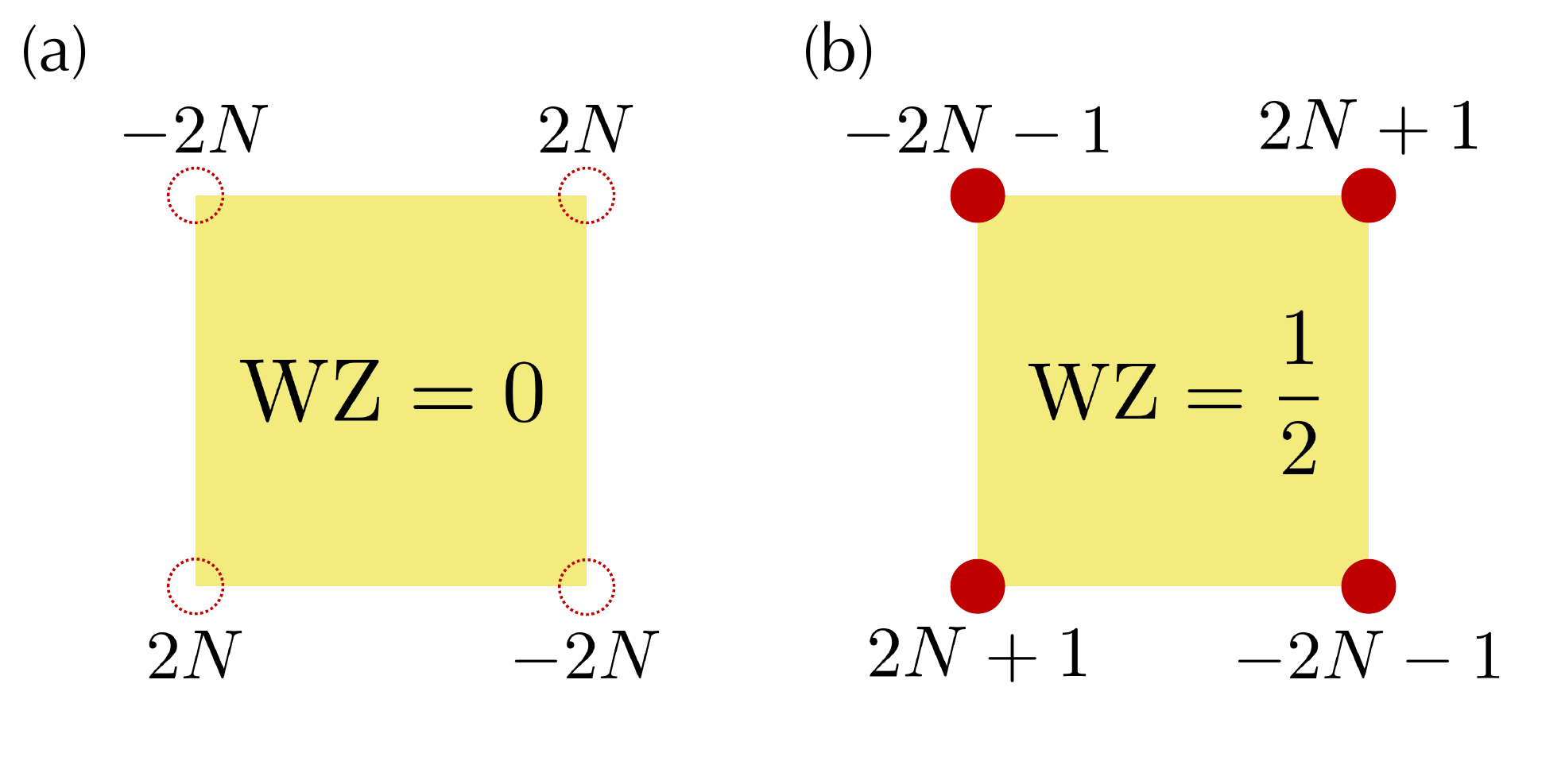} 
\caption{Wess-Zumino (WZ) term and corner zero modes. The number of the zero modes in the extended Hermitian Hamiltonian $\tilde H \left( k_x, k_y \right)$ with four-fold-rotation symmetry is shown as the even and odd integers at each corner. (a) and (b) correspond to the trivial and nontrivial WZ terms, respectively.}
	\label{fig: WZ}
\end{figure}

Then, the changes in the WZ terms 
\begin{align}
\Delta {\rm WZ}_\nu := \oint_0^1 ds \frac{d}{ds} {\rm WZ} \left[ H_\nu \left( s \right) \right],\quad 
\nu \in \{{\rm l}, {\rm r}, {\rm u}, {\rm d}\},
\end{align}
from the bulk to the vacuum for the four edges satisfy 
\begin{equation}
\Delta {\rm WZ}_{\rm l} 
= - \Delta {\rm WZ}_{\rm u}
= \Delta {\rm WZ}_{\rm r} 
= -\Delta {\rm WZ}_{\rm d}.
\end{equation}
Here, the vacuum Hamiltonian $\tilde H_{\rm vac}$ is assumed to be in common for all the edges. This structure gives a constraint on the three-dimensional winding numbers in Eq.~\eqref{eq:W3} of the four corners: $W_3$ of the upper-right corner is given as
\begin{equation}
W_3 = \Delta {\rm WZ}_{\rm r} - \Delta {\rm WZ}_{\rm u} = 2 \Delta {\rm WZ}_{\rm r} \equiv -2 {\rm WZ} \left[ H \right]
\end{equation} 
modulo 2. This implies that if the quantized WZ term of the bulk is nontrivial (i.e., ${\rm WZ} \left[ H \right] = 1/2$), the three-dimensional winding number $W_3$ of the four corners should be odd, especially nonzero, and hence the extended Hermitian Hamiltonian should have zero modes localized at the corners. See Fig.~\ref{fig: WZ} for possible quartets of the numbers of the corner zero modes accompanied by the trivial and nontrivial bulk WZ terms. Since the presence of the zero modes in the extended Hermitian Hamiltonian leads to the skin effect in the non-Hermitian Hamiltonian~\cite{OKSS-20}, the bulk WZ term leads to the corner skin effect. Similar $\mathbb{Z}_{2}$ quantization for corner zero modes was recently discussed for Hermitian second-order topological superconductors~\cite{Tiwari-20}.

It should be noted that the nontrivial WZ term does not always imply the corner skin effect. Suppose a real line gap is open and the Chern number ${\rm Ch}$ is well defined for each band. Then, under rotation-type symmetry in Eq.~(\ref{eq:C4Sym}), or more generally Eq.~(\ref{eq:NH_rot_sym}), with a common unitary matrix $U = V$, we have the equality
\begin{equation}
    2\mathrm{WZ} \equiv \mathrm{Ch}
    \label{eq:wz_ch}
\end{equation}
modulo $2$. This is because the inverse of the Green's function $G^{-1} \left( \bm{k}, \omega \right) := \ii \omega - H \left( \bm{k} \right)$ plays the role of the Hamiltonian $H \left( \bm{k}, s \right)$ introduced before, and the Chern number is given as the three-dimensional winding number 
\begin{equation}
    {\rm Ch} = \frac{1}{24\pi^2} \oint_{\left[0,2\pi\right]^2 \times [-\infty,\infty]} {\rm tr} \left[ GdG^{-1} \right]^3
\end{equation}
of $G^{-1} \left( \bm{k}, \omega \right)$. The condition $U=V$ is crucial in Eq.~\eqref{eq:wz_ch}; for $U \neq V$, the gluing condition $U \left[ G^{-1} \left(\bm{k}, \omega=0 \right) \right]^{\dag} V^{-1} = G^{-1} \left( \bm{k}, \omega=0 \right)$ at $\omega=0$ does not hold. As a corollary, Hermitian Hamiltonians always satisfy Eq.~\eqref{eq:wz_ch} since Hermiticity is equivalent to the trivial rotation (i.e., $c_{1} \bm{k} = \bm{k}$) with $U = V = 1$. Thus, Chern insulators can also have the nontrivial WZ term regardless of the presence or absence of the corner skin modes.

On the other hand, even though the non-Hermitian model in Eq.~\eqref{eq: NH-BBH-2D} takes the nontrivial WZ term ${\rm WZ}=1/2$ for $\left| \gamma \right| < \left|\lambda \right|$, it has a nonzero real line gap except for $\left| \gamma \right| = \left|\lambda \right|$, and the Chern number vanishes. While this difference between the WZ term and the Chern number may seem like a contradiction, we do not actually have any contradictions. First, Eq.~\eqref{eq:wz_ch} is not always true for generic unitary matrices $U$ and $V$. In fact, the non-Hermitian model in Eq.~\eqref{eq: NH-BBH-2D}, for which we have $U \neq V$ [see Eq.~(\ref{eq: NH-BBH-C4-rotation})], does not satisfy Eq.~\eqref{eq:wz_ch}. Moreover, we inevitably have an obstacle to having a continuous path from the non-Hermitian model in Eq.~\eqref{eq: NH-BBH-2D} to a Hermitian Hamiltonian while keeping the real line gap and rotation-type symmetry in Eq.~\eqref{eq: NH-BBH-C4-rotation}: if a Hermitian Hamiltonian $H \left( \bm{k} \right) = H^{\dag} \left( \bm{k} \right)$ respects Eq.~\eqref{eq: NH-BBH-C4-rotation}, it is subject to the constraint $-\ii \sigma_{y} H \left( \bm{k}_{*} \right) = H \left( \bm{k}_{*} \right)$ and hence vanishes [i.e., $H \left( \bm{k}_{*} \right) = 0$] at the symmetric points $\bm{k}_{*} = \left( 0, 0 \right), \left( \pi, \pi \right)$, meaning closing of both point and line gaps. This fact implies intrinsic non-Hermitian topology of the model in Eq.~\eqref{eq: NH-BBH-2D}.

%%%%%%%%%%
\subsection{Non-Bloch band theory}
	\label{sec: SOSE-band}

While symmetry can protect skin effects, it can also vanish skin effects. Prime examples include spatial-inversion (parity) symmetry
\begin{equation}
\mathcal{P} H \left( \bm{k} \right) \mathcal{P}^{-1}
= H \left( - \bm{k} \right)
	\label{eq: inversion}
\end{equation}
with a unitary matrix $\mathcal{P}$ respecting $\mathcal{P}^{2} = 1$, and transposition-associated mirror symmetry
\begin{equation}
\mathcal{M}_{i} H^{T} \left( \bm{k} \right) \mathcal{M}_{i}^{-1}
= H \left( m_{i} \bm{k} \right)
	\label{eq: transposition-mirror}
\end{equation}
with a unitary matrix $\mathcal{M}_{i}$ respecting $\mathcal{M}_{i}^{2} = 1$. Here, $m_{i}$ denotes a reflection that changes $k_{i}$ into $-k_{i}$; in two dimensions, for example, we have $m_{x} \left( k_{x}, k_{y} \right) = \left( -k_{x}, k_{y} \right)$ and $m_{y} \left( k_{x}, k_{y} \right) = \left( k_{x}, -k_{y} \right)$. Our model with the corner skin modes indeed respects these symmetry with $\mathcal{P} = \sigma_{y}$ [i.e., Eq.~(\ref{eq: NH-BBH-inversion})], $\mathcal{M}_{x} = \sigma_{x}$ [i.e., Eq.~(\ref{eq: NH-BBH-transposition-x})], and $\mathcal{M}_{y} = \sigma_{z}$ [i.e., Eq.~(\ref{eq: NH-BBH-transposition-y})]. In one dimension, no skin effect occurs in the presence of these symmetry~\cite{KSUS-19}. This is compatible with vanishing winding number in Eq.~(\ref{eq: winding}) in the presence of these symmetry. Even in higher dimensions, mirror-type symmetry in Eq.~(\ref{eq: transposition-mirror}) vanishes the winding number and the consequent skin effect along the $i$ direction.

The absence of the skin effect in one dimension can be shown on the basis of Eq.~(\ref{eq: Yao-Wang-Yokomizo-Murakami}), which is the salient result of the non-Bloch band theory~\cite{YW-18-SSH, YM-19}. We begin with the characteristic equation
\begin{equation}
\det \left[ H \left( \beta \right) - E \right] = 0.	
	\label{eq: characteristic}
\end{equation}
In terms of $H \left( \beta \right)$, spatial-inversion symmetry in Eq.~(\ref{eq: inversion}) imposes
\begin{equation}
\mathcal{P} H \left( \beta \right) \mathcal{P}^{-1}
= H\,( \beta^{-1} ),
\end{equation}
and hence leads to
\begin{equation}
\det \left[ H\,( \beta^{-1} ) - E \right] = 0.
\end{equation}
This equation implies that $\beta^{-1}$ is another solution to the characteristic equation~(\ref{eq: characteristic}) if $\beta$ is a solution. Because of the assumption $\left| \beta_{1} \right| \leq \left| \beta_{2} \right| \leq \cdots \left| \beta_{2M} \right|$, we then have
\begin{equation}
\beta_{2M-i+1} = \beta_{i}^{-1}\quad
\left( i = 1, 2, \cdots, M\right).
\end{equation}
Now, using Eq.~(\ref{eq: Yao-Wang-Yokomizo-Murakami}), we finally have 
\begin{equation}
\left| \beta_{M} \right| = \left| \beta_{M+1} \right| = 1,
\end{equation}
showing that continuum bands are formed by delocalized eigenstates. Similarly, transposition-associated symmetry in Eq.~(\ref{eq: transposition-mirror}) also leads to the absence of the skin effect in one dimension.

Importantly, the above discussion is not directly applicable in higher dimensions. This is because the non-Bloch band theory in Refs.~\cite{YW-18-SSH, YM-19}, especially Eq.~(\ref{eq: Yao-Wang-Yokomizo-Murakami}), is inapplicable under the open boundary conditions along more than one direction. Remarkably, the higher-order skin effect requires modification of the non-Bloch band theory in higher dimensions. In fact, if Eq.~(\ref{eq: Yao-Wang-Yokomizo-Murakami}) were valid even in higher dimensions, transposition-associated mirror symmetry in Eq.~(\ref{eq: transposition-mirror}) leads to the absence of the skin effect along the $i$ direction. However, this would contradict the emergence of the corner skin effect in our two-dimensional model with Eq.~(\ref{eq: transposition-mirror}) for both $x$ and $y$ directions. Hence, the non-Bloch band theory is indeed modified in higher dimensions.

Nevertheless, it is naturally expected that Eq.~(\ref{eq: Yao-Wang-Yokomizo-Murakami}) is valid for an extensive number of eigenstates even in higher dimensions. This is consistent with delocalization of the $\mathcal{O}\,( L^2 )$ bulk modes in our model. On the other hand, the $\mathcal{O}\,( L )$ corner skin modes cannot be described by the current non-Bloch band theory. It is thus important to develop a non-Bloch band theory in higher dimensions in a general manner, which we leave for future work.

\begin{figure*}[t]
\centering
\includegraphics[width=172mm]{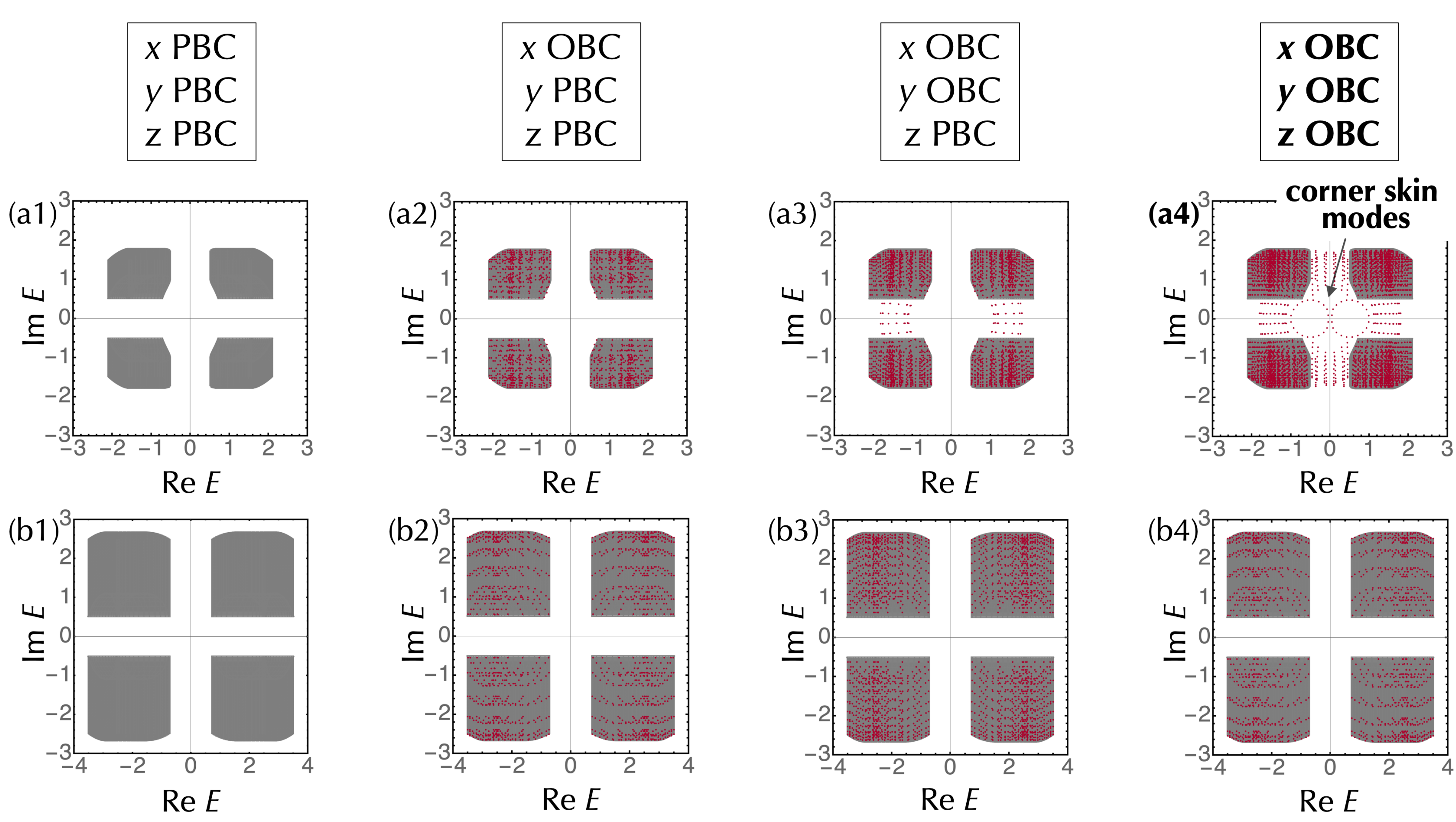} 
\caption{Third-order non-Hermitian skin effect. The complex spectra of the non-Hermitian model in three dimensions [Eq.~(\ref{eq: NH-BBH-3D})] are shown for $10 \times 10 \times 10$ sites. The parameters are given as $\lambda = 1.0$, as well as (a1, a2, a3, a4)~$\gamma = 0.5$ or (b1, b2, b3, b4)~$\gamma = 1.5$. The open boundary conditions are imposed along none of the directions for (a1, b1), only along the $x$ direction for (a2, b2), only along the $x$ and $y$ directions for (a3, b3), and along all the directions for (a4, b4). The spectra for the periodic boundary conditions are shown as the grey regions, while the spectra for the open boundary conditions are shown as the red dots. The corner skin modes appear under the open boundary conditions along all the directions, as shown in (a4).}
	\label{fig: spectrum-TOSE}
\end{figure*}

\begin{figure}[t]
\centering
\includegraphics[width=86mm]{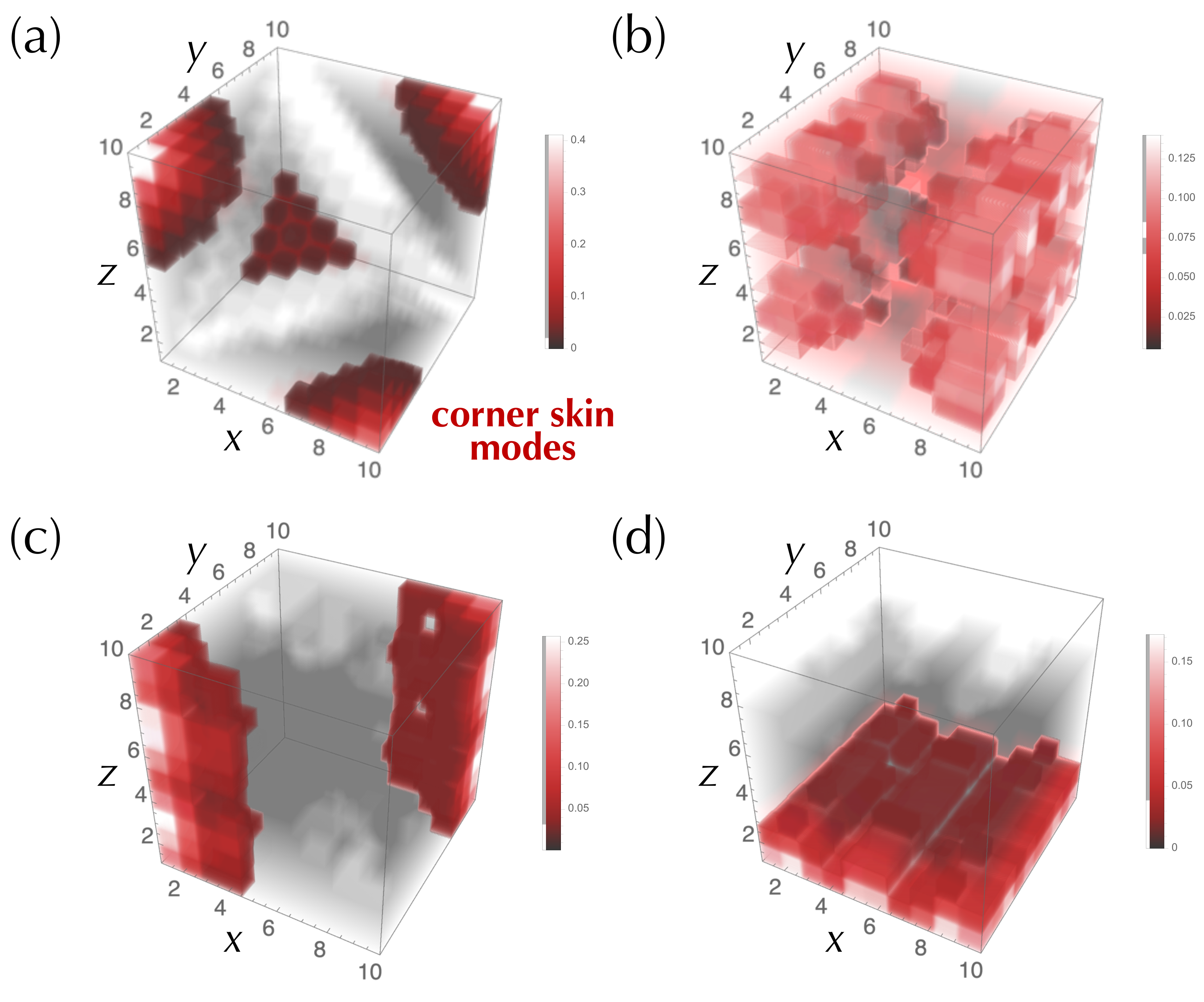} 
\caption{Wavefunctions for the third-order non-Hermitian skin effect. Under the open boundary conditions along all the directions, eigenstates of the non-Hermitian model in three dimensions [Eq.~(\ref{eq: NH-BBH-3D})] are shown for $L = 10$, $\gamma = 0.5$, and $\lambda = 1.0$. (a)~Corner skin modes ($E=-0.050 - 0.086\ii$). (b)~Delocalized bulk modes ($E = -1.49 - 1.34\ii$). (c)~Edge modes ($E=-1.15-0.36\ii$). (d)~Surface modes ($E = -0.13 - 1.20\ii$).}
	\label{fig: wavefunction-TOSE}
\end{figure}

%%%%%%%%%%
\section{Third-order non-Hermitian skin effect}
	\label{sec: TOSE}

The non-Hermitian skin effect can even have the third-order nature in three dimensions. For the third-order non-Hermitian skin effect, $\mathcal{O}\,( L )$ corner skin modes emerge from all the $\mathcal{O}\,( L^3 )$ modes. We provide a model that exhibits the third-order skin effect. The Bloch Hamiltonian reads
\begin{eqnarray}
&&H \left( \bm{k} \right) = \ii \lambda \left( \sin k_{y} \right) \sigma_{x} + \ii \left( \gamma + \lambda \cos k_{y} \right) \sigma_{y} \nonumber \\
&&\qquad\qquad\quad + \ii \lambda \left( \sin k_{x} \right) \sigma_{z} + \left( \gamma + \lambda \cos k_{x} \right) \tau_{z} \nonumber \\
&&\qquad\qquad\quad + \lambda \left( \sin k_{z} \right) \tau_{y} + \left( \gamma + \lambda \cos k_{z} \right) \tau_{x},
	\label{eq: NH-BBH-3D}
\end{eqnarray}
where $\gamma$ and $\lambda$ are real parameters, and $\sigma_{i}$'s and $\tau_{i}$'s ($i = x, y, z$) are Pauli matrices. The extended Hermitian Hamiltonian reads
\begin{eqnarray}
\tilde{H}_{\rm BBH} \left( \bm{k} \right) 
&=& \begin{pmatrix}
0 & H \left( \bm{k} \right) \\
H^{\dag} \left( \bm{k} \right) & 0 
\end{pmatrix} \nonumber \\
&=& -\lambda \left( \sin k_{y} \right) \rho_{y} \sigma_{x} - \left( \gamma + \lambda \cos k_{y} \right) \rho_{y} \sigma_{y} \nonumber \\
&&~ - \lambda \left( \sin k_{x} \right) \rho_{y} \sigma_{z} + \left( \gamma + \lambda \cos k_{x} \right) \rho_{x} \tau_{z} \nonumber \\
&&~ + \lambda \left( \sin k_{z} \right) \rho_{x} \tau_{y} + \left( \gamma + \lambda \cos k_{z} \right) \rho_{x} \tau_{x},\qquad
	\label{eq: BBH-3D}
\end{eqnarray}
where $\rho_{i}$'s ($i = x, y, z$) are Pauli matrices that account for the additional degrees of freedom. Similarly to the second-order topological insulator, $\tilde{H}_{\rm BBH} \left( \bm{k} \right)$ is a prototypical example of a third-order topological insulator that was first proposed by Benalcazar, Bernevig, and Hughes~\cite{BBH-17}. It can exhibit zero-energy modes localized at the corners under the open boundary conditions along all the three directions, although no boundary modes appear under other boundary conditions.

The Hermitian model $\tilde{H}_{\rm BBH} \left( \bm{k} \right)$ respects spatial-inversion symmetry:
\begin{eqnarray}
\left( \rho_{y} \sigma_{y} \tau_{y} \right) \tilde{H}_{\rm BBH} \left( \bm{k} \right) \left( \rho_{y} \sigma_{y} \tau_{y} \right)^{-1} &=& \tilde{H}_{\rm BBH} \left( - \bm{k} \right).
\end{eqnarray}
Correspondingly, $H \left( \bm{k} \right)$ respects
\begin{eqnarray}
\left( \sigma_{y} \tau_{y} \right) H^{\dag} \left( \bm{k} \right) \left( \sigma_{y} \tau_{y} \right)^{-1} &=& - H \left( - \bm{k} \right).
\end{eqnarray}
Moreover, $\tilde{H}_{\rm BBH} \left( \bm{k} \right)$ respects mirror symmetry:
\begin{eqnarray}
&&\left( \rho_{x} \sigma_{z} \right) \tilde{H}_{\rm BBH} \left( k_{x}, k_{y}, k_{z} \right) \left( \rho_{x} \sigma_{z} \right)^{-1} \nonumber \\
&&\qquad\qquad\qquad=\tilde{H}_{\rm BBH} \left( - k_{x}, k_{y}, k_{z} \right), \\
&&\left( \rho_{x} \sigma_{x} \right) \tilde{H}_{\rm BBH} \left( k_{x}, k_{y}, k_{z} \right) \left( \rho_{x} \sigma_{x} \right)^{-1} \nonumber \\
&&\qquad\qquad\qquad=\tilde{H}_{\rm BBH} \left( k_{x}, -k_{y}, k_{z} \right), \\
&&\left( \rho_{y} \tau_{y} \right) \tilde{H}_{\rm BBH} \left( k_{x}, k_{y}, k_{z} \right) \left( \rho_{y} \tau_{y} \right)^{-1} \nonumber \\
&&\qquad\qquad\qquad=\tilde{H}_{\rm BBH} \left( k_{x}, k_{y}, -k_{z} \right).
\end{eqnarray}
Correspondingly, $H \left( \bm{k} \right)$ respects
\begin{eqnarray}
\sigma_{z} H^{\dag} \left( k_{x}, k_{y}, k_{z} \right) \sigma_{z}^{-1} 
&=& H \left( - k_{x}, k_{y}, k_{z} \right), \\
\sigma_{x} H^{\dag} \left( k_{x}, k_{y}, k_{z} \right) \sigma_{x}^{-1} 
&=& H \left( k_{x}, -k_{y}, k_{z} \right), \\
\tau_{y} H^{\dag} \left( k_{x}, k_{y}, k_{z} \right) \tau_{y} ^{-1} 
&=& - H \left( k_{x}, k_{y}, -k_{z} \right).
\end{eqnarray}
Such spatial symmetry plays a crucial role in the third-order skin effect.

The third-order topological insulator $\tilde{H}_{\rm BBH} \left( \bm{k} \right)$ exhibits zero-energy corner modes for $\left| \gamma/\lambda \right| < 1$. Correspondingly, corner skin modes appear in the non-Hermitian model $H \left( \bm{k} \right)$ with open boundaries along all the directions. In Fig.~\ref{fig: spectrum-TOSE}, we show the numerically obtained spectra for various boundary conditions. Under the periodic boundary conditions, no skin effect occurs, and all the eigenstates are delocalized through the bulk. The bulk forms four bands and their spectrum is given as
\begin{eqnarray}
&&E \left( \bm{k} \right)
= \pm \sqrt{\left( \gamma + \lambda \cos k_{x} \right)^{2} + \left( \gamma + \lambda \cos k_{z} \right)^{2} + \lambda^{2} \sin^{2} k_{z}} \nonumber \\
&&\qquad\pm \ii \sqrt{\lambda^{2} \sin^{2} k_{x} + \left( \gamma + \lambda \cos k_{y}\right)^{2} + \lambda^{2} \sin^{2} k_{y}}.\qquad
\end{eqnarray}
Under the open boundary conditions along all the directions, an extensive number of the eigenstates remain delocalized and form the bulk bands [Fig.~\ref{fig: wavefunction-TOSE}\,(b)]. However, some of the eigenstates exhibit the skin effect and are localized at the four corners [Figs.~\ref{fig: spectrum-TOSE}\,(a4) and \ref{fig: wavefunction-TOSE}\,(a)]. For the conventional skin effect, there appear $\mathcal{O}\,( L^{3} )$ skin modes in a three-dimensional system with the system size $L \times L \times L$; for the third-order skin effect, by contrast, only $\mathcal{O} \left( L \right)$ skin modes appear at the corners. This also contrasts with zero-energy corner modes in Hermitian third-order topological insulators, the number of which is $\mathcal{O} \left( 1 \right)$. Thus, the third-order non-Hermitian skin effect gives rise to a new type of boundary physics in three dimensions.

Finally, it is notable that the three-dimensional model in Eq.~(\ref{eq: NH-BBH-3D}) exhibits different types of boundary modes in addition to the corner skin modes. As shown in Fig.~\ref{fig: spectrum-TOSE}\,(a3), gapless modes appear as long as the open boundary conditions are imposed for both $x$ and $y$ directions. These gapless modes appear even though the periodic boundary conditions are imposed along the $z$ direction. Their spectrum crosses $\mathrm{Im}\,E = 0$, i.e., the imaginary line gap is closed. Consistently, they are localized at the corners on the $xy$ plane, but delocalized along the $z$ direction [Fig.~\ref{fig: wavefunction-TOSE}\,(c)]. Moreover, other gapless modes appear as long as the open boundary conditions are imposed for the $z$ direction [Fig.~\ref{fig: spectrum-TOSE}\,(a4)]. Their spectrum crosses $\mathrm{Re}\,E = 0$, i.e., the real line gap is closed. These gapless modes are localized on the surfaces perpendicular to the $z$ axis [Fig.~\ref{fig: wavefunction-TOSE}\,(d)].

%%%%% Conclusion %%%%%
\section{Conclusion}
	\label{sec: conclusion}

In this work, we have discovered the higher-order non-Hermitian skin effect. It leads to new types of boundary physics, which may further give rise to new non-Hermitian topological phenomena. In two dimensions, the second-order skin effect accompanies $\mathcal{O}\,( L )$ corner skin modes in contrast to $\mathcal{O}\,( L^{2} )$ skin modes in the conventional (first-order) skin effect. This also contrasts with $\mathcal{O}\,( 1 )$ corner zero modes in Hermitian second-order topological insulators. Similarly, in three dimensions, the third-order skin effect accompanies $\mathcal{O}\,( L )$ corner skin modes in contrast to $\mathcal{O}\,( L^{3} )$ skin modes in the conventional (first-order) skin effect and $\mathcal{O}\,( 1 )$ corner zero modes in Hermitian third-order topological insulators. These higher-order skin effects originate from intrinsic non-Hermitian topology protected by spatial symmetry. Furthermore, they imply modification of the conventional non-Bloch band theory in higher dimensions. 

It merits further research to develop a non-Bloch band theory that works even in higher dimensions. Moreover, the higher-order skin effect is a new non-Hermitian phenomenon that originates from spatial symmetry. It is also worthwhile to further explore unique phenomena and functionalities that arise from the interplay of non-Hermiticity and spatial symmetry.

%%%%% Acknowledgement %%%%%
\section*{Acknowledgment}

K.K. thanks Ching Hua Lee and Daichi Nakamura for helpful discussions. K.S. thanks Shin Hayashi and Mayuko Yamashita for useful discussions on the WZ term. This work was supported by JST CREST Grant No.~JPMJCR19T2. K.K. was supported by KAKENHI Grant No.~JP19J21927 from the Japan Society for the Promotion of Science (JSPS). M.S. was supported by KAKENHI Grant No.~JP20H00131 from the JSPS. K.S. was supported by Grant No.~JPMJPR18L4 from PRESTO, JST.

%%%%% Note %%%%%
\medskip
{\it Note added.\,---\,}After completion of this work, we became aware of a recent related work~\cite{Okugawa-Takahashi-Yokomizo-20}.

%%%%% Appendix %%%%%
\appendix

\section{Exact corner skin modes}
	\label{appendix: SOSE}

We exactly solve the non-Hermitian Hamiltonian in Eq.~(\ref{eq: NH-BBH-2D}) with open boundaries along both $x$ and $y$ directions. In particular, we obtain the corner skin modes in an analytical manner. Let an eigenenergy be $E \in \mathbb{C}$, and the component of the corresponding eigenstate at the lattice site $\left( m, n \right) \in \left[ 1, L \right]^{2}$ be $\vec{\psi} \left( m, n \right) \in \mathbb{C}^{2}$. Because of periodicity of the bulk, as well as transposition-associated mirror symmetry in Eqs.~(\ref{eq: NH-BBH-transposition-x}) and (\ref{eq: NH-BBH-transposition-y}), $\vec{\psi} \left( m, n \right)$ can be described as
\begin{eqnarray}
&&\vec{\psi} \left( m, n \right)
= \beta_{x}^{m} \beta_{y}^{n} \vec{v}_{++}
+ \beta_{x}^{m} \beta_{y}^{L+1-n} \vec{v}_{+-} \nonumber \\
&&\qquad +\beta_{x}^{L+1-m} \beta_{y}^{n} \vec{v}_{-+}
+ \beta_{x}^{L+1-m} \beta_{y}^{L+1-n} \vec{v}_{--}
	\label{aeq: eigen-ansatz}
\end{eqnarray}
with $\beta_{x}, \beta_{y} \in \mathbb{C}$ and $\vec{v}_{\pm\pm} \in \mathbb{C}^{2}$. The normalization of $\vec{\psi} \left( m, n \right)$ requires $\left| \beta_{x} \right| \leq 1$ and $\left| \beta_{y} \right| \leq 1$. 

In the bulk, the Schr\"odinger equation reads
\begin{eqnarray}
&&M \vec{\psi} \left( m, n \right)
+ T_{x+} \vec{\psi} \left( m-1, n \right)
+ T_{x-} \vec{\psi} \left( m+1, n \right) \nonumber \\
&&+ T_{y+} \vec{\psi} \left( m, n-1 \right) 
+ T_{y-} \vec{\psi} \left( m, n+1 \right)
= E\,\vec{\psi} \left( m, n \right)\qquad
\end{eqnarray}
with
\begin{eqnarray}
M &:=& -\ii \gamma + \gamma \sigma_{y},\\
T_{x\pm} &:=& \frac{\ii \lambda \left( -1 \pm \sigma_{z} \right)}{2},
    \label{aeq: hopping-x} \\
T_{y\pm} &:=& \frac{\lambda \left( \sigma_{y} \pm \ii \sigma_{x} \right)}{2}.
    \label{aeq: hopping-y}
\end{eqnarray}
With Eq.~(\ref{aeq: eigen-ansatz}), the bulk equation leads to 
\begin{equation}
H\,( \beta_{x}^{\pm 1}, \beta_{y}^{\pm 1} )\,\vec{v}_{\pm\pm} = E\,\vec{v}_{\pm\pm},
	\label{aeq: bulk-eq}
\end{equation}
where $H \left( \beta_{x}, \beta_{y} \right)$ is the bulk Hamiltonian
\begin{equation}
H \left( \beta_{x}, \beta_{y} \right)
= -\ii \begin{pmatrix}
\gamma + \lambda \beta_{x} & \gamma + \lambda \beta_{y} \\
- \gamma - \lambda \beta_{y}^{-1} & \gamma + \lambda \beta_{x}^{-1}
\end{pmatrix}.
\end{equation}

At the boundaries, on the other hand, the Schr\"odinger equation reads
\begin{eqnarray}
T_{x+} \vec{\psi} \left( 0, n \right) &=& 0 \\
T_{x-} \vec{\psi} \left( L+1, n \right) &=& 0 \\
T_{y+} \vec{\psi} \left( m, 0 \right) &=& 0 \\
T_{y-} \vec{\psi} \left( m, L+1 \right) &=& 0
\end{eqnarray}
with $m, n = 1, 2, \cdots, L$. With Eq.~(\ref{aeq: eigen-ansatz}), these boundary equations reduce to
\begin{eqnarray}
T_{x+} \left( \vec{v}_{++} + \beta_{x}^{L+1} \vec{v}_{-+} \right) &=& T_{x+} \left( \vec{v}_{+-} + \beta_{x}^{L+1} \vec{v}_{--} \right) = 0,\qquad\quad \\
T_{x-} \left( \beta_{x}^{L+1} \vec{v}_{++} + \vec{v}_{-+} \right) &=& T_{x-} \left( \beta_{x}^{L+1} \vec{v}_{+-} + \vec{v}_{--} \right) = 0,\qquad\quad \\
T_{y+} \left( \vec{v}_{++} + \beta_{y}^{L+1} \vec{v}_{+-} \right) &=& T_{y-} \left( \vec{v}_{-+} + \beta_{y}^{L+1} \vec{v}_{--} \right) = 0,\qquad\quad \\
T_{y+} \left( \beta_{y}^{L+1} \vec{v}_{++} + \vec{v}_{+-} \right) &=& T_{y-} \left( \beta_{y}^{L+1} \vec{v}_{-+} + \vec{v}_{--} \right) = 0.
\end{eqnarray}
Now, we express $\vec{v}_{\pm\pm}$ as $\vec{v}_{\pm\pm} = \left( a_{\pm\pm}~~b_{\pm\pm} \right)^{T}$. Then, these boundary equations reduce to
\begin{eqnarray}
b_{++} + \beta_{x}^{L+1} b_{-+} = b_{+-} + \beta_{x}^{L+1} b_{--} &=& 0, \\
\beta_{x}^{L+1} a_{++} + a_{-+} = \beta_{x}^{L+1} a_{+-} + a_{--} &=& 0, \\
a_{++} + \beta_{y}^{L+1} a_{+-} = a_{-+} + \beta_{y}^{L+1} a_{--} &=& 0, \\
\beta_{y}^{L+1} b_{++} + b_{+-} = \beta_{y}^{L+1} b_{-+} + b_{--} &=& 0,
\end{eqnarray}
which are further simplified to
\begin{eqnarray}
\frac{a_{+-}}{a_{++}} &=& 
\left( \frac{b_{+-}}{b_{++}} \right)^{-1} = - \beta_{y}^{-L-1}, \label{aeq: ab+-} \\
\frac{a_{-+}}{a_{++}} &=& 
\left( \frac{b_{-+}}{b_{++}} \right)^{-1} = - \beta_{x}^{L+1}, \label{aeq: ab-+} \\
\frac{a_{--}}{a_{++}} &=& 
\left( \frac{b_{--}}{b_{++}} \right)^{-1} = \beta_{x}^{L+1} \beta_{y}^{-L-1}. \label{aeq: ab--}
\end{eqnarray}

Meanwhile, since $\vec{v}_{++}$ ($\vec{v}_{-+}$) is an eigenstate of $H \left( \beta_{x}, \beta_{y} \right)$ [$H \left( \beta_{x}^{-1}, \beta_{y} \right)$] from Eq.~(\ref{aeq: bulk-eq}), we have
\begin{eqnarray}
\left( \gamma + \lambda \beta_{x} - \ii E \right) a_{++} + \left( \gamma + \lambda \beta_{y} \right) b_{++} &=& 0, \\
\left( \gamma + \lambda \beta_{x}^{-1} - \ii E \right) a_{-+} + \left( \gamma + \lambda \beta_{y} \right) b_{-+} &=& 0.
\end{eqnarray}
Using Eq.~(\ref{aeq: ab-+}), we have
\begin{equation}
\begin{pmatrix}
\gamma + \lambda \beta_{x} - \ii E & \gamma + \lambda \beta_{y} \\
\beta_{x}^{L+1} \left( \gamma + \lambda \beta_{x}^{-1} - \ii E \right) & \beta_{x}^{-L-1} \left( \gamma + \lambda \beta_{y} \right)
\end{pmatrix} \begin{pmatrix}
a_{++} \\ b_{++} 
\end{pmatrix} = 0.
	\label{aeq: a++b++1}
\end{equation}
To have a nontrivial solution $\left( a_{++}~~b_{++} \right) \neq 0$, the determinant of the coefficient matrix should vanish, which results in
\begin{equation}
\beta_{y} = - \frac{\gamma}{\lambda}
	\label{aeq: betay1}
\end{equation}
or 
\begin{equation}
\frac{\ii E - \gamma}{\lambda} = \frac{\beta_{x}^{L} - \beta_{x}^{-L}}{\beta_{x}^{L+1} - \beta_{x}^{-L-1}}.
	\label{aeq: En-betax1}
\end{equation}
Similarly, since $\vec{v}_{++}$ ($\vec{v}_{+-}$) is an eigenstate of $H \left( \beta_{x}, \beta_{y} \right)$ [$H \left( \beta_{x}, \beta_{y}^{-1} \right)$] from Eq.~(\ref{aeq: bulk-eq}), we have
\begin{equation}
E = -\ii \left( \gamma + \lambda \beta_{x} \right)
	\label{aeq: En-betax2}
\end{equation}
or 
\begin{equation}
- \frac{\gamma}{\lambda} = \frac{\beta_{y}^{L} - \beta_{y}^{-L}}{\beta_{y}^{L+1} - \beta_{y}^{-L-1}}.
	\label{aeq: betay2}
\end{equation}
Furthermore, since $\vec{v}_{++}$ ($\vec{v}_{--}$) is an eigenstate of $H \left( \beta_{x}, \beta_{y} \right)$ [$H \left( \beta_{x}^{-1}, \beta_{y}^{-1} \right)$] from Eq.~(\ref{aeq: bulk-eq}), we have
\begin{equation}
\begin{pmatrix}
\gamma + \lambda \beta_{x} - \ii E & \gamma + \lambda \beta_{y} \\
\beta_{x}^{2 \left( L+1 \right)} \left( \gamma + \lambda \beta_{y} \right) & -\beta_{y}^{2 \left( L+1 \right)} \left( \gamma + \lambda \beta_{x} - \ii E \right)
\end{pmatrix}
\begin{pmatrix}
a_{++} \\ b_{++} 
\end{pmatrix} = 0,
	\label{aeq: a++b++2}
\end{equation}
resulting in
\begin{equation}
\beta_{x}^{2 \left( L+1 \right)} \left( \gamma + \lambda \beta_{y} \right)^{2}
= - \beta_{y}^{2 \left( L+1 \right)} \left( \gamma + \lambda \beta_{x} - \ii E\right)^{2}.
\end{equation}
Importantly, we need 
\begin{equation}
\left| \beta_{x} \right| = \left| \beta_{y} \right|
	\label{aeq: betax-betay}
\end{equation}
so that the above equation will hold for sufficiently large $L$.

The corner skin modes are described by Eq.~(\ref{aeq: En-betax2}). If Eq.~(\ref{aeq: En-betax1}) holds in addition to Eq.~(\ref{aeq: En-betax2}), we have $\beta_{x} = \pm 1$. Then, we also have $E = -\ii \left( \gamma \pm \lambda \right)$ and $\beta_{y} = - \lambda/\gamma, - \gamma/\lambda$, which further leads to $\left| \gamma \right| = \left| \lambda \right|$ from Eq.~(\ref{aeq: betax-betay}). Hence, we have Eq.~(\ref{aeq: betay1}) as long as Eq.~(\ref{aeq: En-betax2}) and $\left| \gamma \right| \neq \left| \lambda \right|$ hold. Because of the normalization condition $\left| \beta_{y} \right| < 1$, we need
\begin{equation}
\left| \frac{\gamma}{\lambda} \right| < 1.
\end{equation}
Equations~(\ref{aeq: betay1}) and (\ref{aeq: En-betax2}) lead to
\begin{equation}
H\left( \beta_{x}, \beta_{y} \right) - E
= -\ii \begin{pmatrix}
0 & 0 \\
\left( \lambda^2 - \gamma^2 \right)/\gamma & - \lambda \left( \beta_{x} - \beta_{x}^{-1} \right)
\end{pmatrix},
\end{equation}
\begin{equation}
H \left( \beta_{x}^{-1}, \beta_{y}^{-1} \right) - E
= \ii \begin{pmatrix}
\lambda \left( \beta_{x} - \beta_{x}^{-1} \right) & \left( \lambda^2 - \gamma^2 \right)/\gamma \\
0 & 0
\end{pmatrix}.
\end{equation}
Since $\left( a_{++}~~b_{++} \right)$ and $\left( a_{--}~~b_{--} \right) \propto\,(\,\beta_{x}^{2 \left( L+1 \right)} a_{++}~~\beta_{y}^{2 \left( L+1 \right)} b_{++}\,)$ are eigenstates of $H\left( \beta_{x}, \beta_{y} \right)$ and $H\,( \beta_{x}^{-1}, \beta_{y}^{-1} )$, respectively, we have
\begin{equation}
\begin{pmatrix}
\left( \lambda^2 - \gamma^2 \right)/\gamma & -\lambda \left( \beta_{x} - \beta_{x}^{-1} \right) \\
\beta_{x}^{2 \left( L+1 \right)} \lambda \left( \beta_{x} - \beta_{x}^{-1} \right) & \beta_{y}^{2 \left( L+1 \right)}  \left( \lambda^2 - \gamma^2 \right)/\gamma
\end{pmatrix} \begin{pmatrix}
a_{++} \\ b_{++} 
\end{pmatrix} = 0,
\end{equation}
which leads to
\begin{equation}
\left( \beta_{x} - \beta_{x}^{-1} \right)^{2} \left( \frac{\beta_{x}}{\beta_{y}} \right)^{2 \left( L + 1 \right)}
= - \left( \frac{\lambda^{2} - \gamma^{2}}{\lambda \gamma} \right)^{2}.
\end{equation}
To have this equation for sufficiently large $L$, we need $\left| \beta_{x}/\beta_{y} \right| = 1$, i.e., Eq.~(\ref{aeq: betax-betay}). Furthermore, the phase of $\beta_{x}$ is quantized by this equation. Since we have $\left| \beta_{x} \right| = \left| \beta_{y} \right| \neq 1$, the eigenstates are localized at the corners, and the skin effect occurs. The spectrum of these corner skin modes is given as
\begin{equation}
E = -\ii \gamma \left( 1 + e^{\ii \theta} \right),\quad
\theta \in \BZ,
\end{equation}
and their number is $2L$.

On the other hand, the eigenstates described by Eq.~(\ref{aeq: betay2}) are delocalized through the bulk. With $\beta_{y} = e^{\ii k_{y}}$, Eq.~(\ref{aeq: betay2}) reduces to
\begin{equation}
- \frac{\gamma}{\lambda} = \frac{\sin \left( k_{y} L \right)}{\sin \left( k_{y} \left( L+1 \right) \right)},
\end{equation}
which quantizes the wavenumber $k_{y} \in \BZ$. In fact, we have $L$ real solutions in $k_{y} \in \left[ 0, \pi \right]$ for $\left| \gamma/\lambda \right| > 1$; all the $2L^{2}$ eigenstates do not exhibit the skin effect and delocalized through the bulk. For $\left| \gamma/\lambda \right| < 1$, on the other hand, we have $L-1$ real solutions in $k_{y} \in \left[ 0, \pi \right]$; the corresponding $2L \left( L-1 \right)$ eigenstates are delocalized, while the other $2L$ eigenstates are the corner skin modes.

%%%%%
\section{Edge modes}
	\label{appendix: edge}

\subsection{Open boundary conditions along the $y$ direction}
    \label{appendix: edge-OBCy}

We consider the non-Hermitian Hamiltonian in Eq.~(\ref{eq: NH-BBH-2D}), imposing the open boundary conditions along the $y$ direction and the periodic boundary conditions along the $x$ direction. The Schr\"odinger equation is given as
\begin{equation}
M_{k_{x}} \vec{\psi} \left( n \right)
+ T_{y+} \vec{\psi} \left( n-1 \right) 
+ T_{y-} \vec{\psi} \left( n+1 \right) = E\,\vec{\psi} \left( n \right)     \label{aeq: bulk-OBCy}
\end{equation}
in the bulk ($n = 2, 3, \cdots, L-1$), and 
\begin{eqnarray}
M_{k_{x}} \vec{\psi} \left( 1 \right)
+ T_{y-} \vec{\psi} \left( 2 \right) &=& E\,\vec{\psi} \left( 1 \right)
        \label{aeq: edge1-OBCy} \\
M_{k_{x}} \vec{\psi} \left( L \right)
+ T_{y+} \vec{\psi} \left( L-1 \right) &=& E\,\vec{\psi} \left( L \right)
        \label{aeq: edge2-OBCy}
\end{eqnarray}
at the edges. Here, $T_{y\pm}$ is defined as Eq.~(\ref{aeq: hopping-y}), and $M_{k_{x}}$ is defined as
\begin{equation}
M_{k_{x}} := -\ii \left( \gamma + \lambda \cos k_{x} \right) + \lambda \left( \sin k_{x} \right) \sigma_{z} + \gamma \sigma_{y}.
\end{equation}
When $\vec{\psi} \left( 0 \right)$ and $\vec{\psi} \left( L+1 \right)$ are respectively defined by the bulk equations~(\ref{aeq: bulk-OBCy}) for $n=1$ and $n=L$, the boundary conditions in Eqs.~(\ref{aeq: edge1-OBCy}) and (\ref{aeq: edge2-OBCy}) reduce to
\begin{equation}
    T_{y+} \vec{\psi} \left( 0 \right) 
    = T_{y-} \vec{\psi} \left( L+1 \right) 
    = 0.
    \label{aeq: edge3-OBCy}
\end{equation}

Now, suppose $\beta_{y}$ is a solution to the characteristic equation $\det \left[ H \left( \beta_{y} \right) - E \right] = 0$ for an eigenenergy $E \in \mathbb{C}$, where the bulk Hamiltonian $H \left( \beta_{y} \right)$ is given as
\begin{equation}
    H \left( \beta_{y} \right) = M_{k_{x}} + \beta_{y} T_{y-} + \beta_{y}^{-1} T_{y+}.
\end{equation}
Because of transposition-associated mirror symmetry in Eq.~(\ref{eq: NH-BBH-transposition-y}), $\beta_{y}^{-1}$ is another solution to the characteristic equation for the same eigenenergy $E$. Hence, the corresponding eigenstate is generally expanded as
\begin{equation}
    \vec{\psi} \left( n \right) = \beta_{y}^{n} \vec{c}_{+} + \beta_{y}^{L+1-n} \vec{c}_{-}
\end{equation}
with $\left| \beta_{y} \right| \leq 1$ and $\vec{c}_{\pm} \in \mathbb{C}^{2}$. The boundary conditions in Eq.~(\ref{aeq: edge3-OBCy}) further reduce to
\begin{equation}
    T_{y+} \left( \vec{c}_{+} + \beta_{y}^{L+1} \vec{c}_{-} \right) 
    = T_{y-} \left( \beta_{y}^{L+1} \vec{c}_{+} + \vec{c}_{-} \right)
    = 0.
\end{equation}
Thus, for $\left| \beta_{y} \right| < 1$ and sufficiently large $L$, we need
\begin{equation}
    \vec{c}_{+} \simeq \begin{pmatrix}
    0 \\ 1
    \end{pmatrix},\quad
    \vec{c}_{-} \simeq \begin{pmatrix}
    1 \\ 0
    \end{pmatrix}.
\end{equation}
We note that this is not necessarily required for $\left| \beta_{y} \right| = 1$. Since $\beta_{y}^{\pm} \vec{c}_{\pm}$ is an eigenstate of the bulk Hamiltonian $H \left( \beta_{y} \right)$, we finally have
\begin{eqnarray}
    E &=& - \ii \gamma - \ii \lambda e^{-\ii k_{x}}, \\
    \beta_{y} &=& - \frac{\gamma}{\lambda}.
\end{eqnarray}
For the appearance of these edge modes, we need the normalization condition $\left| \beta_{y} \right| < 1$, i.e.,
\begin{equation}
    \left| \frac{\gamma}{\lambda} \right| < 1.
\end{equation}
The obtained analytical results are consistent with the numerical results in Fig.~\ref{fig: spectrum-SOSE}.

\subsection{Open boundary conditions along the $x$ direction}
    \label{appendix: edge-OBCx}

We next consider the non-Hermitian Hamiltonian in Eq.~(\ref{eq: NH-BBH-2D}), imposing the open boundary conditions along the $x$ direction and the periodic boundary conditions along the $y$ direction. The Schr\"odinger equation is given as
\begin{equation}
M_{k_{y}} \vec{\psi} \left( n \right)
+ T_{x+} \vec{\psi} \left( n-1 \right) 
+ T_{x-} \vec{\psi} \left( n+1 \right) = E\,\vec{\psi} \left( n \right)     
    \label{aeq: bulk-OBCx}
\end{equation}
in the bulk ($n = 1, 2, \cdots, L$), and 
\begin{equation}
    T_{x+} \vec{\psi} \left( 0 \right) 
    = T_{x-} \vec{\psi} \left( L+1 \right) 
    = 0.
    \label{aeq: edge3-OBCx}
\end{equation}
at the edges. Here, $T_{x\pm}$ is defined as Eq.~(\ref{aeq: hopping-x}), and $M_{k_{y}}$ is defined as
\begin{equation}
M_{k_{y}} := -\ii \gamma + \left( \gamma + \lambda \cos k_{y} \right) \sigma_{y} + \lambda \left( \sin k_{y} \right) \sigma_{x}.
\end{equation}
The bulk Hamiltonian $H \left( \beta_{x} \right)$ is given as
\begin{equation}
    H \left( \beta_{x} \right) = M_{k_{y}} + \beta_{x} T_{x-} + \beta_{x}^{-1} T_{x+}.
\end{equation}

Similarly to Sec.~\ref{appendix: edge-OBCy}, an eigenstate is generally expanded as
\begin{equation}
    \vec{\psi} \left( n \right) = \beta_{x}^{n} \vec{c}_{+} + \beta_{x}^{L+1-n} \vec{c}_{-}
\end{equation}
with $\left| \beta_{x} \right| \leq 1$ and $\vec{c}_{\pm} \in \mathbb{C}^{2}$. The boundary conditions in Eq.~(\ref{aeq: edge3-OBCx}) further reduce to
\begin{equation}
    T_{x+} \left( \vec{c}_{+} + \beta_{x}^{L+1} \vec{c}_{-} \right) 
    = T_{x-} \left( \beta_{x}^{L+1} \vec{c}_{+} + \vec{c}_{-} \right)
    = 0.
\end{equation}
Thus, for $\left| \beta_{x} \right| < 1$ and sufficiently large $L$, we need
\begin{equation}
    \vec{c}_{+} \simeq \begin{pmatrix}
    1 \\ 0
    \end{pmatrix},\quad
    \vec{c}_{-} \simeq \begin{pmatrix}
    0 \\ 1
    \end{pmatrix}.
\end{equation}
Since $\beta_{x}^{\pm} \vec{c}_{\pm}$ is an eigenstate of the bulk Hamiltonian $H \left( \beta_{x} \right)$, we have
\begin{eqnarray}
    E &=& - \ii \gamma - \ii \lambda \beta_{x}, \\
    e^{-\ii k_{y}} &=& - \frac{\gamma}{\lambda}.
            \label{aeq: OBCx-cond}
\end{eqnarray}
To satisfy Eq.~(\ref{aeq: OBCx-cond}), we need $e^{-\ii k_{y}} \in \mathbb{R}$, i.e., $k_{y} = 0, \pi$. For $k_{y} = 0$ ($k_{y} = \pi$), Eq.~(\ref{aeq: OBCx-cond}) leads to $\gamma = - \lambda$ ($\gamma = \lambda$). Thus, in contrast to Sec.~\ref{appendix: edge-OBCy}, the parameters $\gamma$ and $\lambda$ should be fine-tuned for the appearance of the edge modes. For these fine-tuned parameters, we have $M_{k_{y} = 0, \pi} = -\ii \gamma$, and hence $\beta_{x} = 0$ and $E = -\ii \gamma$. All the $2L$ eigenstates of $H \left( \beta_{x} \right)$ with $k_{y} = 0, \pi$ belong to the same eigenenergy and form an exceptional point. A half of the eigenstates are localized at the left edge and the other half of them are localized at the right edge. The obtained analytical results are consistent with the numerical results in Fig.~\ref{fig: spectrum-SOSE}.

\bibliography{NH-topo}

\begin{thebibliography}{129}%
\makeatletter
\providecommand \@ifxundefined [1]{%
 \@ifx{#1\undefined}
}%
\providecommand \@ifnum [1]{%
 \ifnum #1\expandafter \@firstoftwo
 \else \expandafter \@secondoftwo
 \fi
}%
\providecommand \@ifx [1]{%
 \ifx #1\expandafter \@firstoftwo
 \else \expandafter \@secondoftwo
 \fi
}%
\providecommand \natexlab [1]{#1}%
\providecommand \enquote  [1]{``#1''}%
\providecommand \bibnamefont  [1]{#1}%
\providecommand \bibfnamefont [1]{#1}%
\providecommand \citenamefont [1]{#1}%
\providecommand \href@noop [0]{\@secondoftwo}%
\providecommand \href [0]{\begingroup \@sanitize@url \@href}%
\providecommand \@href[1]{\@@startlink{#1}\@@href}%
\providecommand \@@href[1]{\endgroup#1\@@endlink}%
\providecommand \@sanitize@url [0]{\catcode `\\12\catcode `\$12\catcode
  `\&12\catcode `\#12\catcode `\^12\catcode `\_12\catcode `\%12\relax}%
\providecommand \@@startlink[1]{}%
\providecommand \@@endlink[0]{}%
\providecommand \url  [0]{\begingroup\@sanitize@url \@url }%
\providecommand \@url [1]{\endgroup\@href {#1}{\urlprefix }}%
\providecommand \urlprefix  [0]{URL }%
\providecommand \Eprint [0]{\href }%
\providecommand \doibase [0]{http://dx.doi.org/}%
\providecommand \selectlanguage [0]{\@gobble}%
\providecommand \bibinfo  [0]{\@secondoftwo}%
\providecommand \bibfield  [0]{\@secondoftwo}%
\providecommand \translation [1]{[#1]}%
\providecommand \BibitemOpen [0]{}%
\providecommand \bibitemStop [0]{}%
\providecommand \bibitemNoStop [0]{.\EOS\space}%
\providecommand \EOS [0]{\spacefactor3000\relax}%
\providecommand \BibitemShut  [1]{\csname bibitem#1\endcsname}%
\let\auto@bib@innerbib\@empty
%</preamble>
\bibitem [{\citenamefont {Hasan}\ and\ \citenamefont
  {Kane}(2010)}]{Kane-review}%
  \BibitemOpen
  \bibfield  {author} {\bibinfo {author} {\bibfnamefont {M.~Z.}\ \bibnamefont
  {Hasan}}\ and\ \bibinfo {author} {\bibfnamefont {C.~L.}\ \bibnamefont
  {Kane}},\ }\bibfield  {title} {\enquote {\bibinfo {title} {{Colloquium:
  Topological insulators}},}\ }\href@noop {} {\bibfield  {journal} {\bibinfo
  {journal} {Rev. Mod. Phys.}\ }\textbf {\bibinfo {volume} {82}},\ \bibinfo
  {pages} {3045} (\bibinfo {year} {2010})}\BibitemShut {NoStop}%
\bibitem [{\citenamefont {Qi}\ and\ \citenamefont
  {Zhang}(2011)}]{Zhang-review}%
  \BibitemOpen
  \bibfield  {author} {\bibinfo {author} {\bibfnamefont {X.-L.}\ \bibnamefont
  {Qi}}\ and\ \bibinfo {author} {\bibfnamefont {S.-C.}\ \bibnamefont {Zhang}},\
  }\bibfield  {title} {\enquote {\bibinfo {title} {{Topological insulators and
  superconductors}},}\ }\href@noop {} {\bibfield  {journal} {\bibinfo
  {journal} {{Rev. Mod. Phys.}}\ }\textbf {\bibinfo {volume} {83}},\ \bibinfo
  {pages} {1057} (\bibinfo {year} {2011})}\BibitemShut {NoStop}%
\bibitem [{\citenamefont {Chiu}\ \emph {et~al.}(2016)\citenamefont {Chiu},
  \citenamefont {Teo}, \citenamefont {Schnyder},\ and\ \citenamefont
  {Ryu}}]{Schnyder-Ryu-review}%
  \BibitemOpen
  \bibfield  {author} {\bibinfo {author} {\bibfnamefont {C.-K.}\ \bibnamefont
  {Chiu}}, \bibinfo {author} {\bibfnamefont {J.~C.~Y.}\ \bibnamefont {Teo}},
  \bibinfo {author} {\bibfnamefont {A.~P.}\ \bibnamefont {Schnyder}}, \ and\
  \bibinfo {author} {\bibfnamefont {S.}~\bibnamefont {Ryu}},\ }\bibfield
  {title} {\enquote {\bibinfo {title} {{Classification of topological quantum
  matter with symmetries}},}\ }\href@noop {} {\bibfield  {journal} {\bibinfo
  {journal} {Rev. Mod. Phys.}\ }\textbf {\bibinfo {volume} {88}},\ \bibinfo
  {pages} {035005} (\bibinfo {year} {2016})}\BibitemShut {NoStop}%
\bibitem [{\citenamefont {Su}\ \emph {et~al.}(1979)\citenamefont {Su},
  \citenamefont {Schrieffer},\ and\ \citenamefont {Heeger}}]{SSH-79}%
  \BibitemOpen
  \bibfield  {author} {\bibinfo {author} {\bibfnamefont {W.~P.}\ \bibnamefont
  {Su}}, \bibinfo {author} {\bibfnamefont {J.~R.}\ \bibnamefont {Schrieffer}},
  \ and\ \bibinfo {author} {\bibfnamefont {A.~J.}\ \bibnamefont {Heeger}},\
  }\bibfield  {title} {\enquote {\bibinfo {title} {{Solitons in
  Polyacetylene}},}\ }\href@noop {} {\bibfield  {journal} {\bibinfo  {journal}
  {Phys. Rev. Lett.}\ }\textbf {\bibinfo {volume} {42}},\ \bibinfo {pages}
  {1698} (\bibinfo {year} {1979})}\BibitemShut {NoStop}%
\bibitem [{\citenamefont {Haldane}(1988)}]{Haldane-88}%
  \BibitemOpen
  \bibfield  {author} {\bibinfo {author} {\bibfnamefont {F.~D.~M.}\
  \bibnamefont {Haldane}},\ }\bibfield  {title} {\enquote {\bibinfo {title}
  {{Model for a Quantum Hall Effect without Landau Levels: Condensed-Matter
  Realization of the ``Parity Anomaly''}},}\ }\href@noop {} {\bibfield
  {journal} {\bibinfo  {journal} {Phys. Rev. Lett.}\ }\textbf {\bibinfo
  {volume} {61}},\ \bibinfo {pages} {2015} (\bibinfo {year}
  {1988})}\BibitemShut {NoStop}%
\bibitem [{\citenamefont {Kane}\ and\ \citenamefont
  {Mele}(2005{\natexlab{a}})}]{Kane-Mele-05-QSH}%
  \BibitemOpen
  \bibfield  {author} {\bibinfo {author} {\bibfnamefont {C.~L.}\ \bibnamefont
  {Kane}}\ and\ \bibinfo {author} {\bibfnamefont {E.~J.}\ \bibnamefont
  {Mele}},\ }\bibfield  {title} {\enquote {\bibinfo {title} {{Quantum Spin Hall
  Effect in Graphene}},}\ }\href@noop {} {\bibfield  {journal} {\bibinfo
  {journal} {Phys. Rev. Lett.}\ }\textbf {\bibinfo {volume} {95}},\ \bibinfo
  {pages} {226801} (\bibinfo {year} {2005}{\natexlab{a}})}\BibitemShut
  {NoStop}%
\bibitem [{\citenamefont {Kane}\ and\ \citenamefont
  {Mele}(2005{\natexlab{b}})}]{Kane-Mele-05-Z2}%
  \BibitemOpen
  \bibfield  {author} {\bibinfo {author} {\bibfnamefont {C.~L.}\ \bibnamefont
  {Kane}}\ and\ \bibinfo {author} {\bibfnamefont {E.~J.}\ \bibnamefont
  {Mele}},\ }\bibfield  {title} {\enquote {\bibinfo {title} {{$Z_{\rm 2}$
  Topological Order and the Quantum Spin Hall Effect}},}\ }\href@noop {}
  {\bibfield  {journal} {\bibinfo  {journal} {{Phys. Rev. Lett.}}\ }\textbf
  {\bibinfo {volume} {95}},\ \bibinfo {pages} {146802} (\bibinfo {year}
  {2005}{\natexlab{b}})}\BibitemShut {NoStop}%
\bibitem [{\citenamefont {Benalcazar}\ \emph
  {et~al.}(2017{\natexlab{a}})\citenamefont {Benalcazar}, \citenamefont
  {Bernevig},\ and\ \citenamefont {Hughes}}]{BBH-17}%
  \BibitemOpen
  \bibfield  {author} {\bibinfo {author} {\bibfnamefont {W.~A.}\ \bibnamefont
  {Benalcazar}}, \bibinfo {author} {\bibfnamefont {B.~A.}\ \bibnamefont
  {Bernevig}}, \ and\ \bibinfo {author} {\bibfnamefont {T.~L.}\ \bibnamefont
  {Hughes}},\ }\bibfield  {title} {\enquote {\bibinfo {title} {{Quantized
  electric multipole insulators}},}\ }\href@noop {} {\bibfield  {journal}
  {\bibinfo  {journal} {Science}\ }\textbf {\bibinfo {volume} {357}},\ \bibinfo
  {pages} {61} (\bibinfo {year} {2017}{\natexlab{a}})}\BibitemShut {NoStop}%
\bibitem [{\citenamefont {Benalcazar}\ \emph
  {et~al.}(2017{\natexlab{b}})\citenamefont {Benalcazar}, \citenamefont
  {Bernevig},\ and\ \citenamefont {Hughes}}]{BBH-17B}%
  \BibitemOpen
  \bibfield  {author} {\bibinfo {author} {\bibfnamefont {W.~A.}\ \bibnamefont
  {Benalcazar}}, \bibinfo {author} {\bibfnamefont {B.~A.}\ \bibnamefont
  {Bernevig}}, \ and\ \bibinfo {author} {\bibfnamefont {T.~L.}\ \bibnamefont
  {Hughes}},\ }\bibfield  {title} {\enquote {\bibinfo {title} {{Electric
  multipole moments, topological multipole moment pumping, and chiral hinge
  states in crystalline insulators}},}\ }\href@noop {} {\bibfield  {journal}
  {\bibinfo  {journal} {Phys. Rev. B}\ }\textbf {\bibinfo {volume} {96}},\
  \bibinfo {pages} {245115} (\bibinfo {year} {2017}{\natexlab{b}})}\BibitemShut
  {NoStop}%
\bibitem [{\citenamefont {Langbehn}\ \emph {et~al.}(2017)\citenamefont
  {Langbehn}, \citenamefont {Peng}, \citenamefont {Trifunovic}, \citenamefont
  {von Oppen},\ and\ \citenamefont {Brouwer}}]{Langbehn-17}%
  \BibitemOpen
  \bibfield  {author} {\bibinfo {author} {\bibfnamefont {J.}~\bibnamefont
  {Langbehn}}, \bibinfo {author} {\bibfnamefont {Y.}~\bibnamefont {Peng}},
  \bibinfo {author} {\bibfnamefont {L.}~\bibnamefont {Trifunovic}}, \bibinfo
  {author} {\bibfnamefont {F.}~\bibnamefont {von Oppen}}, \ and\ \bibinfo
  {author} {\bibfnamefont {P.~W.}\ \bibnamefont {Brouwer}},\ }\bibfield
  {title} {\enquote {\bibinfo {title} {{Reflection-Symmetric Second-Order
  Topological Insulators and Superconductors}},}\ }\href@noop {} {\bibfield
  {journal} {\bibinfo  {journal} {Phys. Rev. Lett.}\ }\textbf {\bibinfo
  {volume} {119}},\ \bibinfo {pages} {246401} (\bibinfo {year}
  {2017})}\BibitemShut {NoStop}%
\bibitem [{\citenamefont {Song}\ \emph {et~al.}(2017)\citenamefont {Song},
  \citenamefont {Fang},\ and\ \citenamefont {Fang}}]{Song-17}%
  \BibitemOpen
  \bibfield  {author} {\bibinfo {author} {\bibfnamefont {Z.}~\bibnamefont
  {Song}}, \bibinfo {author} {\bibfnamefont {Z.}~\bibnamefont {Fang}}, \ and\
  \bibinfo {author} {\bibfnamefont {C.}~\bibnamefont {Fang}},\ }\bibfield
  {title} {\enquote {\bibinfo {title} {{$\left( d-2 \right)$-Dimensional Edge
  States of Rotation Symmetry Protected Topological States}},}\ }\href@noop {}
  {\bibfield  {journal} {\bibinfo  {journal} {Phys. Rev. Lett.}\ }\textbf
  {\bibinfo {volume} {119}},\ \bibinfo {pages} {246402} (\bibinfo {year}
  {2017})}\BibitemShut {NoStop}%
\bibitem [{\citenamefont {Fang}\ and\ \citenamefont {Fu}(2019)}]{Fang-Fu-19}%
  \BibitemOpen
  \bibfield  {author} {\bibinfo {author} {\bibfnamefont {C.}~\bibnamefont
  {Fang}}\ and\ \bibinfo {author} {\bibfnamefont {L.}~\bibnamefont {Fu}},\
  }\bibfield  {title} {\enquote {\bibinfo {title} {{New classes of topological
  crystalline insulators having surface rotation anomaly}},}\ }\href@noop {}
  {\bibfield  {journal} {\bibinfo  {journal} {Sci. Adv.}\ }\textbf {\bibinfo
  {volume} {5}},\ \bibinfo {pages} {eaat2374} (\bibinfo {year}
  {2019})}\BibitemShut {NoStop}%
\bibitem [{\citenamefont {Kunst}\ \emph
  {et~al.}(2018{\natexlab{a}})\citenamefont {Kunst}, \citenamefont {van
  Miert},\ and\ \citenamefont {Bergholtz}}]{Kunst-18}%
  \BibitemOpen
  \bibfield  {author} {\bibinfo {author} {\bibfnamefont {F.~K.}\ \bibnamefont
  {Kunst}}, \bibinfo {author} {\bibfnamefont {G.}~\bibnamefont {van Miert}}, \
  and\ \bibinfo {author} {\bibfnamefont {E.~J.}\ \bibnamefont {Bergholtz}},\
  }\bibfield  {title} {\enquote {\bibinfo {title} {{Lattice models with exactly
  solvable topological hinge and corner states}},}\ }\href@noop {} {\bibfield
  {journal} {\bibinfo  {journal} {Phys. Rev. B}\ }\textbf {\bibinfo {volume}
  {97}},\ \bibinfo {pages} {241405(R)} (\bibinfo {year}
  {2018}{\natexlab{a}})}\BibitemShut {NoStop}%
\bibitem [{\citenamefont {Shindler}\ \emph
  {et~al.}(2018{\natexlab{a}})\citenamefont {Shindler}, \citenamefont {Cook},
  \citenamefont {Vergniory}, \citenamefont {Wang}, \citenamefont {Parkin},
  \citenamefont {Bernevig},\ and\ \citenamefont {Neupert}}]{Schindler-18S}%
  \BibitemOpen
  \bibfield  {author} {\bibinfo {author} {\bibfnamefont {F.}~\bibnamefont
  {Shindler}}, \bibinfo {author} {\bibfnamefont {A.~M.}\ \bibnamefont {Cook}},
  \bibinfo {author} {\bibfnamefont {M.~G.}\ \bibnamefont {Vergniory}}, \bibinfo
  {author} {\bibfnamefont {Z.}~\bibnamefont {Wang}}, \bibinfo {author}
  {\bibfnamefont {S.~S.~P.}\ \bibnamefont {Parkin}}, \bibinfo {author}
  {\bibfnamefont {B.~A.}\ \bibnamefont {Bernevig}}, \ and\ \bibinfo {author}
  {\bibfnamefont {T.}~\bibnamefont {Neupert}},\ }\bibfield  {title} {\enquote
  {\bibinfo {title} {{Higher-order topological insulators}},}\ }\href@noop {}
  {\bibfield  {journal} {\bibinfo  {journal} {Sci. Adv.}\ }\textbf {\bibinfo
  {volume} {4}},\ \bibinfo {pages} {eaat0346} (\bibinfo {year}
  {2018}{\natexlab{a}})}\BibitemShut {NoStop}%
\bibitem [{\citenamefont {Khalaf}(2018)}]{Khalaf-18B}%
  \BibitemOpen
  \bibfield  {author} {\bibinfo {author} {\bibfnamefont {E.}~\bibnamefont
  {Khalaf}},\ }\bibfield  {title} {\enquote {\bibinfo {title} {{Higher-order
  topological insulators and superconductors protected by inversion
  symmetry}},}\ }\href@noop {} {\bibfield  {journal} {\bibinfo  {journal}
  {Phys. Rev. B}\ }\textbf {\bibinfo {volume} {97}},\ \bibinfo {pages} {205136}
  (\bibinfo {year} {2018})}\BibitemShut {NoStop}%
\bibitem [{\citenamefont {Khalaf}\ \emph {et~al.}(2018)\citenamefont {Khalaf},
  \citenamefont {Po}, \citenamefont {Vishwanath},\ and\ \citenamefont
  {Watanabe}}]{Khalaf-18X}%
  \BibitemOpen
  \bibfield  {author} {\bibinfo {author} {\bibfnamefont {E.}~\bibnamefont
  {Khalaf}}, \bibinfo {author} {\bibfnamefont {H.~C.}\ \bibnamefont {Po}},
  \bibinfo {author} {\bibfnamefont {A.}~\bibnamefont {Vishwanath}}, \ and\
  \bibinfo {author} {\bibfnamefont {H.}~\bibnamefont {Watanabe}},\ }\bibfield
  {title} {\enquote {\bibinfo {title} {{Symmetry Indicators and Anomalous
  Surface States of Topological Crystalline Insulators}},}\ }\href@noop {}
  {\bibfield  {journal} {\bibinfo  {journal} {Phys. Rev. X}\ }\textbf {\bibinfo
  {volume} {8}},\ \bibinfo {pages} {031070} (\bibinfo {year}
  {2018})}\BibitemShut {NoStop}%
\bibitem [{\citenamefont {Yan}\ \emph {et~al.}(2018)\citenamefont {Yan},
  \citenamefont {Song},\ and\ \citenamefont {Wang}}]{Yan-18}%
  \BibitemOpen
  \bibfield  {author} {\bibinfo {author} {\bibfnamefont {Z.}~\bibnamefont
  {Yan}}, \bibinfo {author} {\bibfnamefont {F.}~\bibnamefont {Song}}, \ and\
  \bibinfo {author} {\bibfnamefont {Z.}~\bibnamefont {Wang}},\ }\bibfield
  {title} {\enquote {\bibinfo {title} {{Majorana Corner Modes in a
  High-Temperature Platform}},}\ }\href@noop {} {\bibfield  {journal} {\bibinfo
   {journal} {Phys. Rev. Lett.}\ }\textbf {\bibinfo {volume} {121}},\ \bibinfo
  {pages} {096803} (\bibinfo {year} {2018})}\BibitemShut {NoStop}%
\bibitem [{\citenamefont {Matsugatani}\ and\ \citenamefont
  {Watanabe}(2018)}]{Matsugatani-18}%
  \BibitemOpen
  \bibfield  {author} {\bibinfo {author} {\bibfnamefont {A.}~\bibnamefont
  {Matsugatani}}\ and\ \bibinfo {author} {\bibfnamefont {H.}~\bibnamefont
  {Watanabe}},\ }\bibfield  {title} {\enquote {\bibinfo {title} {{Connecting
  higher-order topological insulators to lower-dimensional topological
  insulators}},}\ }\href@noop {} {\bibfield  {journal} {\bibinfo  {journal}
  {Phys. Rev. B}\ }\textbf {\bibinfo {volume} {98}},\ \bibinfo {pages} {205129}
  (\bibinfo {year} {2018})}\BibitemShut {NoStop}%
\bibitem [{\citenamefont {Trifunovic}\ and\ \citenamefont
  {Brouwer}(2019)}]{Trifunovic-19}%
  \BibitemOpen
  \bibfield  {author} {\bibinfo {author} {\bibfnamefont {L.}~\bibnamefont
  {Trifunovic}}\ and\ \bibinfo {author} {\bibfnamefont {P.~W.}\ \bibnamefont
  {Brouwer}},\ }\bibfield  {title} {\enquote {\bibinfo {title} {{Higher-Order
  Bulk-Boundary Correspondence for Topological Crystalline Phases}},}\
  }\href@noop {} {\bibfield  {journal} {\bibinfo  {journal} {Phys. Rev. X}\
  }\textbf {\bibinfo {volume} {9}},\ \bibinfo {pages} {011012} (\bibinfo {year}
  {2019})}\BibitemShut {NoStop}%
\bibitem [{\citenamefont {Liu}\ \emph {et~al.}(2019{\natexlab{a}})\citenamefont
  {Liu}, \citenamefont {Vishwanath},\ and\ \citenamefont {Khalaf}}]{Liu-19}%
  \BibitemOpen
  \bibfield  {author} {\bibinfo {author} {\bibfnamefont {S.}~\bibnamefont
  {Liu}}, \bibinfo {author} {\bibfnamefont {A.}~\bibnamefont {Vishwanath}}, \
  and\ \bibinfo {author} {\bibfnamefont {E.}~\bibnamefont {Khalaf}},\
  }\bibfield  {title} {\enquote {\bibinfo {title} {{Shift Insulators:
  Rotation-Protected Two-Dimensional Topological Crystalline Insulators}},}\
  }\href@noop {} {\bibfield  {journal} {\bibinfo  {journal} {Phys. Rev. X}\
  }\textbf {\bibinfo {volume} {9}},\ \bibinfo {pages} {031003} (\bibinfo {year}
  {2019}{\natexlab{a}})}\BibitemShut {NoStop}%
\bibitem [{\citenamefont {Benalcazar}\ \emph {et~al.}(2019)\citenamefont
  {Benalcazar}, \citenamefont {Li},\ and\ \citenamefont
  {Hughes}}]{Benalcazar-19}%
  \BibitemOpen
  \bibfield  {author} {\bibinfo {author} {\bibfnamefont {W.~A.}\ \bibnamefont
  {Benalcazar}}, \bibinfo {author} {\bibfnamefont {T.}~\bibnamefont {Li}}, \
  and\ \bibinfo {author} {\bibfnamefont {T.~L.}\ \bibnamefont {Hughes}},\
  }\bibfield  {title} {\enquote {\bibinfo {title} {{Quantization of fractional
  corner charge in ${C}_{n}$-symmetric higher-order topological crystalline
  insulators}},}\ }\href@noop {} {\bibfield  {journal} {\bibinfo  {journal}
  {Phys. Rev. B}\ }\textbf {\bibinfo {volume} {99}},\ \bibinfo {pages} {245151}
  (\bibinfo {year} {2019})}\BibitemShut {NoStop}%
\bibitem [{\citenamefont {Serra-Garcia}\ \emph {et~al.}(2018)\citenamefont
  {Serra-Garcia}, \citenamefont {Peri}, \citenamefont {S\"usstrunk},
  \citenamefont {Bilal}, \citenamefont {Larsen}, \citenamefont {Villanueva},\
  and\ \citenamefont {Huber}}]{Serra-Garcia-18}%
  \BibitemOpen
  \bibfield  {author} {\bibinfo {author} {\bibfnamefont {M.}~\bibnamefont
  {Serra-Garcia}}, \bibinfo {author} {\bibfnamefont {V.}~\bibnamefont {Peri}},
  \bibinfo {author} {\bibfnamefont {R.}~\bibnamefont {S\"usstrunk}}, \bibinfo
  {author} {\bibfnamefont {O.~R.}\ \bibnamefont {Bilal}}, \bibinfo {author}
  {\bibfnamefont {T.}~\bibnamefont {Larsen}}, \bibinfo {author} {\bibfnamefont
  {L.~G.}\ \bibnamefont {Villanueva}}, \ and\ \bibinfo {author} {\bibfnamefont
  {S.~D.}\ \bibnamefont {Huber}},\ }\bibfield  {title} {\enquote {\bibinfo
  {title} {Observation of a phononic quadrupole topological insulator},}\
  }\href@noop {} {\bibfield  {journal} {\bibinfo  {journal} {Nature}\ }\textbf
  {\bibinfo {volume} {555}},\ \bibinfo {pages} {342} (\bibinfo {year}
  {2018})}\BibitemShut {NoStop}%
\bibitem [{\citenamefont {Imhof}\ \emph {et~al.}(2018)\citenamefont {Imhof},
  \citenamefont {Berger}, \citenamefont {Bayer}, \citenamefont {Brehm},
  \citenamefont {Molenkamp}, \citenamefont {Kiessling}, \citenamefont
  {Schindler}, \citenamefont {Lee}, \citenamefont {Greiter}, \citenamefont
  {Neupert},\ and\ \citenamefont {Thomale}}]{Imhof-18}%
  \BibitemOpen
  \bibfield  {author} {\bibinfo {author} {\bibfnamefont {S.}~\bibnamefont
  {Imhof}}, \bibinfo {author} {\bibfnamefont {C.}~\bibnamefont {Berger}},
  \bibinfo {author} {\bibfnamefont {F.}~\bibnamefont {Bayer}}, \bibinfo
  {author} {\bibfnamefont {J.}~\bibnamefont {Brehm}}, \bibinfo {author}
  {\bibfnamefont {L.~W.}\ \bibnamefont {Molenkamp}}, \bibinfo {author}
  {\bibfnamefont {T.}~\bibnamefont {Kiessling}}, \bibinfo {author}
  {\bibfnamefont {F.}~\bibnamefont {Schindler}}, \bibinfo {author}
  {\bibfnamefont {C.~H.}\ \bibnamefont {Lee}}, \bibinfo {author} {\bibfnamefont
  {M.}~\bibnamefont {Greiter}}, \bibinfo {author} {\bibfnamefont
  {T.}~\bibnamefont {Neupert}}, \ and\ \bibinfo {author} {\bibfnamefont
  {R.}~\bibnamefont {Thomale}},\ }\bibfield  {title} {\enquote {\bibinfo
  {title} {{Topolectrical-circuit realization of topological corner modes}},}\
  }\href@noop {} {\bibfield  {journal} {\bibinfo  {journal} {Nat. Phys.}\
  }\textbf {\bibinfo {volume} {14}},\ \bibinfo {pages} {925} (\bibinfo {year}
  {2018})}\BibitemShut {NoStop}%
\bibitem [{\citenamefont {C.~W.~Peterson}\ \emph {et~al.}(2018)\citenamefont
  {C.~W.~Peterson}, \citenamefont {Hughes},\ and\ \citenamefont
  {Bahl}}]{Peterson-18}%
  \BibitemOpen
  \bibfield  {author} {\bibinfo {author} {\bibfnamefont {W.~A.~Benalcazar}\
  \bibnamefont {C.~W.~Peterson}}, \bibinfo {author} {\bibfnamefont {T.~L.}\
  \bibnamefont {Hughes}}, \ and\ \bibinfo {author} {\bibfnamefont
  {G.}~\bibnamefont {Bahl}},\ }\bibfield  {title} {\enquote {\bibinfo {title}
  {A quantized microwave quadrupole insulator with topologically protected
  corner states},}\ }\href@noop {} {\bibfield  {journal} {\bibinfo  {journal}
  {Nature}\ }\textbf {\bibinfo {volume} {555}},\ \bibinfo {pages} {346}
  (\bibinfo {year} {2018})}\BibitemShut {NoStop}%
\bibitem [{\citenamefont {Shindler}\ \emph
  {et~al.}(2018{\natexlab{b}})\citenamefont {Shindler}, \citenamefont {Wang},
  \citenamefont {Vergniory}, \citenamefont {Cook}, \citenamefont {Murani},
  \citenamefont {Sengupta}, \citenamefont {Kasumov}, \citenamefont {Deblock},
  \citenamefont {Jeon}, \citenamefont {Drozdov}, \citenamefont {Bouchiat},
  \citenamefont {Gu\'eron}, \citenamefont {Yazdani}, \citenamefont {Bernevig},\
  and\ \citenamefont {Neupert}}]{Schindler-18N}%
  \BibitemOpen
  \bibfield  {author} {\bibinfo {author} {\bibfnamefont {F.}~\bibnamefont
  {Shindler}}, \bibinfo {author} {\bibfnamefont {Z.}~\bibnamefont {Wang}},
  \bibinfo {author} {\bibfnamefont {M.~G.}\ \bibnamefont {Vergniory}}, \bibinfo
  {author} {\bibfnamefont {A.~M.}\ \bibnamefont {Cook}}, \bibinfo {author}
  {\bibfnamefont {A.}~\bibnamefont {Murani}}, \bibinfo {author} {\bibfnamefont
  {S.}~\bibnamefont {Sengupta}}, \bibinfo {author} {\bibfnamefont {A.~Yu.}\
  \bibnamefont {Kasumov}}, \bibinfo {author} {\bibfnamefont {R.}~\bibnamefont
  {Deblock}}, \bibinfo {author} {\bibfnamefont {S.}~\bibnamefont {Jeon}},
  \bibinfo {author} {\bibfnamefont {I.}~\bibnamefont {Drozdov}}, \bibinfo
  {author} {\bibfnamefont {H.}~\bibnamefont {Bouchiat}}, \bibinfo {author}
  {\bibfnamefont {S.}~\bibnamefont {Gu\'eron}}, \bibinfo {author}
  {\bibfnamefont {A.}~\bibnamefont {Yazdani}}, \bibinfo {author} {\bibfnamefont
  {B.~A.}\ \bibnamefont {Bernevig}}, \ and\ \bibinfo {author} {\bibfnamefont
  {T.}~\bibnamefont {Neupert}},\ }\bibfield  {title} {\enquote {\bibinfo
  {title} {{Higher-order topology in bismuth}},}\ }\href@noop {} {\bibfield
  {journal} {\bibinfo  {journal} {Nat. Phys.}\ }\textbf {\bibinfo {volume}
  {14}},\ \bibinfo {pages} {918} (\bibinfo {year}
  {2018}{\natexlab{b}})}\BibitemShut {NoStop}%
\bibitem [{\citenamefont {Xue}\ \emph {et~al.}(2019)\citenamefont {Xue},
  \citenamefont {Yang}, \citenamefont {Gao}, \citenamefont {Chong},\ and\
  \citenamefont {Zhang}}]{Xue-19}%
  \BibitemOpen
  \bibfield  {author} {\bibinfo {author} {\bibfnamefont {H.}~\bibnamefont
  {Xue}}, \bibinfo {author} {\bibfnamefont {Y.}~\bibnamefont {Yang}}, \bibinfo
  {author} {\bibfnamefont {F.}~\bibnamefont {Gao}}, \bibinfo {author}
  {\bibfnamefont {Y.}~\bibnamefont {Chong}}, \ and\ \bibinfo {author}
  {\bibfnamefont {B.}~\bibnamefont {Zhang}},\ }\bibfield  {title} {\enquote
  {\bibinfo {title} {{Acoustic higher-order topological insulator on a kagome
  lattice}},}\ }\href@noop {} {\bibfield  {journal} {\bibinfo  {journal} {Nat.
  Mater.}\ }\textbf {\bibinfo {volume} {18}},\ \bibinfo {pages} {108} (\bibinfo
  {year} {2019})}\BibitemShut {NoStop}%
\bibitem [{\citenamefont {Ni}\ \emph {et~al.}(2019)\citenamefont {Ni},
  \citenamefont {Weiner}, \citenamefont {Al\`{u}},\ and\ \citenamefont
  {Khanikaev}}]{Ni-19}%
  \BibitemOpen
  \bibfield  {author} {\bibinfo {author} {\bibfnamefont {X.}~\bibnamefont
  {Ni}}, \bibinfo {author} {\bibfnamefont {M.}~\bibnamefont {Weiner}}, \bibinfo
  {author} {\bibfnamefont {A.}~\bibnamefont {Al\`{u}}}, \ and\ \bibinfo
  {author} {\bibfnamefont {A.~B.}\ \bibnamefont {Khanikaev}},\ }\bibfield
  {title} {\enquote {\bibinfo {title} {{Observation of higher-order topological
  acoustic states protected by generalized chiral symmetry}},}\ }\href@noop {}
  {\bibfield  {journal} {\bibinfo  {journal} {Nat. Mater.}\ }\textbf {\bibinfo
  {volume} {18}},\ \bibinfo {pages} {113} (\bibinfo {year} {2019})}\BibitemShut
  {NoStop}%
\bibitem [{\citenamefont {Mittal}\ \emph {et~al.}(2019)\citenamefont {Mittal},
  \citenamefont {Orre}, \citenamefont {Zhu}, \citenamefont {Gorlach},
  \citenamefont {Poddubny},\ and\ \citenamefont {Hafezi}}]{Mittal-19}%
  \BibitemOpen
  \bibfield  {author} {\bibinfo {author} {\bibfnamefont {S.}~\bibnamefont
  {Mittal}}, \bibinfo {author} {\bibfnamefont {V.~V.}\ \bibnamefont {Orre}},
  \bibinfo {author} {\bibfnamefont {G.}~\bibnamefont {Zhu}}, \bibinfo {author}
  {\bibfnamefont {M.~A.}\ \bibnamefont {Gorlach}}, \bibinfo {author}
  {\bibfnamefont {A.}~\bibnamefont {Poddubny}}, \ and\ \bibinfo {author}
  {\bibfnamefont {M.}~\bibnamefont {Hafezi}},\ }\bibfield  {title} {\enquote
  {\bibinfo {title} {{Photonic quadrupole topological phases}},}\ }\href@noop
  {} {\bibfield  {journal} {\bibinfo  {journal} {Nat. Photon.}\ }\textbf
  {\bibinfo {volume} {13}},\ \bibinfo {pages} {692} (\bibinfo {year}
  {2019})}\BibitemShut {NoStop}%
\bibitem [{\citenamefont {Hassan}\ \emph {et~al.}(2019)\citenamefont {Hassan},
  \citenamefont {Kunst}, \citenamefont {Moritz}, \citenamefont {Andler},
  \citenamefont {Bergholtz},\ and\ \citenamefont {Bourennane}}]{Hassan-19}%
  \BibitemOpen
  \bibfield  {author} {\bibinfo {author} {\bibfnamefont {A.~El}\ \bibnamefont
  {Hassan}}, \bibinfo {author} {\bibfnamefont {F.~K.}\ \bibnamefont {Kunst}},
  \bibinfo {author} {\bibfnamefont {A.}~\bibnamefont {Moritz}}, \bibinfo
  {author} {\bibfnamefont {G.}~\bibnamefont {Andler}}, \bibinfo {author}
  {\bibfnamefont {E.~J.}\ \bibnamefont {Bergholtz}}, \ and\ \bibinfo {author}
  {\bibfnamefont {M.}~\bibnamefont {Bourennane}},\ }\bibfield  {title}
  {\enquote {\bibinfo {title} {{Corner states of light in photonic
  waveguides}},}\ }\href@noop {} {\bibfield  {journal} {\bibinfo  {journal}
  {Nat. Photon.}\ }\textbf {\bibinfo {volume} {13}},\ \bibinfo {pages} {697}
  (\bibinfo {year} {2019})}\BibitemShut {NoStop}%
\bibitem [{\citenamefont {Peterson}\ \emph {et~al.}(2020)\citenamefont
  {Peterson}, \citenamefont {Li}, \citenamefont {Benalcazar}, \citenamefont
  {Hughes},\ and\ \citenamefont {Bahl}}]{Peterson-20}%
  \BibitemOpen
  \bibfield  {author} {\bibinfo {author} {\bibfnamefont {C.~W.}\ \bibnamefont
  {Peterson}}, \bibinfo {author} {\bibfnamefont {T.}~\bibnamefont {Li}},
  \bibinfo {author} {\bibfnamefont {W.~A.}\ \bibnamefont {Benalcazar}},
  \bibinfo {author} {\bibfnamefont {T.~L.}\ \bibnamefont {Hughes}}, \ and\
  \bibinfo {author} {\bibfnamefont {G.}~\bibnamefont {Bahl}},\ }\bibfield
  {title} {\enquote {\bibinfo {title} {{A fractional corner anomaly reveals
  higher-order topology}},}\ }\href@noop {} {\bibfield  {journal} {\bibinfo
  {journal} {Science}\ }\textbf {\bibinfo {volume} {368}},\ \bibinfo {pages}
  {1114} (\bibinfo {year} {2020})}\BibitemShut {NoStop}%
\bibitem [{\citenamefont {Ota}\ \emph {et~al.}(2020)\citenamefont {Ota},
  \citenamefont {Takata}, \citenamefont {Ozawa}, \citenamefont {Amo},
  \citenamefont {Jia}, \citenamefont {Kante}, \citenamefont {Notomi},
  \citenamefont {Arakawa},\ and\ \citenamefont {Iwamoto}}]{Ota-review}%
  \BibitemOpen
  \bibfield  {author} {\bibinfo {author} {\bibfnamefont {Y.}~\bibnamefont
  {Ota}}, \bibinfo {author} {\bibfnamefont {K.}~\bibnamefont {Takata}},
  \bibinfo {author} {\bibfnamefont {T.}~\bibnamefont {Ozawa}}, \bibinfo
  {author} {\bibfnamefont {A.}~\bibnamefont {Amo}}, \bibinfo {author}
  {\bibfnamefont {Z.}~\bibnamefont {Jia}}, \bibinfo {author} {\bibfnamefont
  {B.}~\bibnamefont {Kante}}, \bibinfo {author} {\bibfnamefont
  {M.}~\bibnamefont {Notomi}}, \bibinfo {author} {\bibfnamefont
  {Y.}~\bibnamefont {Arakawa}}, \ and\ \bibinfo {author} {\bibfnamefont
  {S.}~\bibnamefont {Iwamoto}},\ }\bibfield  {title} {\enquote {\bibinfo
  {title} {{Active topological photonics}},}\ }\href@noop {} {\bibfield
  {journal} {\bibinfo  {journal} {Nanophotonics}\ }\textbf {\bibinfo {volume}
  {9}},\ \bibinfo {pages} {547} (\bibinfo {year} {2020})}\BibitemShut {NoStop}%
\bibitem [{\citenamefont {Bergholtz}\ \emph {et~al.}()\citenamefont
  {Bergholtz}, \citenamefont {Budich},\ and\ \citenamefont
  {Kunst}}]{Bergholtz-review}%
  \BibitemOpen
  \bibfield  {author} {\bibinfo {author} {\bibfnamefont {E.~J.}\ \bibnamefont
  {Bergholtz}}, \bibinfo {author} {\bibfnamefont {J.~C.}\ \bibnamefont
  {Budich}}, \ and\ \bibinfo {author} {\bibfnamefont {F.~K.}\ \bibnamefont
  {Kunst}},\ }\href@noop {} {\enquote {\bibinfo {title} {{Exceptional Topology
  of Non-Hermitian Systems}},}\ }\bibinfo {note}
  {{arXiv:1912.10048}}\BibitemShut {NoStop}%
\bibitem [{\citenamefont {Konotop}\ \emph {et~al.}(2016)\citenamefont
  {Konotop}, \citenamefont {Yang},\ and\ \citenamefont
  {Zezyulin}}]{Konotop-review}%
  \BibitemOpen
  \bibfield  {author} {\bibinfo {author} {\bibfnamefont {V.~V.}\ \bibnamefont
  {Konotop}}, \bibinfo {author} {\bibfnamefont {J.}~\bibnamefont {Yang}}, \
  and\ \bibinfo {author} {\bibfnamefont {D.~A.}\ \bibnamefont {Zezyulin}},\
  }\bibfield  {title} {\enquote {\bibinfo {title} {{Nonlinear waves in
  $\mathcal{PT}$-symmetric systems}},}\ }\href@noop {} {\bibfield  {journal}
  {\bibinfo  {journal} {Rev. Mod. Phys.}\ }\textbf {\bibinfo {volume} {88}},\
  \bibinfo {pages} {035002} (\bibinfo {year} {2016})}\BibitemShut {NoStop}%
\bibitem [{\citenamefont {El-Ganainy}\ \emph {et~al.}(2018)\citenamefont
  {El-Ganainy}, \citenamefont {Makris}, \citenamefont {Khajavikhan},
  \citenamefont {Musslimani}, \citenamefont {Rotter},\ and\ \citenamefont
  {Christodoulides}}]{Christodoulides-review}%
  \BibitemOpen
  \bibfield  {author} {\bibinfo {author} {\bibfnamefont {R.}~\bibnamefont
  {El-Ganainy}}, \bibinfo {author} {\bibfnamefont {K.~G.}\ \bibnamefont
  {Makris}}, \bibinfo {author} {\bibfnamefont {M.}~\bibnamefont {Khajavikhan}},
  \bibinfo {author} {\bibfnamefont {Z.~H.}\ \bibnamefont {Musslimani}},
  \bibinfo {author} {\bibfnamefont {S.}~\bibnamefont {Rotter}}, \ and\ \bibinfo
  {author} {\bibfnamefont {D.~N.}\ \bibnamefont {Christodoulides}},\ }\bibfield
   {title} {\enquote {\bibinfo {title} {{Non-Hermitian physics and PT
  symmetry}},}\ }\href@noop {} {\bibfield  {journal} {\bibinfo  {journal} {Nat.
  Phys.}\ }\textbf {\bibinfo {volume} {14}},\ \bibinfo {pages} {11} (\bibinfo
  {year} {2018})}\BibitemShut {NoStop}%
\bibitem [{\citenamefont {Rudner}\ and\ \citenamefont
  {Levitov}(2009)}]{Rudner-09}%
  \BibitemOpen
  \bibfield  {author} {\bibinfo {author} {\bibfnamefont {M.~S.}\ \bibnamefont
  {Rudner}}\ and\ \bibinfo {author} {\bibfnamefont {L.~S.}\ \bibnamefont
  {Levitov}},\ }\bibfield  {title} {\enquote {\bibinfo {title} {{Topological
  Transition in a Non-Hermitian Quantum Walk}},}\ }\href@noop {} {\bibfield
  {journal} {\bibinfo  {journal} {Phys. Rev. Lett.}\ }\textbf {\bibinfo
  {volume} {102}},\ \bibinfo {pages} {065703} (\bibinfo {year}
  {2009})}\BibitemShut {NoStop}%
\bibitem [{\citenamefont {Sato}\ \emph {et~al.}(2012)\citenamefont {Sato},
  \citenamefont {Hasebe}, \citenamefont {Esaki},\ and\ \citenamefont
  {Kohmoto}}]{Sato-11}%
  \BibitemOpen
  \bibfield  {author} {\bibinfo {author} {\bibfnamefont {M.}~\bibnamefont
  {Sato}}, \bibinfo {author} {\bibfnamefont {K.}~\bibnamefont {Hasebe}},
  \bibinfo {author} {\bibfnamefont {K.}~\bibnamefont {Esaki}}, \ and\ \bibinfo
  {author} {\bibfnamefont {M.}~\bibnamefont {Kohmoto}},\ }\bibfield  {title}
  {\enquote {\bibinfo {title} {{Time-Reversal Symmetry in Non-Hermitian
  Systems}},}\ }\href@noop {} {\bibfield  {journal} {\bibinfo  {journal} {Prog.
  Theor. Phys.}\ }\textbf {\bibinfo {volume} {127}},\ \bibinfo {pages} {937}
  (\bibinfo {year} {2012})}\BibitemShut {NoStop}%
\bibitem [{\citenamefont {Esaki}\ \emph {et~al.}(2011)\citenamefont {Esaki},
  \citenamefont {Sato}, \citenamefont {Hasebe},\ and\ \citenamefont
  {Kohmoto}}]{Esaki-11}%
  \BibitemOpen
  \bibfield  {author} {\bibinfo {author} {\bibfnamefont {K.}~\bibnamefont
  {Esaki}}, \bibinfo {author} {\bibfnamefont {M.}~\bibnamefont {Sato}},
  \bibinfo {author} {\bibfnamefont {K.}~\bibnamefont {Hasebe}}, \ and\ \bibinfo
  {author} {\bibfnamefont {M.}~\bibnamefont {Kohmoto}},\ }\bibfield  {title}
  {\enquote {\bibinfo {title} {{Edge states and topological phases in
  non-Hermitian systems}},}\ }\href@noop {} {\bibfield  {journal} {\bibinfo
  {journal} {Phys. Rev. B}\ }\textbf {\bibinfo {volume} {84}},\ \bibinfo
  {pages} {205128} (\bibinfo {year} {2011})}\BibitemShut {NoStop}%
\bibitem [{\citenamefont {Hu}\ and\ \citenamefont {Hughes}(2011)}]{Hu-11}%
  \BibitemOpen
  \bibfield  {author} {\bibinfo {author} {\bibfnamefont {Y.~C.}\ \bibnamefont
  {Hu}}\ and\ \bibinfo {author} {\bibfnamefont {T.~L.}\ \bibnamefont
  {Hughes}},\ }\bibfield  {title} {\enquote {\bibinfo {title} {{Absence of
  topological insulator phases in non-Hermitian $\textit{PT}$-symmetric
  Hamiltonians}},}\ }\href@noop {} {\bibfield  {journal} {\bibinfo  {journal}
  {Phys. Rev. B}\ }\textbf {\bibinfo {volume} {84}},\ \bibinfo {pages} {153101}
  (\bibinfo {year} {2011})}\BibitemShut {NoStop}%
\bibitem [{\citenamefont {Schomerus}(2013)}]{Schomerus-13}%
  \BibitemOpen
  \bibfield  {author} {\bibinfo {author} {\bibfnamefont {H.}~\bibnamefont
  {Schomerus}},\ }\bibfield  {title} {\enquote {\bibinfo {title}
  {{Topologically protected midgap states in complex photonic lattices}},}\
  }\href@noop {} {\bibfield  {journal} {\bibinfo  {journal} {Opt. Lett.}\
  }\textbf {\bibinfo {volume} {38}},\ \bibinfo {pages} {1912} (\bibinfo {year}
  {2013})}\BibitemShut {NoStop}%
\bibitem [{\citenamefont {Malzard}\ \emph {et~al.}(2015)\citenamefont
  {Malzard}, \citenamefont {Poli},\ and\ \citenamefont
  {Schomerus}}]{Malzard-15}%
  \BibitemOpen
  \bibfield  {author} {\bibinfo {author} {\bibfnamefont {S.}~\bibnamefont
  {Malzard}}, \bibinfo {author} {\bibfnamefont {C.}~\bibnamefont {Poli}}, \
  and\ \bibinfo {author} {\bibfnamefont {H.}~\bibnamefont {Schomerus}},\
  }\bibfield  {title} {\enquote {\bibinfo {title} {{Topologically Protected
  Defect States in Open Photonic Systems with Non-Hermitian Charge-Conjugation
  and Parity-Time Symmetry}},}\ }\href@noop {} {\bibfield  {journal} {\bibinfo
  {journal} {Phys. Rev. Lett.}\ }\textbf {\bibinfo {volume} {115}},\ \bibinfo
  {pages} {200402} (\bibinfo {year} {2015})}\BibitemShut {NoStop}%
\bibitem [{\citenamefont {Lee}(2016)}]{Lee-16}%
  \BibitemOpen
  \bibfield  {author} {\bibinfo {author} {\bibfnamefont {T.~E.}\ \bibnamefont
  {Lee}},\ }\bibfield  {title} {\enquote {\bibinfo {title} {{Anomalous Edge
  State in a Non-Hermitian Lattice}},}\ }\href@noop {} {\bibfield  {journal}
  {\bibinfo  {journal} {Phys. Rev. Lett.}\ }\textbf {\bibinfo {volume} {116}},\
  \bibinfo {pages} {133903} (\bibinfo {year} {2016})}\BibitemShut {NoStop}%
\bibitem [{\citenamefont {Leykam}\ \emph {et~al.}(2017)\citenamefont {Leykam},
  \citenamefont {Bliokh}, \citenamefont {Huang}, \citenamefont {Chong},\ and\
  \citenamefont {Nori}}]{Leykam-17}%
  \BibitemOpen
  \bibfield  {author} {\bibinfo {author} {\bibfnamefont {D.}~\bibnamefont
  {Leykam}}, \bibinfo {author} {\bibfnamefont {K.~Y.}\ \bibnamefont {Bliokh}},
  \bibinfo {author} {\bibfnamefont {C.}~\bibnamefont {Huang}}, \bibinfo
  {author} {\bibfnamefont {Y.~D.}\ \bibnamefont {Chong}}, \ and\ \bibinfo
  {author} {\bibfnamefont {F.}~\bibnamefont {Nori}},\ }\bibfield  {title}
  {\enquote {\bibinfo {title} {{Edge Modes, Degeneracies, and Topological
  Numbers in Non-Hermitian Systems}},}\ }\href@noop {} {\bibfield  {journal}
  {\bibinfo  {journal} {Phys. Rev. Lett.}\ }\textbf {\bibinfo {volume} {118}},\
  \bibinfo {pages} {040401} (\bibinfo {year} {2017})}\BibitemShut {NoStop}%
\bibitem [{\citenamefont {Xu}\ \emph {et~al.}(2017)\citenamefont {Xu},
  \citenamefont {Wang},\ and\ \citenamefont {Duan}}]{Xu-17}%
  \BibitemOpen
  \bibfield  {author} {\bibinfo {author} {\bibfnamefont {Y.}~\bibnamefont
  {Xu}}, \bibinfo {author} {\bibfnamefont {S.-T.}\ \bibnamefont {Wang}}, \ and\
  \bibinfo {author} {\bibfnamefont {L.-M.}\ \bibnamefont {Duan}},\ }\bibfield
  {title} {\enquote {\bibinfo {title} {{Weyl Exceptional Rings in a
  Three-Dimensional Dissipative Cold Atomic Gas}},}\ }\href@noop {} {\bibfield
  {journal} {\bibinfo  {journal} {Phys. Rev. Lett.}\ }\textbf {\bibinfo
  {volume} {118}},\ \bibinfo {pages} {045701} (\bibinfo {year}
  {2017})}\BibitemShut {NoStop}%
\bibitem [{\citenamefont {Xiong}(2018)}]{Xiong-18}%
  \BibitemOpen
  \bibfield  {author} {\bibinfo {author} {\bibfnamefont {Y.}~\bibnamefont
  {Xiong}},\ }\bibfield  {title} {\enquote {\bibinfo {title} {{Why does bulk
  boundary correspondence fail in some non-hermitian topological models}},}\
  }\href@noop {} {\bibfield  {journal} {\bibinfo  {journal} {J. Phys. Commun.}\
  }\textbf {\bibinfo {volume} {2}},\ \bibinfo {pages} {035043} (\bibinfo {year}
  {2018})}\BibitemShut {NoStop}%
\bibitem [{\citenamefont {Shen}\ \emph {et~al.}(2018)\citenamefont {Shen},
  \citenamefont {Zhen},\ and\ \citenamefont {Fu}}]{Shen-18}%
  \BibitemOpen
  \bibfield  {author} {\bibinfo {author} {\bibfnamefont {H.}~\bibnamefont
  {Shen}}, \bibinfo {author} {\bibfnamefont {B.}~\bibnamefont {Zhen}}, \ and\
  \bibinfo {author} {\bibfnamefont {L.}~\bibnamefont {Fu}},\ }\bibfield
  {title} {\enquote {\bibinfo {title} {{Topological Band Theory for
  Non-Hermitian Hamiltonians}},}\ }\href@noop {} {\bibfield  {journal}
  {\bibinfo  {journal} {Phys. Rev. Lett.}\ }\textbf {\bibinfo {volume} {120}},\
  \bibinfo {pages} {146402} (\bibinfo {year} {2018})}\BibitemShut {NoStop}%
\bibitem [{\citenamefont {Kozii}\ and\ \citenamefont {Fu}()}]{Kozii-17}%
  \BibitemOpen
  \bibfield  {author} {\bibinfo {author} {\bibfnamefont {V.}~\bibnamefont
  {Kozii}}\ and\ \bibinfo {author} {\bibfnamefont {L.}~\bibnamefont {Fu}},\
  }\href@noop {} {\enquote {\bibinfo {title} {{Non-Hermitian Topological Theory
  of Finite-Lifetime Quasiparticles: Prediction of Bulk Fermi Arc Due to
  Exceptional Point}},}\ }\bibinfo {note} {{arXiv:1708.05841}}\BibitemShut
  {NoStop}%
\bibitem [{\citenamefont {Takata}\ and\ \citenamefont
  {Notomi}(2018)}]{Takata-18}%
  \BibitemOpen
  \bibfield  {author} {\bibinfo {author} {\bibfnamefont {K.}~\bibnamefont
  {Takata}}\ and\ \bibinfo {author} {\bibfnamefont {M.}~\bibnamefont
  {Notomi}},\ }\bibfield  {title} {\enquote {\bibinfo {title} {{Photonic
  Topological Insulating Phase Induced Solely by Gain and Loss}},}\ }\href@noop
  {} {\bibfield  {journal} {\bibinfo  {journal} {Phys. Rev. Lett.}\ }\textbf
  {\bibinfo {volume} {121}},\ \bibinfo {pages} {213902} (\bibinfo {year}
  {2018})}\BibitemShut {NoStop}%
\bibitem [{\citenamefont {Martinez~Alvarez}\ \emph {et~al.}(2018)\citenamefont
  {Martinez~Alvarez}, \citenamefont {Barrios~Vargas},\ and\ \citenamefont
  {Foa~Torres}}]{MartinezAlvarez-18}%
  \BibitemOpen
  \bibfield  {author} {\bibinfo {author} {\bibfnamefont {V.~M.}\ \bibnamefont
  {Martinez~Alvarez}}, \bibinfo {author} {\bibfnamefont {J.~E.}\ \bibnamefont
  {Barrios~Vargas}}, \ and\ \bibinfo {author} {\bibfnamefont {L.~E.~F.}\
  \bibnamefont {Foa~Torres}},\ }\bibfield  {title} {\enquote {\bibinfo {title}
  {{Non-Hermitian robust edge states in one dimension: Anomalous localization
  and eigenspace condensation at exceptional points}},}\ }\href@noop {}
  {\bibfield  {journal} {\bibinfo  {journal} {Phys. Rev. B}\ }\textbf {\bibinfo
  {volume} {97}},\ \bibinfo {pages} {121401(R)} (\bibinfo {year}
  {2018})}\BibitemShut {NoStop}%
\bibitem [{\citenamefont {Gong}\ \emph {et~al.}(2018)\citenamefont {Gong},
  \citenamefont {Ashida}, \citenamefont {Kawabata}, \citenamefont {Takasan},
  \citenamefont {Higashikawa},\ and\ \citenamefont {Ueda}}]{Gong-18}%
  \BibitemOpen
  \bibfield  {author} {\bibinfo {author} {\bibfnamefont {Z.}~\bibnamefont
  {Gong}}, \bibinfo {author} {\bibfnamefont {Y.}~\bibnamefont {Ashida}},
  \bibinfo {author} {\bibfnamefont {K.}~\bibnamefont {Kawabata}}, \bibinfo
  {author} {\bibfnamefont {K.}~\bibnamefont {Takasan}}, \bibinfo {author}
  {\bibfnamefont {S.}~\bibnamefont {Higashikawa}}, \ and\ \bibinfo {author}
  {\bibfnamefont {M.}~\bibnamefont {Ueda}},\ }\bibfield  {title} {\enquote
  {\bibinfo {title} {{Topological Phases of Non-Hermitian Systems}},}\
  }\href@noop {} {\bibfield  {journal} {\bibinfo  {journal} {Phys. Rev. X}\
  }\textbf {\bibinfo {volume} {8}},\ \bibinfo {pages} {031079} (\bibinfo {year}
  {2018})}\BibitemShut {NoStop}%
\bibitem [{\citenamefont {Kawabata}\ \emph
  {et~al.}(2019{\natexlab{a}})\citenamefont {Kawabata}, \citenamefont
  {Higashikawa}, \citenamefont {Gong}, \citenamefont {Ashida},\ and\
  \citenamefont {Ueda}}]{Kawabata-19}%
  \BibitemOpen
  \bibfield  {author} {\bibinfo {author} {\bibfnamefont {K.}~\bibnamefont
  {Kawabata}}, \bibinfo {author} {\bibfnamefont {S.}~\bibnamefont
  {Higashikawa}}, \bibinfo {author} {\bibfnamefont {Z.}~\bibnamefont {Gong}},
  \bibinfo {author} {\bibfnamefont {Y.}~\bibnamefont {Ashida}}, \ and\ \bibinfo
  {author} {\bibfnamefont {M.}~\bibnamefont {Ueda}},\ }\bibfield  {title}
  {\enquote {\bibinfo {title} {{Topological unification of time-reversal and
  particle-hole symmetries in non-Hermitian physics}},}\ }\href@noop {}
  {\bibfield  {journal} {\bibinfo  {journal} {Nat. Commun.}\ }\textbf {\bibinfo
  {volume} {10}},\ \bibinfo {pages} {297} (\bibinfo {year}
  {2019}{\natexlab{a}})}\BibitemShut {NoStop}%
\bibitem [{\citenamefont {Yao}\ and\ \citenamefont {Wang}(2018)}]{YW-18-SSH}%
  \BibitemOpen
  \bibfield  {author} {\bibinfo {author} {\bibfnamefont {S.}~\bibnamefont
  {Yao}}\ and\ \bibinfo {author} {\bibfnamefont {Z.}~\bibnamefont {Wang}},\
  }\bibfield  {title} {\enquote {\bibinfo {title} {{Edge States and Topological
  Invariants of Non-Hermitian Systems}},}\ }\href@noop {} {\bibfield  {journal}
  {\bibinfo  {journal} {Phys. Rev. Lett.}\ }\textbf {\bibinfo {volume} {121}},\
  \bibinfo {pages} {086803} (\bibinfo {year} {2018})}\BibitemShut {NoStop}%
\bibitem [{\citenamefont {Yao}\ \emph {et~al.}(2018)\citenamefont {Yao},
  \citenamefont {Song},\ and\ \citenamefont {Wang}}]{YSW-18-Chern}%
  \BibitemOpen
  \bibfield  {author} {\bibinfo {author} {\bibfnamefont {S.}~\bibnamefont
  {Yao}}, \bibinfo {author} {\bibfnamefont {F.}~\bibnamefont {Song}}, \ and\
  \bibinfo {author} {\bibfnamefont {Z.}~\bibnamefont {Wang}},\ }\bibfield
  {title} {\enquote {\bibinfo {title} {{Non-Hermitian Chern Bands}},}\
  }\href@noop {} {\bibfield  {journal} {\bibinfo  {journal} {{Phys. Rev.
  Lett.}}\ }\textbf {\bibinfo {volume} {121}},\ \bibinfo {pages} {136802}
  (\bibinfo {year} {2018})}\BibitemShut {NoStop}%
\bibitem [{\citenamefont {Kunst}\ \emph
  {et~al.}(2018{\natexlab{b}})\citenamefont {Kunst}, \citenamefont
  {Edvardsson}, \citenamefont {Budich},\ and\ \citenamefont
  {Bergholtz}}]{Kunst-18NH}%
  \BibitemOpen
  \bibfield  {author} {\bibinfo {author} {\bibfnamefont {F.~K.}\ \bibnamefont
  {Kunst}}, \bibinfo {author} {\bibfnamefont {E.}~\bibnamefont {Edvardsson}},
  \bibinfo {author} {\bibfnamefont {J.~C.}\ \bibnamefont {Budich}}, \ and\
  \bibinfo {author} {\bibfnamefont {E.~J.}\ \bibnamefont {Bergholtz}},\
  }\bibfield  {title} {\enquote {\bibinfo {title} {{Biorthogonal Bulk-Boundary
  Correspondence in Non-Hermitian Systems}},}\ }\href@noop {} {\bibfield
  {journal} {\bibinfo  {journal} {Phys. Rev. Lett.}\ }\textbf {\bibinfo
  {volume} {121}},\ \bibinfo {pages} {026808} (\bibinfo {year}
  {2018}{\natexlab{b}})}\BibitemShut {NoStop}%
\bibitem [{\citenamefont {Kawabata}\ \emph {et~al.}(2018)\citenamefont
  {Kawabata}, \citenamefont {Shiozaki},\ and\ \citenamefont {Ueda}}]{KSU-18}%
  \BibitemOpen
  \bibfield  {author} {\bibinfo {author} {\bibfnamefont {K.}~\bibnamefont
  {Kawabata}}, \bibinfo {author} {\bibfnamefont {K.}~\bibnamefont {Shiozaki}},
  \ and\ \bibinfo {author} {\bibfnamefont {M.}~\bibnamefont {Ueda}},\
  }\bibfield  {title} {\enquote {\bibinfo {title} {{Anomalous helical edge
  states in a non-Hermitian Chern insulator}},}\ }\href@noop {} {\bibfield
  {journal} {\bibinfo  {journal} {Phys. Rev. B}\ }\textbf {\bibinfo {volume}
  {98}},\ \bibinfo {pages} {165148} (\bibinfo {year} {2018})}\BibitemShut
  {NoStop}%
\bibitem [{\citenamefont {McDonald}\ \emph {et~al.}(2018)\citenamefont
  {McDonald}, \citenamefont {Pereg-Barnea},\ and\ \citenamefont
  {Clerk}}]{McDonald-18}%
  \BibitemOpen
  \bibfield  {author} {\bibinfo {author} {\bibfnamefont {A.}~\bibnamefont
  {McDonald}}, \bibinfo {author} {\bibfnamefont {T.}~\bibnamefont
  {Pereg-Barnea}}, \ and\ \bibinfo {author} {\bibfnamefont {A.~A.}\
  \bibnamefont {Clerk}},\ }\bibfield  {title} {\enquote {\bibinfo {title}
  {{Phase-Dependent Chiral Transport and Effective Non-Hermitian Dynamics in a
  Bosonic Kitaev-Majorana Chain}},}\ }\href@noop {} {\bibfield  {journal}
  {\bibinfo  {journal} {Phys. Rev. X}\ }\textbf {\bibinfo {volume} {8}},\
  \bibinfo {pages} {041031} (\bibinfo {year} {2018})}\BibitemShut {NoStop}%
\bibitem [{\citenamefont {Lee}\ and\ \citenamefont {Thomale}(2019)}]{Lee-19}%
  \BibitemOpen
  \bibfield  {author} {\bibinfo {author} {\bibfnamefont {C.~H.}\ \bibnamefont
  {Lee}}\ and\ \bibinfo {author} {\bibfnamefont {R.}~\bibnamefont {Thomale}},\
  }\bibfield  {title} {\enquote {\bibinfo {title} {{Anatomy of skin modes and
  topology in non-Hermitian systems}},}\ }\href@noop {} {\bibfield  {journal}
  {\bibinfo  {journal} {Phys. Rev. B}\ }\textbf {\bibinfo {volume} {99}},\
  \bibinfo {pages} {201103(R)} (\bibinfo {year} {2019})}\BibitemShut {NoStop}%
\bibitem [{\citenamefont {Jin}\ and\ \citenamefont {Song}(2019)}]{Jin-19}%
  \BibitemOpen
  \bibfield  {author} {\bibinfo {author} {\bibfnamefont {L.}~\bibnamefont
  {Jin}}\ and\ \bibinfo {author} {\bibfnamefont {Z.}~\bibnamefont {Song}},\
  }\bibfield  {title} {\enquote {\bibinfo {title} {{Bulk-boundary
  correspondence in a non-Hermitian system in one dimension with chiral
  inversion symmetry}},}\ }\href@noop {} {\bibfield  {journal} {\bibinfo
  {journal} {Phys. Rev. B}\ }\textbf {\bibinfo {volume} {99}},\ \bibinfo
  {pages} {081103(R)} (\bibinfo {year} {2019})}\BibitemShut {NoStop}%
\bibitem [{\citenamefont {Budich}\ \emph {et~al.}(2019)\citenamefont {Budich},
  \citenamefont {Carlstr\"om}, \citenamefont {Kunst},\ and\ \citenamefont
  {Bergholtz}}]{Budich-19}%
  \BibitemOpen
  \bibfield  {author} {\bibinfo {author} {\bibfnamefont {J.~C.}\ \bibnamefont
  {Budich}}, \bibinfo {author} {\bibfnamefont {J.}~\bibnamefont {Carlstr\"om}},
  \bibinfo {author} {\bibfnamefont {F.~K.}\ \bibnamefont {Kunst}}, \ and\
  \bibinfo {author} {\bibfnamefont {E.~J.}\ \bibnamefont {Bergholtz}},\
  }\bibfield  {title} {\enquote {\bibinfo {title} {{Symmetry-protected nodal
  phases in non-Hermitian systems}},}\ }\href@noop {} {\bibfield  {journal}
  {\bibinfo  {journal} {Phys. Rev. B}\ }\textbf {\bibinfo {volume} {99}},\
  \bibinfo {pages} {041406(R)} (\bibinfo {year} {2019})}\BibitemShut {NoStop}%
\bibitem [{\citenamefont {Okugawa}\ and\ \citenamefont
  {Yokoyama}(2019)}]{Okugawa-19}%
  \BibitemOpen
  \bibfield  {author} {\bibinfo {author} {\bibfnamefont {R.}~\bibnamefont
  {Okugawa}}\ and\ \bibinfo {author} {\bibfnamefont {T.}~\bibnamefont
  {Yokoyama}},\ }\bibfield  {title} {\enquote {\bibinfo {title} {{Topological
  exceptional surfaces in non-Hermitian systems with parity-time and
  parity-particle-hole symmetries}},}\ }\href@noop {} {\bibfield  {journal}
  {\bibinfo  {journal} {Phys. Rev. B}\ }\textbf {\bibinfo {volume} {99}},\
  \bibinfo {pages} {041202(R)} (\bibinfo {year} {2019})}\BibitemShut {NoStop}%
\bibitem [{\citenamefont {Liu}\ \emph {et~al.}(2019{\natexlab{b}})\citenamefont
  {Liu}, \citenamefont {Zhang}, \citenamefont {Ai}, \citenamefont {Gong},
  \citenamefont {Kawabata}, \citenamefont {Ueda},\ and\ \citenamefont
  {Nori}}]{Liu-19NHSOTI}%
  \BibitemOpen
  \bibfield  {author} {\bibinfo {author} {\bibfnamefont {T.}~\bibnamefont
  {Liu}}, \bibinfo {author} {\bibfnamefont {Y.-R.}\ \bibnamefont {Zhang}},
  \bibinfo {author} {\bibfnamefont {Q.}~\bibnamefont {Ai}}, \bibinfo {author}
  {\bibfnamefont {Z.}~\bibnamefont {Gong}}, \bibinfo {author} {\bibfnamefont
  {K.}~\bibnamefont {Kawabata}}, \bibinfo {author} {\bibfnamefont
  {M.}~\bibnamefont {Ueda}}, \ and\ \bibinfo {author} {\bibfnamefont
  {F.}~\bibnamefont {Nori}},\ }\bibfield  {title} {\enquote {\bibinfo {title}
  {{Second-Order Topological Phases in Non-Hermitian Systems}},}\ }\href@noop
  {} {\bibfield  {journal} {\bibinfo  {journal} {Phys. Rev. Lett.}\ }\textbf
  {\bibinfo {volume} {122}},\ \bibinfo {pages} {076801} (\bibinfo {year}
  {2019}{\natexlab{b}})}\BibitemShut {NoStop}%
\bibitem [{\citenamefont {Yoshida}\ \emph {et~al.}(2019)\citenamefont
  {Yoshida}, \citenamefont {Peters}, \citenamefont {Kawakami},\ and\
  \citenamefont {Hatsugai}}]{Yoshida-19}%
  \BibitemOpen
  \bibfield  {author} {\bibinfo {author} {\bibfnamefont {T.}~\bibnamefont
  {Yoshida}}, \bibinfo {author} {\bibfnamefont {R.}~\bibnamefont {Peters}},
  \bibinfo {author} {\bibfnamefont {N.}~\bibnamefont {Kawakami}}, \ and\
  \bibinfo {author} {\bibfnamefont {Y.}~\bibnamefont {Hatsugai}},\ }\bibfield
  {title} {\enquote {\bibinfo {title} {{Symmetry-protected exceptional rings in
  two-dimensional correlated systems with chiral symmetry}},}\ }\href@noop {}
  {\bibfield  {journal} {\bibinfo  {journal} {Phys. Rev. B}\ }\textbf {\bibinfo
  {volume} {99}},\ \bibinfo {pages} {121101(R)} (\bibinfo {year}
  {2019})}\BibitemShut {NoStop}%
\bibitem [{\citenamefont {Kimura}\ \emph {et~al.}(2019)\citenamefont {Kimura},
  \citenamefont {Yoshida},\ and\ \citenamefont {Kawakami}}]{Kimura-19}%
  \BibitemOpen
  \bibfield  {author} {\bibinfo {author} {\bibfnamefont {K.}~\bibnamefont
  {Kimura}}, \bibinfo {author} {\bibfnamefont {T.}~\bibnamefont {Yoshida}}, \
  and\ \bibinfo {author} {\bibfnamefont {N.}~\bibnamefont {Kawakami}},\
  }\bibfield  {title} {\enquote {\bibinfo {title} {Chiral-symmetry protected
  exceptional torus in correlated nodal-line semimetals},}\ }\href@noop {}
  {\bibfield  {journal} {\bibinfo  {journal} {{Phys. Rev. B}}\ }\textbf
  {\bibinfo {volume} {100}},\ \bibinfo {pages} {115124} (\bibinfo {year}
  {2019})}\BibitemShut {NoStop}%
\bibitem [{\citenamefont {Zhou}\ \emph {et~al.}(2019)\citenamefont {Zhou},
  \citenamefont {Lee}, \citenamefont {Liu},\ and\ \citenamefont
  {Zhen}}]{Zhou-19}%
  \BibitemOpen
  \bibfield  {author} {\bibinfo {author} {\bibfnamefont {H.}~\bibnamefont
  {Zhou}}, \bibinfo {author} {\bibfnamefont {J.~Y.}\ \bibnamefont {Lee}},
  \bibinfo {author} {\bibfnamefont {S.}~\bibnamefont {Liu}}, \ and\ \bibinfo
  {author} {\bibfnamefont {B.}~\bibnamefont {Zhen}},\ }\bibfield  {title}
  {\enquote {\bibinfo {title} {{Exceptional surfaces in
  $\mathcal{PT}$-symmetric non-Hermitian photonic systems}},}\ }\href@noop {}
  {\bibfield  {journal} {\bibinfo  {journal} {Optica}\ }\textbf {\bibinfo
  {volume} {6}},\ \bibinfo {pages} {190} (\bibinfo {year} {2019})}\BibitemShut
  {NoStop}%
\bibitem [{\citenamefont {Lee}\ \emph {et~al.}(2019)\citenamefont {Lee},
  \citenamefont {Li},\ and\ \citenamefont {Gong}}]{Lee-Li-Gong-19}%
  \BibitemOpen
  \bibfield  {author} {\bibinfo {author} {\bibfnamefont {C.~H.}\ \bibnamefont
  {Lee}}, \bibinfo {author} {\bibfnamefont {L.}~\bibnamefont {Li}}, \ and\
  \bibinfo {author} {\bibfnamefont {J.}~\bibnamefont {Gong}},\ }\bibfield
  {title} {\enquote {\bibinfo {title} {{Hybrid Higher-Order Skin-Topological
  Modes in Nonreciprocal Systems}},}\ }\href@noop {} {\bibfield  {journal}
  {\bibinfo  {journal} {Phys. Rev. Lett.}\ }\textbf {\bibinfo {volume} {123}},\
  \bibinfo {pages} {016805} (\bibinfo {year} {2019})}\BibitemShut {NoStop}%
\bibitem [{\citenamefont {Ezawa}(2019)}]{Ezawa-19}%
  \BibitemOpen
  \bibfield  {author} {\bibinfo {author} {\bibfnamefont {M.}~\bibnamefont
  {Ezawa}},\ }\bibfield  {title} {\enquote {\bibinfo {title} {{Non-Hermitian
  boundary and interface states in nonreciprocal higher-order topological
  metals and electrical circuits}},}\ }\href@noop {} {\bibfield  {journal}
  {\bibinfo  {journal} {Phys. Rev. B}\ }\textbf {\bibinfo {volume} {99}},\
  \bibinfo {pages} {121411(R)} (\bibinfo {year} {2019})}\BibitemShut {NoStop}%
\bibitem [{\citenamefont {Kunst}\ and\ \citenamefont
  {Dwivedi}(2019)}]{Kunst-19}%
  \BibitemOpen
  \bibfield  {author} {\bibinfo {author} {\bibfnamefont {F.~K.}\ \bibnamefont
  {Kunst}}\ and\ \bibinfo {author} {\bibfnamefont {V.}~\bibnamefont
  {Dwivedi}},\ }\bibfield  {title} {\enquote {\bibinfo {title} {{Non-Hermitian
  systems and topology: A transfer-matrix perspective}},}\ }\href@noop {}
  {\bibfield  {journal} {\bibinfo  {journal} {Phys. Rev. B}\ }\textbf {\bibinfo
  {volume} {99}},\ \bibinfo {pages} {245116} (\bibinfo {year}
  {2019})}\BibitemShut {NoStop}%
\bibitem [{\citenamefont {Edvardsson}\ \emph {et~al.}(2019)\citenamefont
  {Edvardsson}, \citenamefont {Kunst},\ and\ \citenamefont
  {Bergholtz}}]{Edvardsson-19}%
  \BibitemOpen
  \bibfield  {author} {\bibinfo {author} {\bibfnamefont {E.}~\bibnamefont
  {Edvardsson}}, \bibinfo {author} {\bibfnamefont {F.~K.}\ \bibnamefont
  {Kunst}}, \ and\ \bibinfo {author} {\bibfnamefont {E.~J.}\ \bibnamefont
  {Bergholtz}},\ }\bibfield  {title} {\enquote {\bibinfo {title}
  {{Non-Hermitian extensions of higher-order topological phases and their
  biorthogonal bulk-boundary correspondence}},}\ }\href@noop {} {\bibfield
  {journal} {\bibinfo  {journal} {Phys. Rev. B}\ }\textbf {\bibinfo {volume}
  {99}},\ \bibinfo {pages} {081302(R)} (\bibinfo {year} {2019})}\BibitemShut
  {NoStop}%
\bibitem [{\citenamefont {Kawabata}\ \emph
  {et~al.}(2019{\natexlab{b}})\citenamefont {Kawabata}, \citenamefont
  {Shiozaki}, \citenamefont {Ueda},\ and\ \citenamefont {Sato}}]{KSUS-19}%
  \BibitemOpen
  \bibfield  {author} {\bibinfo {author} {\bibfnamefont {K.}~\bibnamefont
  {Kawabata}}, \bibinfo {author} {\bibfnamefont {K.}~\bibnamefont {Shiozaki}},
  \bibinfo {author} {\bibfnamefont {M.}~\bibnamefont {Ueda}}, \ and\ \bibinfo
  {author} {\bibfnamefont {M.}~\bibnamefont {Sato}},\ }\bibfield  {title}
  {\enquote {\bibinfo {title} {{Symmetry and Topology in Non-Hermitian
  Physics}},}\ }\href@noop {} {\bibfield  {journal} {\bibinfo  {journal} {Phys.
  Rev. X}\ }\textbf {\bibinfo {volume} {9}},\ \bibinfo {pages} {041015}
  (\bibinfo {year} {2019}{\natexlab{b}})}\BibitemShut {NoStop}%
\bibitem [{\citenamefont {Zhou}\ and\ \citenamefont {Lee}(2019)}]{ZL-19}%
  \BibitemOpen
  \bibfield  {author} {\bibinfo {author} {\bibfnamefont {H.}~\bibnamefont
  {Zhou}}\ and\ \bibinfo {author} {\bibfnamefont {J.~Y.}\ \bibnamefont {Lee}},\
  }\bibfield  {title} {\enquote {\bibinfo {title} {{Periodic table for
  topological bands with non-Hermitian symmetries}},}\ }\href@noop {}
  {\bibfield  {journal} {\bibinfo  {journal} {Phys. Rev. B}\ }\textbf {\bibinfo
  {volume} {99}},\ \bibinfo {pages} {235112} (\bibinfo {year}
  {2019})}\BibitemShut {NoStop}%
\bibitem [{\citenamefont {Herviou}\ \emph
  {et~al.}(2019{\natexlab{a}})\citenamefont {Herviou}, \citenamefont
  {Bardarson},\ and\ \citenamefont {Regnault}}]{Herviou-19}%
  \BibitemOpen
  \bibfield  {author} {\bibinfo {author} {\bibfnamefont {L.}~\bibnamefont
  {Herviou}}, \bibinfo {author} {\bibfnamefont {J.~H.}\ \bibnamefont
  {Bardarson}}, \ and\ \bibinfo {author} {\bibfnamefont {N.}~\bibnamefont
  {Regnault}},\ }\bibfield  {title} {\enquote {\bibinfo {title} {{Defining a
  bulk-edge correspondence for non-Hermitian Hamiltonians via singular-value
  decomposition}},}\ }\href@noop {} {\bibfield  {journal} {\bibinfo  {journal}
  {Phys. Rev. A}\ }\textbf {\bibinfo {volume} {99}},\ \bibinfo {pages} {052118}
  (\bibinfo {year} {2019}{\natexlab{a}})}\BibitemShut {NoStop}%
\bibitem [{\citenamefont {Zhang}\ \emph {et~al.}(2019)\citenamefont {Zhang},
  \citenamefont {L\'opez}, \citenamefont {Cheng}, \citenamefont {Liu},\ and\
  \citenamefont {Christensen}}]{Zhang-19NHSOTI}%
  \BibitemOpen
  \bibfield  {author} {\bibinfo {author} {\bibfnamefont {Z.}~\bibnamefont
  {Zhang}}, \bibinfo {author} {\bibfnamefont {M.~Rosendo}\ \bibnamefont
  {L\'opez}}, \bibinfo {author} {\bibfnamefont {Y.}~\bibnamefont {Cheng}},
  \bibinfo {author} {\bibfnamefont {X.}~\bibnamefont {Liu}}, \ and\ \bibinfo
  {author} {\bibfnamefont {J.}~\bibnamefont {Christensen}},\ }\bibfield
  {title} {\enquote {\bibinfo {title} {{Non-Hermitian Sonic Second-Order
  Topological Insulator}},}\ }\href@noop {} {\bibfield  {journal} {\bibinfo
  {journal} {Phys. Rev. Lett.}\ }\textbf {\bibinfo {volume} {122}},\ \bibinfo
  {pages} {195501} (\bibinfo {year} {2019})}\BibitemShut {NoStop}%
\bibitem [{\citenamefont {Zirnstein}\ \emph {et~al.}()\citenamefont
  {Zirnstein}, \citenamefont {Refael},\ and\ \citenamefont
  {Rosenow}}]{Zirnstein-19}%
  \BibitemOpen
  \bibfield  {author} {\bibinfo {author} {\bibfnamefont {H.-G.}\ \bibnamefont
  {Zirnstein}}, \bibinfo {author} {\bibfnamefont {G.}~\bibnamefont {Refael}}, \
  and\ \bibinfo {author} {\bibfnamefont {B.}~\bibnamefont {Rosenow}},\
  }\href@noop {} {\enquote {\bibinfo {title} {{Bulk-boundary correspondence for
  non-Hermitian Hamiltonians via Green functions}},}\ }\bibinfo {note}
  {{arXiv:1901.11241}}\BibitemShut {NoStop}%
\bibitem [{\citenamefont {Borgnia}\ \emph {et~al.}(2020)\citenamefont
  {Borgnia}, \citenamefont {Kruchkov},\ and\ \citenamefont
  {Slager}}]{Borgnia-19}%
  \BibitemOpen
  \bibfield  {author} {\bibinfo {author} {\bibfnamefont {D.~S.}\ \bibnamefont
  {Borgnia}}, \bibinfo {author} {\bibfnamefont {A.~J.}\ \bibnamefont
  {Kruchkov}}, \ and\ \bibinfo {author} {\bibfnamefont {R.-J.}\ \bibnamefont
  {Slager}},\ }\bibfield  {title} {\enquote {\bibinfo {title} {{Non-Hermitian
  Boundary Modes and Topology}},}\ }\href@noop {} {\bibfield  {journal}
  {\bibinfo  {journal} {Phys. Rev. Lett.}\ }\textbf {\bibinfo {volume} {124}},\
  \bibinfo {pages} {056802} (\bibinfo {year} {2020})}\BibitemShut {NoStop}%
\bibitem [{\citenamefont {Kawabata}\ \emph
  {et~al.}(2019{\natexlab{c}})\citenamefont {Kawabata}, \citenamefont
  {Bessho},\ and\ \citenamefont {Sato}}]{KBS-19}%
  \BibitemOpen
  \bibfield  {author} {\bibinfo {author} {\bibfnamefont {K.}~\bibnamefont
  {Kawabata}}, \bibinfo {author} {\bibfnamefont {T.}~\bibnamefont {Bessho}}, \
  and\ \bibinfo {author} {\bibfnamefont {M.}~\bibnamefont {Sato}},\ }\bibfield
  {title} {\enquote {\bibinfo {title} {{Classification of Exceptional Points
  and Non-Hermitian Topological Semimetals}},}\ }\href@noop {} {\bibfield
  {journal} {\bibinfo  {journal} {Phys. Rev. Lett.}\ }\textbf {\bibinfo
  {volume} {123}},\ \bibinfo {pages} {066405} (\bibinfo {year}
  {2019}{\natexlab{c}})}\BibitemShut {NoStop}%
\bibitem [{\citenamefont {Yokomizo}\ and\ \citenamefont
  {Murakami}(2019)}]{YM-19}%
  \BibitemOpen
  \bibfield  {author} {\bibinfo {author} {\bibfnamefont {K.}~\bibnamefont
  {Yokomizo}}\ and\ \bibinfo {author} {\bibfnamefont {S.}~\bibnamefont
  {Murakami}},\ }\bibfield  {title} {\enquote {\bibinfo {title} {{Non-Bloch
  Band Theory of Non-Hermitian Systems}},}\ }\href@noop {} {\bibfield
  {journal} {\bibinfo  {journal} {Phys. Rev. Lett.}\ }\textbf {\bibinfo
  {volume} {123}},\ \bibinfo {pages} {066404} (\bibinfo {year}
  {2019})}\BibitemShut {NoStop}%
\bibitem [{\citenamefont {Luo}\ and\ \citenamefont {Zhang}(2019)}]{Luo-19}%
  \BibitemOpen
  \bibfield  {author} {\bibinfo {author} {\bibfnamefont {X.-W.}\ \bibnamefont
  {Luo}}\ and\ \bibinfo {author} {\bibfnamefont {C.}~\bibnamefont {Zhang}},\
  }\bibfield  {title} {\enquote {\bibinfo {title} {{Higher-Order Topological
  Corner States Induced by Gain and Loss}},}\ }\href@noop {} {\bibfield
  {journal} {\bibinfo  {journal} {Phys. Rev. Lett.}\ }\textbf {\bibinfo
  {volume} {123}},\ \bibinfo {pages} {073601} (\bibinfo {year}
  {2019})}\BibitemShut {NoStop}%
\bibitem [{\citenamefont {McClarty}\ and\ \citenamefont
  {Rau}(2019)}]{McClarty-19}%
  \BibitemOpen
  \bibfield  {author} {\bibinfo {author} {\bibfnamefont {P.~A.}\ \bibnamefont
  {McClarty}}\ and\ \bibinfo {author} {\bibfnamefont {J.~G.}\ \bibnamefont
  {Rau}},\ }\bibfield  {title} {\enquote {\bibinfo {title} {{Non-Hermitian
  topology of spontaneous magnon decay}},}\ }\href@noop {} {\bibfield
  {journal} {\bibinfo  {journal} {Phys. Rev. B}\ }\textbf {\bibinfo {volume}
  {100}},\ \bibinfo {pages} {100405(R)} (\bibinfo {year} {2019})}\BibitemShut
  {NoStop}%
\bibitem [{\citenamefont {Okuma}\ and\ \citenamefont {Sato}(2019)}]{Okuma-19}%
  \BibitemOpen
  \bibfield  {author} {\bibinfo {author} {\bibfnamefont {N.}~\bibnamefont
  {Okuma}}\ and\ \bibinfo {author} {\bibfnamefont {M.}~\bibnamefont {Sato}},\
  }\bibfield  {title} {\enquote {\bibinfo {title} {{Topological Phase
  Transition Driven by Infinitesimal Instability: Majorana Fermions in
  Non-Hermitian Spintronics}},}\ }\href@noop {} {\bibfield  {journal} {\bibinfo
   {journal} {Phys. Rev. Lett.}\ }\textbf {\bibinfo {volume} {123}},\ \bibinfo
  {pages} {097701} (\bibinfo {year} {2019})}\BibitemShut {NoStop}%
\bibitem [{\citenamefont {Song}\ \emph
  {et~al.}(2019{\natexlab{a}})\citenamefont {Song}, \citenamefont {Yao},\ and\
  \citenamefont {Wang}}]{Song-19-Lindblad}%
  \BibitemOpen
  \bibfield  {author} {\bibinfo {author} {\bibfnamefont {F.}~\bibnamefont
  {Song}}, \bibinfo {author} {\bibfnamefont {S.}~\bibnamefont {Yao}}, \ and\
  \bibinfo {author} {\bibfnamefont {Z.}~\bibnamefont {Wang}},\ }\bibfield
  {title} {\enquote {\bibinfo {title} {{Non-Hermitian Skin Effect and Chiral
  Damping in Open Quantum Systems}},}\ }\href@noop {} {\bibfield  {journal}
  {\bibinfo  {journal} {Phys. Rev. Lett.}\ }\textbf {\bibinfo {volume} {123}},\
  \bibinfo {pages} {170401} (\bibinfo {year} {2019}{\natexlab{a}})}\BibitemShut
  {NoStop}%
\bibitem [{\citenamefont {Song}\ \emph
  {et~al.}(2019{\natexlab{b}})\citenamefont {Song}, \citenamefont {Yao},\ and\
  \citenamefont {Wang}}]{Song-19-real}%
  \BibitemOpen
  \bibfield  {author} {\bibinfo {author} {\bibfnamefont {F.}~\bibnamefont
  {Song}}, \bibinfo {author} {\bibfnamefont {S.}~\bibnamefont {Yao}}, \ and\
  \bibinfo {author} {\bibfnamefont {Z.}~\bibnamefont {Wang}},\ }\bibfield
  {title} {\enquote {\bibinfo {title} {{Non-Hermitian Topological Invariants in
  Real Space}},}\ }\href@noop {} {\bibfield  {journal} {\bibinfo  {journal}
  {Phys. Rev. Lett.}\ }\textbf {\bibinfo {volume} {123}},\ \bibinfo {pages}
  {246801} (\bibinfo {year} {2019}{\natexlab{b}})}\BibitemShut {NoStop}%
\bibitem [{\citenamefont {Bergholtz}\ and\ \citenamefont
  {Budich}(2019)}]{Bergholtz-19}%
  \BibitemOpen
  \bibfield  {author} {\bibinfo {author} {\bibfnamefont {E.~J.}\ \bibnamefont
  {Bergholtz}}\ and\ \bibinfo {author} {\bibfnamefont {J.~C.}\ \bibnamefont
  {Budich}},\ }\bibfield  {title} {\enquote {\bibinfo {title} {{Non-Hermitian
  Weyl physics in topological insulator ferromagnet junctions}},}\ }\href@noop
  {} {\bibfield  {journal} {\bibinfo  {journal} {Phys. Rev. Research}\ }\textbf
  {\bibinfo {volume} {1}},\ \bibinfo {pages} {012003(R)} (\bibinfo {year}
  {2019})}\BibitemShut {NoStop}%
\bibitem [{\citenamefont {Rui}\ \emph {et~al.}(2019)\citenamefont {Rui},
  \citenamefont {Hirschmann},\ and\ \citenamefont {Schnyder}}]{Rui-19}%
  \BibitemOpen
  \bibfield  {author} {\bibinfo {author} {\bibfnamefont {W.~B.}\ \bibnamefont
  {Rui}}, \bibinfo {author} {\bibfnamefont {M.~M.}\ \bibnamefont {Hirschmann}},
  \ and\ \bibinfo {author} {\bibfnamefont {A.~P.}\ \bibnamefont {Schnyder}},\
  }\bibfield  {title} {\enquote {\bibinfo {title} {{$\mathcal{PT}$-symmetric
  non-Hermitian Dirac semimetals}},}\ }\href@noop {} {\bibfield  {journal}
  {\bibinfo  {journal} {Phys. Rev. B}\ }\textbf {\bibinfo {volume} {100}},\
  \bibinfo {pages} {245116} (\bibinfo {year} {2019})}\BibitemShut {NoStop}%
\bibitem [{\citenamefont {Schomerus}(2020)}]{Schomerus-20}%
  \BibitemOpen
  \bibfield  {author} {\bibinfo {author} {\bibfnamefont {H.}~\bibnamefont
  {Schomerus}},\ }\bibfield  {title} {\enquote {\bibinfo {title}
  {{Nonreciprocal response theory of non-Hermitian mechanical metamaterials:
  Response phase transition from the skin effect of zero modes}},}\ }\href@noop
  {} {\bibfield  {journal} {\bibinfo  {journal} {Phys. Rev. Research}\ }\textbf
  {\bibinfo {volume} {2}},\ \bibinfo {pages} {013058} (\bibinfo {year}
  {2020})}\BibitemShut {NoStop}%
\bibitem [{\citenamefont {Imura}\ and\ \citenamefont
  {Takane}(2019)}]{Imura-19}%
  \BibitemOpen
  \bibfield  {author} {\bibinfo {author} {\bibfnamefont {K.-I.}\ \bibnamefont
  {Imura}}\ and\ \bibinfo {author} {\bibfnamefont {Y.}~\bibnamefont {Takane}},\
  }\bibfield  {title} {\enquote {\bibinfo {title} {{Generalized bulk-edge
  correspondence for non-Hermitian topological systems}},}\ }\href@noop {}
  {\bibfield  {journal} {\bibinfo  {journal} {Phys. Rev. B}\ }\textbf {\bibinfo
  {volume} {100}},\ \bibinfo {pages} {165430} (\bibinfo {year}
  {2019})}\BibitemShut {NoStop}%
\bibitem [{\citenamefont {Herviou}\ \emph
  {et~al.}(2019{\natexlab{b}})\citenamefont {Herviou}, \citenamefont
  {Regnault},\ and\ \citenamefont {Bardarson}}]{Herviou-19-ES}%
  \BibitemOpen
  \bibfield  {author} {\bibinfo {author} {\bibfnamefont {L.}~\bibnamefont
  {Herviou}}, \bibinfo {author} {\bibfnamefont {N.}~\bibnamefont {Regnault}}, \
  and\ \bibinfo {author} {\bibfnamefont {J.~H.}\ \bibnamefont {Bardarson}},\
  }\bibfield  {title} {\enquote {\bibinfo {title} {{Entanglement spectrum and
  symmetries in non-Hermitian fermionic non-interacting models}},}\ }\href@noop
  {} {\bibfield  {journal} {\bibinfo  {journal} {SciPost Phys.}\ }\textbf
  {\bibinfo {volume} {7}},\ \bibinfo {pages} {069} (\bibinfo {year}
  {2019}{\natexlab{b}})}\BibitemShut {NoStop}%
\bibitem [{\citenamefont {Chang}\ \emph {et~al.}(2020)\citenamefont {Chang},
  \citenamefont {You}, \citenamefont {Wen},\ and\ \citenamefont
  {Ryu}}]{Chang-20}%
  \BibitemOpen
  \bibfield  {author} {\bibinfo {author} {\bibfnamefont {P.-Y.}\ \bibnamefont
  {Chang}}, \bibinfo {author} {\bibfnamefont {J.-S.}\ \bibnamefont {You}},
  \bibinfo {author} {\bibfnamefont {X.}~\bibnamefont {Wen}}, \ and\ \bibinfo
  {author} {\bibfnamefont {S.}~\bibnamefont {Ryu}},\ }\bibfield  {title}
  {\enquote {\bibinfo {title} {{Entanglement spectrum and entropy in
  topological non-Hermitian systems and nonunitary conformal field theory}},}\
  }\href@noop {} {\bibfield  {journal} {\bibinfo  {journal} {Phys. Rev.
  Research}\ }\textbf {\bibinfo {volume} {2}},\ \bibinfo {pages} {033069}
  (\bibinfo {year} {2020})}\BibitemShut {NoStop}%
\bibitem [{\citenamefont {Zhang}\ \emph {et~al.}(2020)\citenamefont {Zhang},
  \citenamefont {Yang},\ and\ \citenamefont {Fang}}]{Zhang-19}%
  \BibitemOpen
  \bibfield  {author} {\bibinfo {author} {\bibfnamefont {K.}~\bibnamefont
  {Zhang}}, \bibinfo {author} {\bibfnamefont {Z.}~\bibnamefont {Yang}}, \ and\
  \bibinfo {author} {\bibfnamefont {C.}~\bibnamefont {Fang}},\ }\bibfield
  {title} {\enquote {\bibinfo {title} {{Correspondence between Winding Numbers
  and Skin Modes in Non-Hermitian Systems}},}\ }\href@noop {} {\bibfield
  {journal} {\bibinfo  {journal} {Phys. Rev. Lett.}\ }\textbf {\bibinfo
  {volume} {125}},\ \bibinfo {pages} {126402} (\bibinfo {year}
  {2020})}\BibitemShut {NoStop}%
\bibitem [{\citenamefont {Okuma}\ \emph {et~al.}(2020)\citenamefont {Okuma},
  \citenamefont {Kawabata}, \citenamefont {Shiozaki},\ and\ \citenamefont
  {Sato}}]{OKSS-20}%
  \BibitemOpen
  \bibfield  {author} {\bibinfo {author} {\bibfnamefont {N.}~\bibnamefont
  {Okuma}}, \bibinfo {author} {\bibfnamefont {K.}~\bibnamefont {Kawabata}},
  \bibinfo {author} {\bibfnamefont {K.}~\bibnamefont {Shiozaki}}, \ and\
  \bibinfo {author} {\bibfnamefont {M.}~\bibnamefont {Sato}},\ }\bibfield
  {title} {\enquote {\bibinfo {title} {{Topological Origin of Non-Hermitian
  Skin Effects}},}\ }\href@noop {} {\bibfield  {journal} {\bibinfo  {journal}
  {Phys. Rev. Lett.}\ }\textbf {\bibinfo {volume} {124}},\ \bibinfo {pages}
  {086801} (\bibinfo {year} {2020})}\BibitemShut {NoStop}%
\bibitem [{\citenamefont {Longhi}(2020)}]{Longhi-20}%
  \BibitemOpen
  \bibfield  {author} {\bibinfo {author} {\bibfnamefont {S.}~\bibnamefont
  {Longhi}},\ }\bibfield  {title} {\enquote {\bibinfo {title} {{Non-Bloch-Band
  Collapse and Chiral Zener Tunneling}},}\ }\href@noop {} {\bibfield  {journal}
  {\bibinfo  {journal} {Phys. Rev. Lett.}\ }\textbf {\bibinfo {volume} {124}},\
  \bibinfo {pages} {066602} (\bibinfo {year} {2020})}\BibitemShut {NoStop}%
\bibitem [{\citenamefont {Li}\ and\ \citenamefont {Mong}()}]{Li-19}%
  \BibitemOpen
  \bibfield  {author} {\bibinfo {author} {\bibfnamefont {Z.}~\bibnamefont
  {Li}}\ and\ \bibinfo {author} {\bibfnamefont {R.~S.~K.}\ \bibnamefont
  {Mong}},\ }\href@noop {} {\enquote {\bibinfo {title} {{Homotopical
  classification of non-Hermitian band structures}},}\ }\bibinfo {note}
  {{arXiv:1911.02697}}\BibitemShut {NoStop}%
\bibitem [{\citenamefont {Wojcik}\ \emph {et~al.}(2020)\citenamefont {Wojcik},
  \citenamefont {Sun}, \citenamefont {Bzdu\v{s}ek},\ and\ \citenamefont
  {Fan}}]{Wojcik-20}%
  \BibitemOpen
  \bibfield  {author} {\bibinfo {author} {\bibfnamefont {C.~C.}\ \bibnamefont
  {Wojcik}}, \bibinfo {author} {\bibfnamefont {X.-Q.}\ \bibnamefont {Sun}},
  \bibinfo {author} {\bibfnamefont {T.}~\bibnamefont {Bzdu\v{s}ek}}, \ and\
  \bibinfo {author} {\bibfnamefont {S.}~\bibnamefont {Fan}},\ }\bibfield
  {title} {\enquote {\bibinfo {title} {{Homotopy characterization of
  non-Hermitian Hamiltonians}},}\ }\href@noop {} {\bibfield  {journal}
  {\bibinfo  {journal} {Phys. Rev. B}\ }\textbf {\bibinfo {volume} {101}},\
  \bibinfo {pages} {205417} (\bibinfo {year} {2020})}\BibitemShut {NoStop}%
\bibitem [{\citenamefont {Wang}\ \emph {et~al.}(2020)\citenamefont {Wang},
  \citenamefont {Guo},\ and\ \citenamefont {Kou}}]{Wang-20}%
  \BibitemOpen
  \bibfield  {author} {\bibinfo {author} {\bibfnamefont {X.-R.}\ \bibnamefont
  {Wang}}, \bibinfo {author} {\bibfnamefont {C.-X.}\ \bibnamefont {Guo}}, \
  and\ \bibinfo {author} {\bibfnamefont {S.-P.}\ \bibnamefont {Kou}},\
  }\bibfield  {title} {\enquote {\bibinfo {title} {{Defective edge states and
  number-anomalous bulk-boundary correspondence in non-Hermitian topological
  systems}},}\ }\href@noop {} {\bibfield  {journal} {\bibinfo  {journal} {Phys.
  Rev. B}\ }\textbf {\bibinfo {volume} {101}},\ \bibinfo {pages} {121116(R)}
  (\bibinfo {year} {2020})}\BibitemShut {NoStop}%
\bibitem [{\citenamefont {Yoshida}\ \emph {et~al.}(2020)\citenamefont
  {Yoshida}, \citenamefont {Mizoguchi},\ and\ \citenamefont
  {Hatsugai}}]{Yoshida-20}%
  \BibitemOpen
  \bibfield  {author} {\bibinfo {author} {\bibfnamefont {T.}~\bibnamefont
  {Yoshida}}, \bibinfo {author} {\bibfnamefont {T.}~\bibnamefont {Mizoguchi}},
  \ and\ \bibinfo {author} {\bibfnamefont {Y.}~\bibnamefont {Hatsugai}},\
  }\bibfield  {title} {\enquote {\bibinfo {title} {{Mirror skin effect and its
  electric circuit simulation}},}\ }\href@noop {} {\bibfield  {journal}
  {\bibinfo  {journal} {Phys. Rev. Research}\ }\textbf {\bibinfo {volume}
  {2}},\ \bibinfo {pages} {022062} (\bibinfo {year} {2020})}\BibitemShut
  {NoStop}%
\bibitem [{\citenamefont {Scheibner}\ \emph {et~al.}(2020)\citenamefont
  {Scheibner}, \citenamefont {Irvine},\ and\ \citenamefont
  {Vitelli}}]{Scheibner-20}%
  \BibitemOpen
  \bibfield  {author} {\bibinfo {author} {\bibfnamefont {C.}~\bibnamefont
  {Scheibner}}, \bibinfo {author} {\bibfnamefont {W.~T.~M.}\ \bibnamefont
  {Irvine}}, \ and\ \bibinfo {author} {\bibfnamefont {V.}~\bibnamefont
  {Vitelli}},\ }\bibfield  {title} {\enquote {\bibinfo {title} {{Non-Hermitian
  Band Topology and Skin Modes in Active Elastic Media}},}\ }\href@noop {}
  {\bibfield  {journal} {\bibinfo  {journal} {Phys. Rev. Lett.}\ }\textbf
  {\bibinfo {volume} {125}},\ \bibinfo {pages} {118001} (\bibinfo {year}
  {2020})}\BibitemShut {NoStop}%
\bibitem [{\citenamefont {Yokomizo}\ and\ \citenamefont
  {Murakami}(2020)}]{Yokomizo-20}%
  \BibitemOpen
  \bibfield  {author} {\bibinfo {author} {\bibfnamefont {K.}~\bibnamefont
  {Yokomizo}}\ and\ \bibinfo {author} {\bibfnamefont {S.}~\bibnamefont
  {Murakami}},\ }\bibfield  {title} {\enquote {\bibinfo {title} {{Topological
  semimetal phase with exceptional points in one-dimensional non-Hermitian
  systems}},}\ }\href@noop {} {\bibfield  {journal} {\bibinfo  {journal} {Phys.
  Rev. Research}\ }\textbf {\bibinfo {volume} {2}},\ \bibinfo {pages} {043045}
  (\bibinfo {year} {2020})}\BibitemShut {NoStop}%
\bibitem [{\citenamefont {Yi}\ and\ \citenamefont {Yang}()}]{Yi-Yang-20}%
  \BibitemOpen
  \bibfield  {author} {\bibinfo {author} {\bibfnamefont {Y.}~\bibnamefont
  {Yi}}\ and\ \bibinfo {author} {\bibfnamefont {Z.}~\bibnamefont {Yang}},\
  }\href@noop {} {\enquote {\bibinfo {title} {{Non-Hermitian skin modes induced
  by on-site dissipations and chiral tunneling effect}},}\ }\bibinfo {note}
  {{arXiv:2003.02219}}\BibitemShut {NoStop}%
\bibitem [{\citenamefont {Kawabata}\ \emph {et~al.}(2020)\citenamefont
  {Kawabata}, \citenamefont {Okuma},\ and\ \citenamefont {Sato}}]{KOS-20}%
  \BibitemOpen
  \bibfield  {author} {\bibinfo {author} {\bibfnamefont {K.}~\bibnamefont
  {Kawabata}}, \bibinfo {author} {\bibfnamefont {N.}~\bibnamefont {Okuma}}, \
  and\ \bibinfo {author} {\bibfnamefont {M.}~\bibnamefont {Sato}},\ }\bibfield
  {title} {\enquote {\bibinfo {title} {{Non-Bloch band theory of non-Hermitian
  Hamiltonians in the symplectic class}},}\ }\href@noop {} {\bibfield
  {journal} {\bibinfo  {journal} {Phys. Rev. B}\ }\textbf {\bibinfo {volume}
  {101}},\ \bibinfo {pages} {195147} (\bibinfo {year} {2020})}\BibitemShut
  {NoStop}%
\bibitem [{\citenamefont {Terrier}\ and\ \citenamefont
  {Kunst}(2020)}]{Terrier-20}%
  \BibitemOpen
  \bibfield  {author} {\bibinfo {author} {\bibfnamefont {F.}~\bibnamefont
  {Terrier}}\ and\ \bibinfo {author} {\bibfnamefont {F.~K.}\ \bibnamefont
  {Kunst}},\ }\bibfield  {title} {\enquote {\bibinfo {title} {{Dissipative
  analog of four-dimensional quantum Hall physics}},}\ }\href@noop {}
  {\bibfield  {journal} {\bibinfo  {journal} {Phys. Rev. Research}\ }\textbf
  {\bibinfo {volume} {2}},\ \bibinfo {pages} {023364} (\bibinfo {year}
  {2020})}\BibitemShut {NoStop}%
\bibitem [{\citenamefont {Budich}\ and\ \citenamefont
  {Bergholtz}()}]{Budich-20}%
  \BibitemOpen
  \bibfield  {author} {\bibinfo {author} {\bibfnamefont {J.~C.}\ \bibnamefont
  {Budich}}\ and\ \bibinfo {author} {\bibfnamefont {E.~J.}\ \bibnamefont
  {Bergholtz}},\ }\href@noop {} {\enquote {\bibinfo {title} {{Non-Hermitian
  Topological Sensors}},}\ }\bibinfo {note} {{arXiv:2003.13699}}\BibitemShut
  {NoStop}%
\bibitem [{\citenamefont {McDonald}\ and\ \citenamefont
  {Clerk}()}]{McDonald-20}%
  \BibitemOpen
  \bibfield  {author} {\bibinfo {author} {\bibfnamefont {A.}~\bibnamefont
  {McDonald}}\ and\ \bibinfo {author} {\bibfnamefont {A.~A.}\ \bibnamefont
  {Clerk}},\ }\href@noop {} {\enquote {\bibinfo {title}
  {{Exponentially-enhanced quantum sensing with non-Hermitian lattice
  dynamics}},}\ }\bibinfo {note} {{arXiv:2004.00585}}\BibitemShut {NoStop}%
\bibitem [{\citenamefont {Yu}\ \emph {et~al.}()\citenamefont {Yu},
  \citenamefont {Jung},\ and\ \citenamefont {Shvets}}]{Yu-20}%
  \BibitemOpen
  \bibfield  {author} {\bibinfo {author} {\bibfnamefont {Y.}~\bibnamefont
  {Yu}}, \bibinfo {author} {\bibfnamefont {M.}~\bibnamefont {Jung}}, \ and\
  \bibinfo {author} {\bibfnamefont {G.}~\bibnamefont {Shvets}},\ }\href@noop {}
  {\enquote {\bibinfo {title} {{Zero-energy Corner States in a Non-Hermitian
  Quadrupole Insulator}},}\ }\bibinfo {note} {{arXiv:2004.04235}}\BibitemShut
  {NoStop}%
\bibitem [{\citenamefont {Denner}\ \emph {et~al.}()\citenamefont {Denner},
  \citenamefont {Skurativska}, \citenamefont {Schindler}, \citenamefont
  {Fischer}, \citenamefont {Thomale}, \citenamefont {Bzdu\v{s}ek},\ and\
  \citenamefont {Neupert}}]{Denner-20}%
  \BibitemOpen
  \bibfield  {author} {\bibinfo {author} {\bibfnamefont {M.~M.}\ \bibnamefont
  {Denner}}, \bibinfo {author} {\bibfnamefont {A.}~\bibnamefont {Skurativska}},
  \bibinfo {author} {\bibfnamefont {F.}~\bibnamefont {Schindler}}, \bibinfo
  {author} {\bibfnamefont {M.~H.}\ \bibnamefont {Fischer}}, \bibinfo {author}
  {\bibfnamefont {R.}~\bibnamefont {Thomale}}, \bibinfo {author} {\bibfnamefont
  {T.}~\bibnamefont {Bzdu\v{s}ek}}, \ and\ \bibinfo {author} {\bibfnamefont
  {T.}~\bibnamefont {Neupert}},\ }\href@noop {} {\enquote {\bibinfo {title}
  {{Exceptional Topological Insulators}},}\ }\bibinfo {note}
  {{arXiv:2008.01090}}\BibitemShut {NoStop}%
\bibitem [{\citenamefont {Poli}\ \emph {et~al.}(2015)\citenamefont {Poli},
  \citenamefont {Bellec}, \citenamefont {Kuhl}, \citenamefont {Mortessagne},\
  and\ \citenamefont {Schomerus}}]{Poli-15}%
  \BibitemOpen
  \bibfield  {author} {\bibinfo {author} {\bibfnamefont {C.}~\bibnamefont
  {Poli}}, \bibinfo {author} {\bibfnamefont {M.}~\bibnamefont {Bellec}},
  \bibinfo {author} {\bibfnamefont {U.}~\bibnamefont {Kuhl}}, \bibinfo {author}
  {\bibfnamefont {F.}~\bibnamefont {Mortessagne}}, \ and\ \bibinfo {author}
  {\bibfnamefont {H.}~\bibnamefont {Schomerus}},\ }\bibfield  {title} {\enquote
  {\bibinfo {title} {{Selective enhancement of topologically induced interface
  states in a dielectric resonator chain}},}\ }\href@noop {} {\bibfield
  {journal} {\bibinfo  {journal} {Nat. Commun.}\ }\textbf {\bibinfo {volume}
  {6}},\ \bibinfo {pages} {6710} (\bibinfo {year} {2015})}\BibitemShut
  {NoStop}%
\bibitem [{\citenamefont {Zeuner}\ \emph {et~al.}(2015)\citenamefont {Zeuner},
  \citenamefont {Rechtsman}, \citenamefont {Plotnik}, \citenamefont {Lumer},
  \citenamefont {Nolte}, \citenamefont {Rudner}, \citenamefont {Segev},\ and\
  \citenamefont {Szameit}}]{Zeuner-15}%
  \BibitemOpen
  \bibfield  {author} {\bibinfo {author} {\bibfnamefont {J.~M.}\ \bibnamefont
  {Zeuner}}, \bibinfo {author} {\bibfnamefont {M.~C.}\ \bibnamefont
  {Rechtsman}}, \bibinfo {author} {\bibfnamefont {Y.}~\bibnamefont {Plotnik}},
  \bibinfo {author} {\bibfnamefont {Y.}~\bibnamefont {Lumer}}, \bibinfo
  {author} {\bibfnamefont {S.}~\bibnamefont {Nolte}}, \bibinfo {author}
  {\bibfnamefont {M.~S.}\ \bibnamefont {Rudner}}, \bibinfo {author}
  {\bibfnamefont {M.}~\bibnamefont {Segev}}, \ and\ \bibinfo {author}
  {\bibfnamefont {A.}~\bibnamefont {Szameit}},\ }\bibfield  {title} {\enquote
  {\bibinfo {title} {{Observation of a Topological Transition in the Bulk of a
  Non-Hermitian System}},}\ }\href@noop {} {\bibfield  {journal} {\bibinfo
  {journal} {{Phys. Rev. Lett.}}\ }\textbf {\bibinfo {volume} {115}},\ \bibinfo
  {pages} {040402} (\bibinfo {year} {2015})}\BibitemShut {NoStop}%
\bibitem [{\citenamefont {Zhen}\ \emph {et~al.}(2015)\citenamefont {Zhen},
  \citenamefont {Hsu}, \citenamefont {Igarashi}, \citenamefont {Lu},
  \citenamefont {Kaminer}, \citenamefont {Pick}, \citenamefont {Chua},
  \citenamefont {Joannopoulos},\ and\ \citenamefont
  {Solja\u{c}i\'c}}]{Zhen-15}%
  \BibitemOpen
  \bibfield  {author} {\bibinfo {author} {\bibfnamefont {B.}~\bibnamefont
  {Zhen}}, \bibinfo {author} {\bibfnamefont {C.~W.}\ \bibnamefont {Hsu}},
  \bibinfo {author} {\bibfnamefont {Y.}~\bibnamefont {Igarashi}}, \bibinfo
  {author} {\bibfnamefont {L.}~\bibnamefont {Lu}}, \bibinfo {author}
  {\bibfnamefont {I.}~\bibnamefont {Kaminer}}, \bibinfo {author} {\bibfnamefont
  {A.}~\bibnamefont {Pick}}, \bibinfo {author} {\bibfnamefont {S.-L.}\
  \bibnamefont {Chua}}, \bibinfo {author} {\bibfnamefont {J.~D.}\ \bibnamefont
  {Joannopoulos}}, \ and\ \bibinfo {author} {\bibfnamefont {M.}~\bibnamefont
  {Solja\u{c}i\'c}},\ }\bibfield  {title} {\enquote {\bibinfo {title}
  {{Spawning rings of exceptional points out of Dirac cones}},}\ }\href@noop {}
  {\bibfield  {journal} {\bibinfo  {journal} {Nature}\ }\textbf {\bibinfo
  {volume} {525}},\ \bibinfo {pages} {354} (\bibinfo {year}
  {2015})}\BibitemShut {NoStop}%
\bibitem [{\citenamefont {Weimann}\ \emph {et~al.}(2017)\citenamefont
  {Weimann}, \citenamefont {Kremer}, \citenamefont {Plotnik}, \citenamefont
  {Lumer}, \citenamefont {Nolte}, \citenamefont {Makris}, \citenamefont
  {Segev}, \citenamefont {Rechtsman},\ and\ \citenamefont
  {Szameit}}]{Weimann-17}%
  \BibitemOpen
  \bibfield  {author} {\bibinfo {author} {\bibfnamefont {S.}~\bibnamefont
  {Weimann}}, \bibinfo {author} {\bibfnamefont {M.}~\bibnamefont {Kremer}},
  \bibinfo {author} {\bibfnamefont {Y.}~\bibnamefont {Plotnik}}, \bibinfo
  {author} {\bibfnamefont {Y.}~\bibnamefont {Lumer}}, \bibinfo {author}
  {\bibfnamefont {S.}~\bibnamefont {Nolte}}, \bibinfo {author} {\bibfnamefont
  {K.~G.}\ \bibnamefont {Makris}}, \bibinfo {author} {\bibfnamefont
  {M.}~\bibnamefont {Segev}}, \bibinfo {author} {\bibfnamefont {M.~C.}\
  \bibnamefont {Rechtsman}}, \ and\ \bibinfo {author} {\bibfnamefont
  {A.}~\bibnamefont {Szameit}},\ }\bibfield  {title} {\enquote {\bibinfo
  {title} {{Topologically protected bound states in photonic
  parity-time-symmetric crystals}},}\ }\href@noop {} {\bibfield  {journal}
  {\bibinfo  {journal} {Nat. Mater.}\ }\textbf {\bibinfo {volume} {16}},\
  \bibinfo {pages} {433} (\bibinfo {year} {2017})}\BibitemShut {NoStop}%
\bibitem [{\citenamefont {Xiao}\ \emph {et~al.}(2017)\citenamefont {Xiao},
  \citenamefont {Zhan}, \citenamefont {Bian}, \citenamefont {Wang},
  \citenamefont {Zhang}, \citenamefont {Wang}, \citenamefont {Li},
  \citenamefont {Mochizuki}, \citenamefont {Kim}, \citenamefont {Kawakami},
  \citenamefont {Yi}, \citenamefont {Obuse}, \citenamefont {Sanders},\ and\
  \citenamefont {Xue}}]{Xiao-17}%
  \BibitemOpen
  \bibfield  {author} {\bibinfo {author} {\bibfnamefont {L.}~\bibnamefont
  {Xiao}}, \bibinfo {author} {\bibfnamefont {X.}~\bibnamefont {Zhan}}, \bibinfo
  {author} {\bibfnamefont {Z.~H.}\ \bibnamefont {Bian}}, \bibinfo {author}
  {\bibfnamefont {K.~K.}\ \bibnamefont {Wang}}, \bibinfo {author}
  {\bibfnamefont {X.}~\bibnamefont {Zhang}}, \bibinfo {author} {\bibfnamefont
  {X.~P.}\ \bibnamefont {Wang}}, \bibinfo {author} {\bibfnamefont
  {J.}~\bibnamefont {Li}}, \bibinfo {author} {\bibfnamefont {K.}~\bibnamefont
  {Mochizuki}}, \bibinfo {author} {\bibfnamefont {D.}~\bibnamefont {Kim}},
  \bibinfo {author} {\bibfnamefont {N.}~\bibnamefont {Kawakami}}, \bibinfo
  {author} {\bibfnamefont {W.}~\bibnamefont {Yi}}, \bibinfo {author}
  {\bibfnamefont {H.}~\bibnamefont {Obuse}}, \bibinfo {author} {\bibfnamefont
  {B.~C.}\ \bibnamefont {Sanders}}, \ and\ \bibinfo {author} {\bibfnamefont
  {P.}~\bibnamefont {Xue}},\ }\bibfield  {title} {\enquote {\bibinfo {title}
  {{Observation of topological edge states in parity-time-symmetric quantum
  walks}},}\ }\href@noop {} {\bibfield  {journal} {\bibinfo  {journal} {Nat.
  Phys.}\ }\textbf {\bibinfo {volume} {13}},\ \bibinfo {pages} {1117} (\bibinfo
  {year} {2017})}\BibitemShut {NoStop}%
\bibitem [{\citenamefont {St-Jean}\ \emph {et~al.}(2017)\citenamefont
  {St-Jean}, \citenamefont {Goblot}, \citenamefont {Galopin}, \citenamefont
  {Lema\^itre}, \citenamefont {Ozawa}, \citenamefont {Gratiet}, \citenamefont
  {Sagnes}, \citenamefont {Bloch},\ and\ \citenamefont {Amo}}]{St-Jean-17}%
  \BibitemOpen
  \bibfield  {author} {\bibinfo {author} {\bibfnamefont {P.}~\bibnamefont
  {St-Jean}}, \bibinfo {author} {\bibfnamefont {V.}~\bibnamefont {Goblot}},
  \bibinfo {author} {\bibfnamefont {E.}~\bibnamefont {Galopin}}, \bibinfo
  {author} {\bibfnamefont {A.}~\bibnamefont {Lema\^itre}}, \bibinfo {author}
  {\bibfnamefont {T.}~\bibnamefont {Ozawa}}, \bibinfo {author} {\bibfnamefont
  {L.~Le}\ \bibnamefont {Gratiet}}, \bibinfo {author} {\bibfnamefont
  {I.}~\bibnamefont {Sagnes}}, \bibinfo {author} {\bibfnamefont
  {J.}~\bibnamefont {Bloch}}, \ and\ \bibinfo {author} {\bibfnamefont
  {A.}~\bibnamefont {Amo}},\ }\bibfield  {title} {\enquote {\bibinfo {title}
  {{Lasing in topological edge states of a one-dimensional lattice}},}\
  }\href@noop {} {\bibfield  {journal} {\bibinfo  {journal} {Nat. Photon.}\
  }\textbf {\bibinfo {volume} {11}},\ \bibinfo {pages} {651} (\bibinfo {year}
  {2017})}\BibitemShut {NoStop}%
\bibitem [{\citenamefont {Parto}\ \emph {et~al.}(2018)\citenamefont {Parto},
  \citenamefont {Wittek}, \citenamefont {Hodaei}, \citenamefont {Harari},
  \citenamefont {Bandres}, \citenamefont {Ren}, \citenamefont {Rechtsman},
  \citenamefont {Segev}, \citenamefont {Christodoulides},\ and\ \citenamefont
  {Khajavikhan}}]{Parto-17}%
  \BibitemOpen
  \bibfield  {author} {\bibinfo {author} {\bibfnamefont {M.}~\bibnamefont
  {Parto}}, \bibinfo {author} {\bibfnamefont {S.}~\bibnamefont {Wittek}},
  \bibinfo {author} {\bibfnamefont {H.}~\bibnamefont {Hodaei}}, \bibinfo
  {author} {\bibfnamefont {G.}~\bibnamefont {Harari}}, \bibinfo {author}
  {\bibfnamefont {M.~A.}\ \bibnamefont {Bandres}}, \bibinfo {author}
  {\bibfnamefont {J.}~\bibnamefont {Ren}}, \bibinfo {author} {\bibfnamefont
  {M.~C.}\ \bibnamefont {Rechtsman}}, \bibinfo {author} {\bibfnamefont
  {M.}~\bibnamefont {Segev}}, \bibinfo {author} {\bibfnamefont {D.~N.}\
  \bibnamefont {Christodoulides}}, \ and\ \bibinfo {author} {\bibfnamefont
  {M.}~\bibnamefont {Khajavikhan}},\ }\bibfield  {title} {\enquote {\bibinfo
  {title} {{Edge-Mode Lasing in 1D Topological Active Arrays}},}\ }\href@noop
  {} {\bibfield  {journal} {\bibinfo  {journal} {Phys. Rev. Lett.}\ }\textbf
  {\bibinfo {volume} {120}},\ \bibinfo {pages} {113901} (\bibinfo {year}
  {2018})}\BibitemShut {NoStop}%
\bibitem [{\citenamefont {Bahari}\ \emph {et~al.}(2017)\citenamefont {Bahari},
  \citenamefont {Ndao}, \citenamefont {Vallini}, \citenamefont {Amili},
  \citenamefont {Fainman},\ and\ \citenamefont {Kant\'e}}]{Bahari-17}%
  \BibitemOpen
  \bibfield  {author} {\bibinfo {author} {\bibfnamefont {B.}~\bibnamefont
  {Bahari}}, \bibinfo {author} {\bibfnamefont {A.}~\bibnamefont {Ndao}},
  \bibinfo {author} {\bibfnamefont {F.}~\bibnamefont {Vallini}}, \bibinfo
  {author} {\bibfnamefont {A.~El}\ \bibnamefont {Amili}}, \bibinfo {author}
  {\bibfnamefont {Y.}~\bibnamefont {Fainman}}, \ and\ \bibinfo {author}
  {\bibfnamefont {B.}~\bibnamefont {Kant\'e}},\ }\bibfield  {title} {\enquote
  {\bibinfo {title} {{Nonreciprocal lasing in topological cavities of arbitrary
  geometries}},}\ }\href@noop {} {\bibfield  {journal} {\bibinfo  {journal}
  {Science}\ }\textbf {\bibinfo {volume} {358}},\ \bibinfo {pages} {636}
  (\bibinfo {year} {2017})}\BibitemShut {NoStop}%
\bibitem [{\citenamefont {Zhao}\ \emph {et~al.}(2018)\citenamefont {Zhao},
  \citenamefont {Miao}, \citenamefont {Teimourpour}, \citenamefont {Malzard},
  \citenamefont {El-Ganainy}, \citenamefont {Schomerus},\ and\ \citenamefont
  {Feng}}]{Zhao-18}%
  \BibitemOpen
  \bibfield  {author} {\bibinfo {author} {\bibfnamefont {H.}~\bibnamefont
  {Zhao}}, \bibinfo {author} {\bibfnamefont {P.}~\bibnamefont {Miao}}, \bibinfo
  {author} {\bibfnamefont {M.~H.}\ \bibnamefont {Teimourpour}}, \bibinfo
  {author} {\bibfnamefont {S.}~\bibnamefont {Malzard}}, \bibinfo {author}
  {\bibfnamefont {R.}~\bibnamefont {El-Ganainy}}, \bibinfo {author}
  {\bibfnamefont {H.}~\bibnamefont {Schomerus}}, \ and\ \bibinfo {author}
  {\bibfnamefont {L.}~\bibnamefont {Feng}},\ }\bibfield  {title} {\enquote
  {\bibinfo {title} {{Topological hybrid silicon microlasers}},}\ }\href@noop
  {} {\bibfield  {journal} {\bibinfo  {journal} {Nat. Commun.}\ }\textbf
  {\bibinfo {volume} {9}},\ \bibinfo {pages} {981} (\bibinfo {year}
  {2018})}\BibitemShut {NoStop}%
\bibitem [{\citenamefont {Zhou}\ \emph {et~al.}(2018)\citenamefont {Zhou},
  \citenamefont {Peng}, \citenamefont {Yoon}, \citenamefont {Hsu},
  \citenamefont {Nelson}, \citenamefont {Fu}, \citenamefont {Joannopoulos},
  \citenamefont {Solja\u{c}i\'c},\ and\ \citenamefont {Zhen}}]{Zhou-18}%
  \BibitemOpen
  \bibfield  {author} {\bibinfo {author} {\bibfnamefont {H.}~\bibnamefont
  {Zhou}}, \bibinfo {author} {\bibfnamefont {C.}~\bibnamefont {Peng}}, \bibinfo
  {author} {\bibfnamefont {Y.}~\bibnamefont {Yoon}}, \bibinfo {author}
  {\bibfnamefont {C.~W.}\ \bibnamefont {Hsu}}, \bibinfo {author} {\bibfnamefont
  {K.~A.}\ \bibnamefont {Nelson}}, \bibinfo {author} {\bibfnamefont
  {L.}~\bibnamefont {Fu}}, \bibinfo {author} {\bibfnamefont {J.~D.}\
  \bibnamefont {Joannopoulos}}, \bibinfo {author} {\bibfnamefont
  {M.}~\bibnamefont {Solja\u{c}i\'c}}, \ and\ \bibinfo {author} {\bibfnamefont
  {B.}~\bibnamefont {Zhen}},\ }\bibfield  {title} {\enquote {\bibinfo {title}
  {{Observation of bulk Fermi arc and polarization half charge from paired
  exceptional points}},}\ }\href@noop {} {\bibfield  {journal} {\bibinfo
  {journal} {Science}\ }\textbf {\bibinfo {volume} {359}},\ \bibinfo {pages}
  {1009} (\bibinfo {year} {2018})}\BibitemShut {NoStop}%
\bibitem [{\citenamefont {Harari}\ \emph {et~al.}(2018)\citenamefont {Harari},
  \citenamefont {Bandres}, \citenamefont {Lumer}, \citenamefont {Rechtsman},
  \citenamefont {Chong}, \citenamefont {Khajavikhan}, \citenamefont
  {Christodoulides},\ and\ \citenamefont {Segev}}]{Harari-18}%
  \BibitemOpen
  \bibfield  {author} {\bibinfo {author} {\bibfnamefont {G.}~\bibnamefont
  {Harari}}, \bibinfo {author} {\bibfnamefont {M.~A.}\ \bibnamefont {Bandres}},
  \bibinfo {author} {\bibfnamefont {Y.}~\bibnamefont {Lumer}}, \bibinfo
  {author} {\bibfnamefont {M.~C.}\ \bibnamefont {Rechtsman}}, \bibinfo {author}
  {\bibfnamefont {Y.~D.}\ \bibnamefont {Chong}}, \bibinfo {author}
  {\bibfnamefont {M.}~\bibnamefont {Khajavikhan}}, \bibinfo {author}
  {\bibfnamefont {D.~N.}\ \bibnamefont {Christodoulides}}, \ and\ \bibinfo
  {author} {\bibfnamefont {M.}~\bibnamefont {Segev}},\ }\bibfield  {title}
  {\enquote {\bibinfo {title} {{Topological insulator laser: Theory}},}\
  }\href@noop {} {\bibfield  {journal} {\bibinfo  {journal} {Science}\ }\textbf
  {\bibinfo {volume} {359}},\ \bibinfo {pages} {eaar4003} (\bibinfo {year}
  {2018})}\BibitemShut {NoStop}%
\bibitem [{\citenamefont {Bandres}\ \emph {et~al.}(2018)\citenamefont
  {Bandres}, \citenamefont {Wittek}, \citenamefont {Harari}, \citenamefont
  {Parto}, \citenamefont {Ren}, \citenamefont {Segev}, \citenamefont
  {Christodoulides},\ and\ \citenamefont {Khajavikhan}}]{Bandres-18}%
  \BibitemOpen
  \bibfield  {author} {\bibinfo {author} {\bibfnamefont {M.~A.}\ \bibnamefont
  {Bandres}}, \bibinfo {author} {\bibfnamefont {S.}~\bibnamefont {Wittek}},
  \bibinfo {author} {\bibfnamefont {G.}~\bibnamefont {Harari}}, \bibinfo
  {author} {\bibfnamefont {M.}~\bibnamefont {Parto}}, \bibinfo {author}
  {\bibfnamefont {J.}~\bibnamefont {Ren}}, \bibinfo {author} {\bibfnamefont
  {M.}~\bibnamefont {Segev}}, \bibinfo {author} {\bibfnamefont
  {D.}~\bibnamefont {Christodoulides}}, \ and\ \bibinfo {author} {\bibfnamefont
  {M.}~\bibnamefont {Khajavikhan}},\ }\bibfield  {title} {\enquote {\bibinfo
  {title} {{Topological insulator laser: Experiments}},}\ }\href@noop {}
  {\bibfield  {journal} {\bibinfo  {journal} {{Science}}\ }\textbf {\bibinfo
  {volume} {359}},\ \bibinfo {pages} {eaar4005} (\bibinfo {year}
  {2018})}\BibitemShut {NoStop}%
\bibitem [{\citenamefont {Cerjan}\ \emph {et~al.}(2019)\citenamefont {Cerjan},
  \citenamefont {Huang}, \citenamefont {Chen}, \citenamefont {Chong},\ and\
  \citenamefont {Rechtsman}}]{Cerjan-19}%
  \BibitemOpen
  \bibfield  {author} {\bibinfo {author} {\bibfnamefont {A.}~\bibnamefont
  {Cerjan}}, \bibinfo {author} {\bibfnamefont {S.}~\bibnamefont {Huang}},
  \bibinfo {author} {\bibfnamefont {K.~P.}\ \bibnamefont {Chen}}, \bibinfo
  {author} {\bibfnamefont {Y.}~\bibnamefont {Chong}}, \ and\ \bibinfo {author}
  {\bibfnamefont {M.~C.}\ \bibnamefont {Rechtsman}},\ }\bibfield  {title}
  {\enquote {\bibinfo {title} {{Experimental realization of a Weyl exceptional
  ring}},}\ }\href@noop {} {\bibfield  {journal} {\bibinfo  {journal} {Nat.
  Photon.}\ }\textbf {\bibinfo {volume} {13}},\ \bibinfo {pages} {623}
  (\bibinfo {year} {2019})}\BibitemShut {NoStop}%
\bibitem [{\citenamefont {Zhao}\ \emph {et~al.}(2019)\citenamefont {Zhao},
  \citenamefont {Qiao}, \citenamefont {Wu}, \citenamefont {Midya},
  \citenamefont {Longhi},\ and\ \citenamefont {Feng}}]{Zhao-19}%
  \BibitemOpen
  \bibfield  {author} {\bibinfo {author} {\bibfnamefont {H.}~\bibnamefont
  {Zhao}}, \bibinfo {author} {\bibfnamefont {X.}~\bibnamefont {Qiao}}, \bibinfo
  {author} {\bibfnamefont {T.}~\bibnamefont {Wu}}, \bibinfo {author}
  {\bibfnamefont {B.}~\bibnamefont {Midya}}, \bibinfo {author} {\bibfnamefont
  {S.}~\bibnamefont {Longhi}}, \ and\ \bibinfo {author} {\bibfnamefont
  {L.}~\bibnamefont {Feng}},\ }\bibfield  {title} {\enquote {\bibinfo {title}
  {{Non-Hermitian topological light steering}},}\ }\href@noop {} {\bibfield
  {journal} {\bibinfo  {journal} {Science}\ }\textbf {\bibinfo {volume}
  {365}},\ \bibinfo {pages} {1163} (\bibinfo {year} {2019})}\BibitemShut
  {NoStop}%
\bibitem [{\citenamefont {Brandenbourger}\ \emph {et~al.}(2019)\citenamefont
  {Brandenbourger}, \citenamefont {Locsin}, \citenamefont {Lerner},\ and\
  \citenamefont {Coulais}}]{Brandenbourger-19-skin-exp}%
  \BibitemOpen
  \bibfield  {author} {\bibinfo {author} {\bibfnamefont {M.}~\bibnamefont
  {Brandenbourger}}, \bibinfo {author} {\bibfnamefont {X.}~\bibnamefont
  {Locsin}}, \bibinfo {author} {\bibfnamefont {E.}~\bibnamefont {Lerner}}, \
  and\ \bibinfo {author} {\bibfnamefont {C.}~\bibnamefont {Coulais}},\
  }\bibfield  {title} {\enquote {\bibinfo {title} {{Non-reciprocal robotic
  metamaterials}},}\ }\href@noop {} {\bibfield  {journal} {\bibinfo  {journal}
  {Nat. Commun.}\ }\textbf {\bibinfo {volume} {10}},\ \bibinfo {pages} {4608}
  (\bibinfo {year} {2019})}\BibitemShut {NoStop}%
\bibitem [{\citenamefont {Ghatak}\ \emph {et~al.}()\citenamefont {Ghatak},
  \citenamefont {Brandenbourger}, \citenamefont {van Wezel},\ and\
  \citenamefont {Coulais}}]{Ghatak-19-skin-exp}%
  \BibitemOpen
  \bibfield  {author} {\bibinfo {author} {\bibfnamefont {A.}~\bibnamefont
  {Ghatak}}, \bibinfo {author} {\bibfnamefont {M.}~\bibnamefont
  {Brandenbourger}}, \bibinfo {author} {\bibfnamefont {J.}~\bibnamefont {van
  Wezel}}, \ and\ \bibinfo {author} {\bibfnamefont {C.}~\bibnamefont
  {Coulais}},\ }\href@noop {} {\enquote {\bibinfo {title} {{Observation of
  non-Hermitian topology and its bulk-edge correspondence}},}\ }\bibinfo {note}
  {{arXiv:1907.11619}}\BibitemShut {NoStop}%
\bibitem [{\citenamefont {Helbig}\ \emph {et~al.}(2020)\citenamefont {Helbig},
  \citenamefont {Hofmann}, \citenamefont {Imhof}, \citenamefont {Abdelghany},
  \citenamefont {Kiessling}, \citenamefont {Molenkamp}, \citenamefont {Lee},
  \citenamefont {Szameit}, \citenamefont {Greiter},\ and\ \citenamefont
  {Thomale}}]{Helbig-19-skin-exp}%
  \BibitemOpen
  \bibfield  {author} {\bibinfo {author} {\bibfnamefont {T.}~\bibnamefont
  {Helbig}}, \bibinfo {author} {\bibfnamefont {T.}~\bibnamefont {Hofmann}},
  \bibinfo {author} {\bibfnamefont {S.}~\bibnamefont {Imhof}}, \bibinfo
  {author} {\bibfnamefont {M.}~\bibnamefont {Abdelghany}}, \bibinfo {author}
  {\bibfnamefont {T.}~\bibnamefont {Kiessling}}, \bibinfo {author}
  {\bibfnamefont {L.~W.}\ \bibnamefont {Molenkamp}}, \bibinfo {author}
  {\bibfnamefont {C.~H.}\ \bibnamefont {Lee}}, \bibinfo {author} {\bibfnamefont
  {A.}~\bibnamefont {Szameit}}, \bibinfo {author} {\bibfnamefont
  {M.}~\bibnamefont {Greiter}}, \ and\ \bibinfo {author} {\bibfnamefont
  {R.}~\bibnamefont {Thomale}},\ }\bibfield  {title} {\enquote {\bibinfo
  {title} {{Generalized bulk-boundary correspondence in non-Hermitian
  topolectrical circuits}},}\ }\href@noop {} {\bibfield  {journal} {\bibinfo
  {journal} {Nat. Phys.}\ }\textbf {\bibinfo {volume} {16}},\ \bibinfo {pages}
  {747} (\bibinfo {year} {2020})}\BibitemShut {NoStop}%
\bibitem [{\citenamefont {Hofmann}\ \emph {et~al.}(2020)\citenamefont
  {Hofmann}, \citenamefont {Helbig}, \citenamefont {Schindler}, \citenamefont
  {Salgo}, \citenamefont {Brzezi\'nska}, \citenamefont {Greiter}, \citenamefont
  {Kiessling}, \citenamefont {Wolf}, \citenamefont {Vollhardt}, \citenamefont
  {Kaba\v{s}i}, \citenamefont {Lee}, \citenamefont {Bilu\v{s}i\'c},
  \citenamefont {Thomale},\ and\ \citenamefont
  {Neupert}}]{Hofmann-19-skin-exp}%
  \BibitemOpen
  \bibfield  {author} {\bibinfo {author} {\bibfnamefont {T.}~\bibnamefont
  {Hofmann}}, \bibinfo {author} {\bibfnamefont {T.}~\bibnamefont {Helbig}},
  \bibinfo {author} {\bibfnamefont {F.}~\bibnamefont {Schindler}}, \bibinfo
  {author} {\bibfnamefont {N.}~\bibnamefont {Salgo}}, \bibinfo {author}
  {\bibfnamefont {M.}~\bibnamefont {Brzezi\'nska}}, \bibinfo {author}
  {\bibfnamefont {M.}~\bibnamefont {Greiter}}, \bibinfo {author} {\bibfnamefont
  {T.}~\bibnamefont {Kiessling}}, \bibinfo {author} {\bibfnamefont
  {D.}~\bibnamefont {Wolf}}, \bibinfo {author} {\bibfnamefont {A.}~\bibnamefont
  {Vollhardt}}, \bibinfo {author} {\bibfnamefont {A.}~\bibnamefont
  {Kaba\v{s}i}}, \bibinfo {author} {\bibfnamefont {C.~H.}\ \bibnamefont {Lee}},
  \bibinfo {author} {\bibfnamefont {A.}~\bibnamefont {Bilu\v{s}i\'c}}, \bibinfo
  {author} {\bibfnamefont {R.}~\bibnamefont {Thomale}}, \ and\ \bibinfo
  {author} {\bibfnamefont {T.}~\bibnamefont {Neupert}},\ }\bibfield  {title}
  {\enquote {\bibinfo {title} {{Reciprocal skin effect and its realization in a
  topolectrical circuit}},}\ }\href@noop {} {\bibfield  {journal} {\bibinfo
  {journal} {Phys. Rev. Research}\ }\textbf {\bibinfo {volume} {2}},\ \bibinfo
  {pages} {023265} (\bibinfo {year} {2020})}\BibitemShut {NoStop}%
\bibitem [{\citenamefont {Xiao}\ \emph {et~al.}(2020)\citenamefont {Xiao},
  \citenamefont {Deng}, \citenamefont {Wang}, \citenamefont {Zhu},
  \citenamefont {Wang}, \citenamefont {Yi},\ and\ \citenamefont
  {Xue}}]{Xiao-19-skin-exp}%
  \BibitemOpen
  \bibfield  {author} {\bibinfo {author} {\bibfnamefont {L.}~\bibnamefont
  {Xiao}}, \bibinfo {author} {\bibfnamefont {T.}~\bibnamefont {Deng}}, \bibinfo
  {author} {\bibfnamefont {K.}~\bibnamefont {Wang}}, \bibinfo {author}
  {\bibfnamefont {G.}~\bibnamefont {Zhu}}, \bibinfo {author} {\bibfnamefont
  {Z.}~\bibnamefont {Wang}}, \bibinfo {author} {\bibfnamefont {W.}~\bibnamefont
  {Yi}}, \ and\ \bibinfo {author} {\bibfnamefont {P.}~\bibnamefont {Xue}},\
  }\bibfield  {title} {\enquote {\bibinfo {title} {{Non-Hermitian bulk-boundary
  correspondence in quantum dynamics}},}\ }\href@noop {} {\bibfield  {journal}
  {\bibinfo  {journal} {Nat. Phys.}\ }\textbf {\bibinfo {volume} {16}},\
  \bibinfo {pages} {761} (\bibinfo {year} {2020})}\BibitemShut {NoStop}%
\bibitem [{\citenamefont {Weidemann}\ \emph {et~al.}(2020)\citenamefont
  {Weidemann}, \citenamefont {Kremer}, \citenamefont {Helbig}, \citenamefont
  {Hofmann}, \citenamefont {Stegmaier}, \citenamefont {Greiter}, \citenamefont
  {Thomale},\ and\ \citenamefont {Szameit}}]{Weidemann-20-skin-exp}%
  \BibitemOpen
  \bibfield  {author} {\bibinfo {author} {\bibfnamefont {S.}~\bibnamefont
  {Weidemann}}, \bibinfo {author} {\bibfnamefont {M.}~\bibnamefont {Kremer}},
  \bibinfo {author} {\bibfnamefont {T.}~\bibnamefont {Helbig}}, \bibinfo
  {author} {\bibfnamefont {T.}~\bibnamefont {Hofmann}}, \bibinfo {author}
  {\bibfnamefont {A.}~\bibnamefont {Stegmaier}}, \bibinfo {author}
  {\bibfnamefont {M.}~\bibnamefont {Greiter}}, \bibinfo {author} {\bibfnamefont
  {R.}~\bibnamefont {Thomale}}, \ and\ \bibinfo {author} {\bibfnamefont
  {A.}~\bibnamefont {Szameit}},\ }\bibfield  {title} {\enquote {\bibinfo
  {title} {{Topological funneling of light}},}\ }\href@noop {} {\bibfield
  {journal} {\bibinfo  {journal} {Science}\ }\textbf {\bibinfo {volume}
  {368}},\ \bibinfo {pages} {311} (\bibinfo {year} {2020})}\BibitemShut
  {NoStop}%
\bibitem [{\citenamefont {Hatano}\ and\ \citenamefont
  {Nelson}(1996)}]{Hatano-Nelson-96}%
  \BibitemOpen
  \bibfield  {author} {\bibinfo {author} {\bibfnamefont {N.}~\bibnamefont
  {Hatano}}\ and\ \bibinfo {author} {\bibfnamefont {D.~R.}\ \bibnamefont
  {Nelson}},\ }\bibfield  {title} {\enquote {\bibinfo {title} {{Localization
  Transitions in Non-Hermitian Quantum Mechanics}},}\ }\href@noop {} {\bibfield
   {journal} {\bibinfo  {journal} {Phys. Rev. Lett.}\ }\textbf {\bibinfo
  {volume} {77}},\ \bibinfo {pages} {570} (\bibinfo {year} {1996})}\BibitemShut
  {NoStop}%
\bibitem [{\citenamefont {Hatano}\ and\ \citenamefont
  {Nelson}(1997)}]{Hatano-Nelson-97}%
  \BibitemOpen
  \bibfield  {author} {\bibinfo {author} {\bibfnamefont {N.}~\bibnamefont
  {Hatano}}\ and\ \bibinfo {author} {\bibfnamefont {D.~R.}\ \bibnamefont
  {Nelson}},\ }\bibfield  {title} {\enquote {\bibinfo {title} {{Vortex pinning
  and non-Hermitian quantum mechanics}},}\ }\href@noop {} {\bibfield  {journal}
  {\bibinfo  {journal} {Phys. Rev. B}\ }\textbf {\bibinfo {volume} {56}},\
  \bibinfo {pages} {8651} (\bibinfo {year} {1997})}\BibitemShut {NoStop}%
\bibitem [{\citenamefont {Wess}\ and\ \citenamefont {Zumino}(1971)}]{WZ-71}%
  \BibitemOpen
  \bibfield  {author} {\bibinfo {author} {\bibfnamefont {J.}~\bibnamefont
  {Wess}}\ and\ \bibinfo {author} {\bibfnamefont {B.}~\bibnamefont {Zumino}},\
  }\bibfield  {title} {\enquote {\bibinfo {title} {{Consequences of anomalous
  ward identities}},}\ }\href@noop {} {\bibfield  {journal} {\bibinfo
  {journal} {Phys. Lett. B}\ }\textbf {\bibinfo {volume} {37}},\ \bibinfo
  {pages} {95} (\bibinfo {year} {1971})}\BibitemShut {NoStop}%
\bibitem [{\citenamefont {Vanderbilt}(2018)}]{Vanderbilt-textbook}%
  \BibitemOpen
  \bibfield  {author} {\bibinfo {author} {\bibfnamefont {D.}~\bibnamefont
  {Vanderbilt}},\ }\href@noop {} {\emph {\bibinfo {title} {{Berry Phases in
  Electronic Structure Theory: Electric Polarization, Orbital Magnetization and
  Topological Insulators}}}}\ (\bibinfo  {publisher} {Cambridge University
  Press, Cambridge},\ \bibinfo {year} {2018})\BibitemShut {NoStop}%
\bibitem [{\citenamefont {Teo}\ and\ \citenamefont {Kane}(2010)}]{Teo-Kane-10}%
  \BibitemOpen
  \bibfield  {author} {\bibinfo {author} {\bibfnamefont {J.~C.~Y.}\
  \bibnamefont {Teo}}\ and\ \bibinfo {author} {\bibfnamefont {C.~L.}\
  \bibnamefont {Kane}},\ }\bibfield  {title} {\enquote {\bibinfo {title}
  {{Topological defects and gapless modes in insulators and
  superconductors}},}\ }\href@noop {} {\bibfield  {journal} {\bibinfo
  {journal} {Phys. Rev. B}\ }\textbf {\bibinfo {volume} {82}},\ \bibinfo
  {pages} {115120} (\bibinfo {year} {2010})}\BibitemShut {NoStop}%
\bibitem [{\citenamefont {Tiwari}\ \emph {et~al.}()\citenamefont {Tiwari},
  \citenamefont {Jahin},\ and\ \citenamefont {Wang}}]{Tiwari-20}%
  \BibitemOpen
  \bibfield  {author} {\bibinfo {author} {\bibfnamefont {A.}~\bibnamefont
  {Tiwari}}, \bibinfo {author} {\bibfnamefont {A.}~\bibnamefont {Jahin}}, \
  and\ \bibinfo {author} {\bibfnamefont {Y.}~\bibnamefont {Wang}},\ }\href@noop
  {} {\enquote {\bibinfo {title} {{Chiral Dirac Superconductors: Second-order
  and Boundary-obstructed Topology}},}\ }\bibinfo {note}
  {{arXiv:2005.12291}}\BibitemShut {NoStop}%
\bibitem [{\citenamefont {Okugawa}\ \emph {et~al.}()\citenamefont {Okugawa},
  \citenamefont {Takahashi},\ and\ \citenamefont
  {Yokomizo}}]{Okugawa-Takahashi-Yokomizo-20}%
  \BibitemOpen
  \bibfield  {author} {\bibinfo {author} {\bibfnamefont {R.}~\bibnamefont
  {Okugawa}}, \bibinfo {author} {\bibfnamefont {R.}~\bibnamefont {Takahashi}},
  \ and\ \bibinfo {author} {\bibfnamefont {K.}~\bibnamefont {Yokomizo}},\
  }\href@noop {} {\enquote {\bibinfo {title} {{Second-order topological
  non-Hermitian skin effects}},}\ }\bibinfo {note}
  {{arXiv:2008.03721}}\BibitemShut {NoStop}%
\end{thebibliography}%

\end{document}